%% file: main_arxiv.tex
\documentclass[longauth]{aa}

\renewcommand\linenumbers{}

\usepackage{graphicx}
\usepackage{comment}
\usepackage[encapsulated]{CJK}
\usepackage{ucs}
\newcommand{\orcid}[1]{\protect\href{https://orcid.org/#1}{\protect\includegraphics[width=8pt]{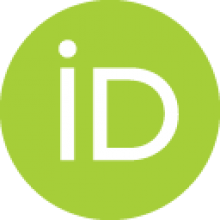}}}

\usepackage[utf8]{inputenc}
\newcommand{\cntext}[1]{\begin{CJK}{UTF8}{gbsn}#1\end{CJK}}

\usepackage[varg]{txfonts}
\usepackage{amsmath,mathtools,mathrsfs}
\usepackage{gensymb}
\usepackage[colorlinks, allcolors=blue, breaklinks=true]{hyperref}
\addto\extrasenglish{

}

\usepackage{placeins}

\usepackage{verbatim}
\usepackage{color}
\usepackage{xspace}
\usepackage{longtable}
\usepackage{multirow}
\usepackage{bm}
\usepackage{amsmath}
\usepackage{tabularx}
\usepackage{bm}
\usepackage{xcolor}
\usepackage{adjustbox}
 
\def\muas{\mu{\rm as}\xspace} 
\def\uas{$\mu$as\xspace}

\def\dd{\mathrm{d}}
\def\m87{M87*}

\def\comrade{\texttt{Comrade}\xspace}
\def\doghit{\texttt{DoG-HIT}\xspace}
\def\moead{\texttt{MOEA/D}\xspace}
\def\ehtim{\texttt{ehtim}\xspace}
\def\difmap{\texttt{Difmap}\xspace}
\def\gpcal{\texttt{GPCAL}\xspace}
\def\lpcal{\texttt{LPCAL}\xspace}
\def\themis{\textsc{Themis}\xspace}
\def\mnet{|m_{\rm net}|\xspace}
\def\mavg{\langle | m | \rangle\xspace}

\newcommand{\norm}[1]{\lVert #1 \rVert}
\newcommand{\avg}[1]{\langle #1 \rangle}

\defcitealias{EHTMWL2017}{M87 MWL2017}
\defcitealias{EHTMWL2018}{M87 MWL2018}

\definecolor{lpcalcolor}{rgb}{0.0, 0.62, 0.451}
\definecolor{gpcalcolor}{rgb}{0.8, 0.475, 0.655}
\definecolor{ehtimcolor}{rgb}{0.941, 0.894, 0.259}
\definecolor{moeadcolor}{rgb}{0.835, 0.369, 0.0}
\definecolor{doghitcolor}{rgb}{0.902, 0.624, 0.0}
\definecolor{comradecolor}{rgb}{0.337, 0.706, 0.914}
\definecolor{themiscolor}{rgb}{0.0, 0.447, 0.698}

\makeatletter

\begin{document}

\nolinenumbers

\title{Horizon-scale variability of \m87 from 2017--2021 EHT observations}

\include{authorlist}

\abstract{
We report three epochs of polarized images of \m87 at 230\,GHz using data from the Event Horizon Telescope (EHT) taken in 2017, 2018, and 2021. The  baseline coverage of the 2021 observations is significantly improved  through the addition of two new EHT stations: the 12\,m Kitt Peak Telescope and the Northern Extended Millimetre Array (NOEMA). All observations result in images dominated by a bright, asymmetric ring with a persistent diameter of $43.9\pm 0.6\,\muas$, consistent with expectations for lensed synchrotron emission encircling the apparent shadow of a supermassive black hole. We find that the total intensity and linear polarization of \m87 vary significantly across the three epochs. Specifically, the azimuthal brightness distribution of the total intensity images varies  from year to year, as expected for a stochastic accretion flow. However, despite a gamma-ray flare erupting in M87 quasi-contemporaneously to the 2018 observations, the 2018 and 2021 images look remarkably similar. The resolved linear polarization fractions in 2018 and 2021 peak at $\sim$5\%, compared to $\sim$15\% in 2017. The spiral polarization pattern on the ring also varies from year to year, including a change in the electric vector position angle helicity in 2021 that could reflect changes in the magnetized accretion flow or an external Faraday screen. The improved 2021 coverage also provides the first EHT constraints on jet emission outside the ring, on scales of $\lesssim 1$\,mas. Overall, these observations provide strong proof of the reliability of the EHT images and probe the dynamic properties of the horizon-scale accretion flow surrounding \m87.
}


\maketitle

\section{Introduction} \label{sec:intro}

For a long time following its initial discovery, the giant elliptical galaxy M87 remained merely an entry in astronomical catalogues \citep{Messier1781}. More than a century later, observations at the Lick Observatory led to the discovery of a `curious straight ray' superimposed on the diffuse emission of the galaxy \citep{Curtis1918}, which decades later was identified as the relativistic jet emanating from the region close to the central supermassive black hole (SMBH). With the advent of radio astronomy and the growing scientific interest in active galactic nuclei (AGNs), M87 became a prime target for observations across the electromagnetic spectrum during the 20th century (see e.g. \citealt{EHTMWL2017} [hereafter \citetalias{EHTMWL2017}], \citealt{EHTMWL2018} [hereafter \citetalias{EHTMWL2018}], and \citealt{Hada2024A&ARv..32....5H} for a review).

At the core of M87 lies the SMBH \m87, and its radio properties have been studied for decades across various frequencies \citep[e.g.][]{Reid1989, Junor1999, Doeleman2012, Hada2016, Mertens2016, Walker2018, Kim2018M87, Lu2023}. In 2019, the Event Horizon Telescope Collaboration (EHTC) produced the first total intensity image of \m87's shadow, using data collected during its initial observing campaign in 2017 \citep{M87p1,M87p2,M87p3,M87p4,M87p5,M87p6}. This was followed by imaging of the linearly polarized emission \citep{M87p7,M87p8} and an analysis of the circular polarization near the event horizon \citep{M87p9}.

The total intensity image of \m87 revealed a ring of $({42\pm3)\,\muas}$ diameter that is brighter in the south \citep{M87p4, M87p6}. Using results from theoretical simulations of \m87's accretion \citep{M87p5}, it was determined that the ring size of \m87 corresponds to a central black hole with a mass of ${(6.5\pm 0.7)\times 10^9 M_\odot}$. These horizon-scale mass measurements are consistent with the mass inferred on much larger scales from stellar velocity dispersion measurements \citep{Gebhardt2009ApJ...700.1690G,Gebhardt2011ApJ...729..119G,M87p6,LiepoldM87mass,SimonM87mass}.

Follow-up observations by the Event Horizon Telescope (EHT) in 2018 \citep{M87_2018p1} verified the ring size, measuring a diameter of $42^{+2}_{-3}\,\muas,$ and confirmed the original interpretation of the ring being due to lensed emission around a SMBH. However, while the ring diameter was stable, the azimuthal structure of the ring evolved significantly. Namely, the angle of the peak brightness shifted by $30\degr{}$ anti-clockwise in 2018. This rotation is consistent with expectations from numerical simulations of \m87 \citep{M87p5}, which show temporal variation in the angle of peak brightness because of intrinsic variability in the accretion flow \citep{M87_2018p2}. Analysis of observations from 2009-2013 with prototype EHT arrays also indicated that the structure of \m87 was consistent with a stable ring with the peak brightness position angle varying from year to year \citep{Wielgus2020}.

Further evidence of the emission seen from \m87 being due to a hot magnetized accretion flow was provided by the linear polarization maps produced by the EHT in 2021 \citep{M87p7,M87p8}. The inner ring was found to be linearly polarized. Most of the linear polarization was concentrated in the south-western portion of the ring in 2017, 
with a polarization fraction reaching $\sim$15\% \citep{M87p7}. The observed polarization fraction is consistent with simulations in which the Faraday rotation internal to the emission region causes the de-polarization of synchrotron radiation \citep{M87p8}.

To probe the magnetic field structure of the ring, the EHT reconstructed its electric vector position angle (EVPA) pattern, observing a largely azimuthal but slightly twisted structure. This pattern is consistent with semi-analytical models that have a strong poloidal magnetic field component \citep{Narayan2021, M87p8} and, ignoring Faraday effects, is predicted for magnetically dominated accretion flows \citep{Chael23}. However, considerable uncertainty remains about internal and external Faraday rotation in \m87, which has been studied at much larger scales using Atacama Large Millimeter/submillimeter Array (ALMA) measurements \citep{Goddi2021}. 

General relativistic magneto-hydrodynamic (GRMHD) simulations predict that the polarized emission should be dynamic around \m87, on timescales as short as weeks. The first hints of variable polarimetric properties were detected by \citet{M87p7}. 
While the observed EVPA was largely stable, mild fluctuations in the fractional polarization were detected. Furthermore, the peak of the polarized emission shifted by $\sim$\,25$\degr{}$ anti-clockwise 
between the images taken on April 5 and 11, 2017. Comparing these polarimetric properties averaged over the four observation days in 2017, \citet{M87p8} were able to constrain various GRMHD models. From this analysis, it was found that weakly magnetized accretion models performed worse than magnetically dominated ones. Assuming \m87 is similar to the magnetically arrested disk (MAD) models used in \citet{M87p8}, it is predicted that the polarization fraction should remain approximately stable for prograde MAD simulations and increase for retrograde systems. Furthermore, ignoring the influence of external Faraday effects, \citet{Chael23} note that the distribution of specific azimuthal polarization modes from \citet{Palumbo2020} may depend on \m87's black hole spin for the preferred models from \citet{M87p8}. Therefore, studying the stability of \m87's EVPA pattern across multiple years will provide valuable insights into the nature of its accretion flow and the central black hole's properties.

This work presents the first 230\,GHz multi-epoch study of \m87's polarimetric variability. We analyse the polarimetric properties of \m87 in 2018 and the total intensity and polarimetric properties of \m87 in 2021. We compare these new results with the properties of \m87 in 2017 to better understand its polarimetric variability. In Sect. \ref{sec:calibration} we provide a brief overview of the 2021 EHT observation campaign and its data properties. Section \ref{sec:imaging} provides the basics of polarimetric very long baseline interferometry (VLBI) imaging and details the different calibration and imaging pipelines used in this work. Section \ref{sec:data_properties} describes the interferometric data properties in 2017, 2018, and 2021. In Sect. \ref{sec:synthetic} we validate the polarimetric imaging pipelines used in this work, demonstrating their reliability for the different array configurations in 2017, 2018, and 2021. The first polarimetric images of \m87 in 2018 and 2021, along with their multi-epoch properties, are presented in Sect. \ref{sec:results}. Additionally, taking Northern Extended Millimetre Array (NOEMA) and Kitt Peak (KP) data from  2021 into account, we discuss the detection of non-trivial closure phases on scales between $200\,\muas$ and $1\,{\rm mas}$, providing the first measurements of \m87's extended structure at 230\,GHz. Finally, the interpretation of the results is presented in Sect. \ref{sec:discussion}, and our conclusions are given in Sect. \ref{sec:summary}.

\begin{figure}[!t]
    \centering
    \includegraphics[width=\columnwidth]{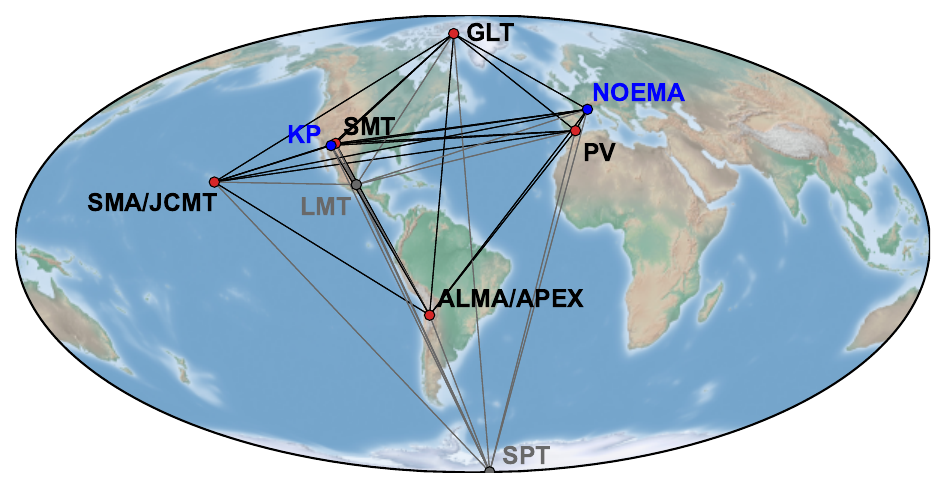}
    \caption{EHT in its 2021 configuration. Compared to the original 2017 array, GLT was added in 2018, and KP and NOEMA joined the EHT for the 2021 campaign (indicated in blue). Baselines from SPT and LMT are greyed out since SPT cannot observe \m87 (only its calibrator 3C\,279), and LMT did not observe in 2021.}
    \label{fig:array}
\end{figure}

 \begin{figure*}[!t]
    \centering
    \includegraphics[width=\linewidth]{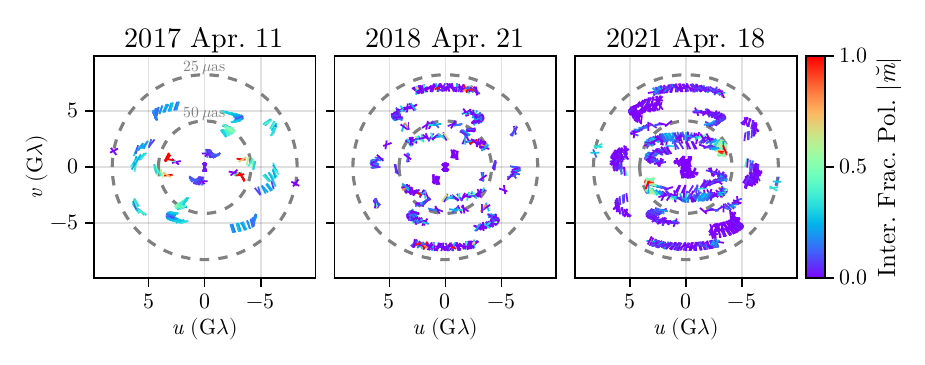}
    \caption{$(u,v)$ coverage of \m87 during the 2017 (left), 2018 (middle), and 2021 (right) campaigns for the band 3 (227.1\,GHz) observations. The ticks show each year's interferometric EVPA and the colour of the observed interferometric fractional linear polarization, after de-biasing for thermal noise and applying leakage corrections. The leakage terms used are the fiducial values from \citet{M87p7} in 2017 and the \themis leakage solutions in 2018 and 2021. The clusters of high polarization fractions in 2018 come from the GLT, which was underperforming in 2018 due to an incomplete commissioning at that time, as described in \citet{eht_memo_koay_2023-l1-02}.}
    \label{fig:polarmetric_cvg}
\end{figure*}

\section{Observations} \label{sec:calibration}

The new \m87 data described and analysed in full polarization in this work were collected as part of the April 9 -- 19 EHT 2021 observing campaign on April 13 and 18. Furthermore, we analysed the \m87 data from April 21, 2018, in full polarization. Previously, these 2018 data were analysed only in total intensity (Stokes $\mathcal{I}$) in \citet{M87_2018p1}.
The 2018 \m87 observations were carried out by ALMA, the Atacama Pathfinder Experiment (APEX), the Greenland Telescope (GLT), the IRAM 30\,m Telescope at Pico Veleta (PV), the \textit{James Clerk Maxwell }Telescope (JCMT), the \textit{Alfonso Serrano} Large Millimeter Telescope  (LMT), the Submillimeter Array (SMA), and the Submillimeter Telescope (SMT).

The 12\,m Kitt Peak Telescope and NOEMA\footnote{After an upgrade from the six-element Plateau de Bure Interferometer, phased NOEMA joined regular VLBI observations with twelve 15 m dishes in 2021. The impact of NOEMA in VLBI observations at 3\,mm from the same year is described in \citet{2023Kim}.} joined EHT observations for the first time in 2021. The LMT did not participate in 2021 but resumed regular EHT observations in 2022.
The 2021 array is displayed in Fig. \ref{fig:array}. The April 18 observations are the focus of this analysis, as they include NOEMA, which did not participate in the April 13 observations due to bad weather. The April 13 data are used for consistency checks, and results are presented in Appendix \ref{appendix:april13_2021}. 64 VLBI scans were recorded on \m87 between 19:20 UT, April 17 and 11:25 UT, April 18 in 2021. Similarly, we have 64 \m87 scans between 19:40 UT, April 12 and 11:40 UT, April 13. In each scan, we integrated for five minutes on the source.

In 2018, the EHT data recording rate was upgraded from 32 gigabits per second (Gbps) to 64\,Gbps, except for the GLT, which used a 32\,Gbps rate in its inaugural observations \citep{2023GLT}. As of 2021, the GLT also uses four Mark-6 units, and thus the new EHT data described in this work marks the first 64 Gbps observations with the GLT. Furthermore, due to photogrammetry and panel adjustments, the GLT 230\,GHz aperture efficiency increased from 22\,\% to 66\,\% after the EHT 2018 observations \citep{2020Koay, eht_memo_koay_2023-l1-02, 2023GLT}. In this work we utilized the upper sideband, and the two new lower sideband frequencies will be analysed in future work. 2021 is also the first EHT observations where the JCMT observed in dual polarization thanks to the new 2 sideband Namakanui receiver \citep{2020JCMT, 2021JCMT}.
The station sensitivities and metadata used for the flux density calibration of the 2021 data are described in \citet{eht_memo_romerocanizales_2025-l1-01}.

The baseband data from both receiver sidebands recorded by each telescope are correlated over four frequency bands centred on 213.1, 215.1, 227.1, and 229.1\,GHz; each with a bandwidth of 1875\,MHz, which we refer to as bands 1–4, respectively. For the 2017 observations, the 227.1\,GHz and 229.1\,GHz bands were referred to as low and high bands, respectively. To accommodate different frequency recording setups at different stations while maintaining a fixed visibility frequency grid with 32 sub-bands, each with 116 channels per band, the \texttt{DiFX} \citep{Deller2007, Deller2011} software was used for correlation with the \textit{outputbands} mode\footnote{\url{https://zenodo.org/record/4319257/files/DiFX_Outputbands_24nov2020.pdf}.}.
This work analyses the two upper sideband frequency bands: the low band (band 3) around 227.1\, GHz and the high band (band 4) around 229.1\,GHz.
Following a linear to circular \texttt{PolConvert} \citep{MartiVidal2016} process, we formed visibilities in a full-polarization circular feed basis: $RR^*, RL^*, LR^*$, and $LL^*$. Combinations of these four correlation products can be used to form the four Stokes parameters, as explained in the next section.
The raw post-correlation visibilities after \texttt{PolConvert} are made publicly available as \textit{L1} releases on the \href{https://eventhorizontelescope.org/for-astronomers/data}{EHT website}\footnote{\url{https://eventhorizontelescope.org/for-astronomers/data}}.

\section{Methods}
\label{sec:imaging}

This study made use of two calibration pipelines for cross verification: \texttt{rPICARD} \citep{2019Janssenproc, Janssen2019} and \texttt{EHT-HOPS} \citep{Blackburn2019}.
Additionally, we employed seven imaging algorithms to ensure the robustness of the results: two CLEAN-based imaging algorithms, \gpcal \citep{Park2021} and \lpcal \citep{Shepherd_1997,M87p4}; three regularized maximum likelihood (RML) algorithms, \doghit \citep{Mueller2022}, \ehtim \citep[\texttt{eht-imaging};][]{Chael2016,Chael2018,M87p4}, and \moead \citep{Mueller2023c,Mus2024a}; and two Bayesian methods, \comrade \citep{Tiede2022} and \themis \citep{Broderick2020}.

\subsection{Overview}
The pairwise correlations between the electric field measurements from two idealized dual-polarized circular feeds ($R_jR_k^*$, $R_jL_k^*$, $L_jR_k^*$, and $L_jL_k^*$) are related to the Stokes visibilities ($\tilde{\mathcal{I}}$, $\tilde{\mathcal{Q}}$, $\tilde{\mathcal{U}}$, and $\tilde{\mathcal{V}}$) through linear algebraic transformations. For a baseline between two stations {\it j} and {\it k}, the true-source coherency matrix can be expressed as\begin{equation}\label{eq:rvis}
\rho_{jk} = \begin{pmatrix} 
\tilde{\mathcal{I}}_{jk} + \tilde{\mathcal{V}}_{jk} & \tilde{\mathcal{Q}}_{jk} + i\tilde{\mathcal{U}}_{jk} \\ 
\tilde{\mathcal{Q}}_{jk} - i\tilde{\mathcal{U}}_{jk} & \tilde{\mathcal{I}}_{jk} - \tilde{\mathcal{V}}_{jk} 
\end{pmatrix} = \begin{pmatrix} 
R_jR_k^* & R_jL_k^* \\ 
L_jR_k^* & L_jL_k^* 
\end{pmatrix}
.\end{equation}

Inverting this equation yields the Stokes visibilities:
\begin{equation}
\begin{pmatrix} 
\tilde{\mathcal{I}}_{jk} \\ \tilde{\mathcal{Q}}_{jk} \\ \tilde{\mathcal{U}}_{jk} \\ \tilde{\mathcal{V}}_{jk} 
\end{pmatrix} = \frac{1}{2} \begin{pmatrix} 
R_jR_k^* + L_jL_k^* \\
R_jL_k^* + L_jR_k^* \\ 
-i(R_jL_k^* - L_jR_k^*) \\
R_jR_k^* - L_jL_k^*
\end{pmatrix}.
\end{equation}
Unfortunately, this simple relation no longer holds exactly for realistic interferometers with atmospheric and instrumental effects.
We used the radio interferometric measurement equation (RIME) formalism to incorporate instrumental and atmospheric effects. The RIME formalism provides a mathematical framework that relates observed visibilities to the true sky brightness distribution while accounting for instrumental and propagation effects \citep{Hamaker1996, Smirnov2011}. The basic form of RIME for a quasi-monochromatic signal is expressed as 
\begin{equation}
\rho^\prime_{jk} = \bm{J}_j \rho_{jk} \bm{J}_k^\dagger
,\end{equation}
where $\rho^\prime_{jk}$ represents the measured visibility or coherency matrix, which is a $2\times2$ matrix that encapsulates four correlations between the voltage signals received from two stations, {\it j} and {\it k}, using dual-polarized feeds, $\rho_{jk}$ is the true-source visibility or coherency matrix that describes the inherent brightness of the source, $\bm{J}$ is the Jones matrix that characterizes the linear transformations that an incoming signal undergoes due to propagation and instrumental effects \citep{Jones1941}, and the $\dagger$ symbol represents the conjugate transpose. Different instrumental and propagation effects are represented by distinct Jones matrices, which are multiplied in a sequence \citep[called a Jones chain;][]{Smirnov2011} that reflects the physical order of these effects along the signal path.

For EHT polarimetric data, the Jones matrix formalism incorporates effects particularly critical for polarization calibration \citep{Thompson2017}. After instrumental calibration and post-processing described above, we parameterized our Jones matrices by 
\begin{equation}\label{eq:jones_decomp}
\bm{J}_{j} = (\bm{\Phi}^{-1}\bm{G} \bm{D} \bm{\Phi})_j.
\end{equation}
Here, $\bm{G}$ is the time-dependent residual instrumental gains matrix, $\bm{\Phi}_j$ is the instrumental feed rotation matrix, $\phi(t)$ is the feed-rotation angle and $\bm{D}$ is the constant instrumental polarization leakage matrix: 
\begin{align}
    \bm{G}_j &= g_{j,R}(t)\cdot {\rm diag}\left[1, \,g_{j,L}(t)/g_{j,R}(t)\right],\label{eqn:gain} \\ 
    \bm{\Phi}_j &= {\rm diag}\left[e^{-i\phi_{i}(t)}, \,e^{i\phi_{i}(t)}\right],\\
    \bm{D}_j &= \begin{pmatrix}
        1 & d_{j, R}\\
        d_{j, L} & 1
    \end{pmatrix}.\label{eqn:leakage]}
\end{align}
In the following, we refer to $g_{j,R}(t)$ as the station-based, time-variable gains, and $g_{j,L}(t)/g_{j,R}(t)$ as the station-based, time-variable gain ratios. The feed rotation angle at site $j$ depends on the source elevation $f_{j, {\rm el}}$, the parallactic angle $f_{j, {\rm par}}$, and an offset $\phi_{j, {\rm offset}}$,  
\begin{equation} 
    \phi_j(t)=f_{j, {\rm el}}\theta_{j, {\rm el}}(t)+f_{j, {\rm par}}\psi_{j, {\rm par}}(t)+\phi_{j, {\rm off}}.
\end{equation}

We analysed the conjugate closure trace products, $\mathscr{C}$ \citep{Broderick2020} for each observation to assess the presence of polarized emission in the data. Conjugate closure trace products are independent of any time-dependent stationized instrumental effects that can be represented as a Jones matrix, including residual instrumental gain errors (Eq. \ref{eqn:gain}) and instrumental polarization leakage errors (Eq. \ref{eqn:leakage]}) and are defined on quadrangles of four stations, ${j}, {k}, {l}$, and ${m}$, as
\begin{equation}
     \mathscr{C}_{jklm} =  \mathcal{T}_{jklm}\mathcal{T}_{jmlk} \label{eqn:closure trace},
\end{equation}  
where $\mathcal{T}_{jklm} = \dfrac{1}{2} \mathrm{tr} \left(\rho_{{jk}}\rho^{-1}_{{lk}}\rho_{{lm}}\rho^{-1}_{{jm}}\right)$ is the closure trace. If no polarized emission is present, $\arg(\mathscr{C})$ will be zero.

\begin{table*}[!t]
\caption{Overview of the imaging methods.}
\centering
\adjustbox{max width=\textwidth}{%
\begin{tabular}{c | c c c c c c c c c c} 
 \hline
\hline
  {Method} & {Image} & \multicolumn{3}{c}{{Data Products}} & {Selfcal} & {Ext.} & {Stokes}  & {Gain} & {D-term}  & {Notes} \\
 \cline{3-5}
  & {model} & $\bm{\mathcal{I}}$ & $\bm{\mathcal{Q,U}}$ & $\bm{\mathcal{V}}$ & & {structure} & {condition} & {ratio} &{estimation} & \\ 
 \hline\hline
 CLEAN\\
  \hline
  \color{lpcalcolor}\lpcal & CLEAN & $RR^*+LL^*$ & $RL^*,LR^*$& &  Stokes & Renorm.  & None  & 1 & Stokes & CLEAN \\
  \difmap & components & & & & $\mathcal{I}$  & intrasites$^\dagger$ & & & $\mathcal{I}$ &  \\ 
   \hline
  \color{gpcalcolor}\gpcal & CLEAN & $RR^*+LL^*$ & $RL^*,LR^*$& &  Stokes & Renorm.  & None  & 1 & Stokes & CLEAN \\
  \difmap & components & & & & $\mathcal{I}$  & intrasites$^\dagger$ & & & $\mathcal{I}$ &  \\ 
  \hline\hline
  RML\\
  \hline
 \color{ehtimcolor}\ehtim & Pixels & CPh, LCA, $|V|$ & $\tilde{\mathcal{P}},m$ & $\tilde{\mathcal{V}} $ &  Stokes & Fix ZBL flux 
 & $\mathcal{Q}^2+\mathcal{U}^2+\mathcal{V}^2 $ & 1 & Iter-& \citet{M87p4} \\
  &  &  &  &  &  $\mathcal{I}$ & by modelling & $\leq  \mathcal{I}^2$& & atively&  reg. weights \\ 
   \hline
\color{moeadcolor}\moead & Pixels & CPh, LCA & $\tilde{\mathcal{P}}$ & $\tilde{\mathcal{V}}$ &  Stokes & Fix ZBL flux & $\mathcal{Q}^2+\mathcal{U}^2+\mathcal{V}^2 $ & 1 & Iter- & Computes \\
 &  &  &  &  &  $\mathcal{I}$ & by modelling & $\leq  \mathcal{I}^2$& & atively &  Pareto front \\ 
  \hline
 \color{doghitcolor}\doghit & Wavelets & CPh, LCA & $\tilde{\mathcal{P}}$ & $\tilde{\mathcal{V}}$ &  Stokes & Fix ZBL flux  & Same  & 1 & Iter-& Compressive \\
  &  &  &  &  &  $\mathcal{I}$ & from amp. & support & & atively &  sensing \\ 
 \hline\hline
 Bayesian\\
 \hline
 \color{comradecolor} \comrade & Fixed & \multicolumn{3}{c}{$RR^*,RL^*,LR^*,LL^*$} &  Simul- & Flags  & $\mathcal{Q}^2+\mathcal{U}^2+\mathcal{V}^2 $  & Any & Simul-& Estimates \\
  & splined raster &  & & & taneous  & intrasites$^\dagger$ & $\leq  \mathcal{I}^2$ & & taneous &  posterior \\ 
  \hline
   \color{themiscolor} \themis & Adaptive & \multicolumn{3}{c}{$RR^*,RL^*,LR^*,LL^*$} &  Simul- & Asymmetric  & $\mathcal{Q}^2+\mathcal{U}^2+\mathcal{V}^2 $  & 1 & Simul-& Estimates \\
  & splined raster &  & & & taneous  & Gaussian & $\leq  \mathcal{I}^2$ & & taneous &  posterior \\ 
\hline
\end{tabular}}
\tablefoot{
$^\dagger$: Intra-sites are defined as the ALMA-APEX, JCMT-SMA, and KP-SMT baselines.\\
ZBL: zero baseline
}
\label{tab:methods}
\end{table*}

Image reconstruction refers to inferring from the measured coherency matrices/visibilities, the set of Stokes parameter maps $\mathcal{I}(\boldsymbol{x})$, $\mathcal{Q}(\boldsymbol{x})$, $\mathcal{U}(\boldsymbol{x})$, $\mathcal{V}(\boldsymbol{x})$, which fully characterize the polarized state of electromagnetic radiation at a given spatial coordinate {\it \textbf{x}=(x, y)}. $\mathcal{I}(\boldsymbol{x})$ gives total intensity, $\mathcal{Q}(\boldsymbol{x})$ measures the difference between horizontal and vertical linear polarization, $\mathcal{U}(\boldsymbol{x})$ quantifies the difference between light polarized at $45^{\circ}$ and $-45^{\circ}$, and $\mathcal{V}(\boldsymbol{x})$ represents the level of circular polarization. The Stokes parameters are related to the Stokes visibilities through the Fourier transform \citep{vancitter,ZERNIKE1938785},
\begin{equation}
    \tilde{\mathcal{I}}_{jk} = \int \mathcal{I}(\bm{x}) e^{2\pi i \bm{u}_{jk} \cdot \bm{x}}\dd x\dd y,
\end{equation}
where $\bm{u}_{jk}$ is the projected baseline between stations {\it j} and {\it k}. In addition, we also defined the total intensity closure phases (CPh), $\psi_{jkl}$, and log-closure amplitudes (LCA), $\mathcal{A}_{jklm}$, which are insensitive to overall gain corruptions \citep[see e.g.][]{RogersCP, blackburn2020}:
\begin{align}
    \psi_{jkl} &= \arg\left[\tilde{\mathcal{I}}'_{jk}\tilde{\mathcal{I}}'_{kl}\tilde{\mathcal{I}}'_{lj}\right]\\
    \mathcal{A}_{jklm} &= \log\left[\frac{
        |\tilde{\mathcal{I}}'_{jk}|\,
        |\tilde{\mathcal{I}}'_{lm}|}{
        |\tilde{\mathcal{I}}'_{jl}|\,
        |\tilde{\mathcal{I}}'_{km}|}\right],
\end{align}
where $\tilde{\mathcal{I}}'_{jk} = \dfrac{1}{2}\,{\rm tr}(\rho'_{jk})$ the approximate Stokes I visibility for baseline $j,\,k$.

For polarization, the complex linear polarization is defined as
\begin{equation}
\mathcal{P} \equiv \mathcal{Q} + i\mathcal{U} = \mathcal{I}|m|e^{2i\chi} ,\label{eqn:evpa_polfraction}
\end{equation}
where $m = (\mathcal{Q} + i\mathcal{U})/\mathcal{I}$ is the linear polarization fraction, and $\chi = 0.5 \arg(\mathcal{P})$ is the EVPA measured east of north on the sky. The circular polarization fraction is given by ${v=\mathcal{V}/\mathcal{I}}$. In the visibility domain, we can define similar interferometric quantities \citep{Johnson2015}, 
\begin{alignat}{3}
&\breve{m} &&= \frac{\tilde{\mathcal{Q}} + i\tilde{\mathcal{U}}}{\tilde{\mathcal{I}}} &&= \frac{2RL^*}{RR^*+LL^*}, \\
&\breve{\chi} && = \dfrac{1}{2}\arg \breve{m} \\
&\breve{v} &&= \frac{\tilde{\mathcal{V}}}{\tilde{\mathcal{I}}} &&= \frac{RR^*-LL^*}{RR^*+LL^*}.
\end{alignat}
Note that $\breve{m},\;\breve{v}$ are not the Fourier transforms of $m,\;v$.

The unresolved (image-integrated) linear and circular polarization fractions, as well as their resolved (image-averaged) counterparts, with $\sum_i$ indicating a sum over all image pixels, are given by
\begin{align}
|m|_{\rm net} &= \frac{1}{\sum_i\mathcal{I}_i}\left[\left(\sum_i\mathcal{Q}_i\right)^2+\left(\sum_i\mathcal{U}_i\right)^2\right]^{1/2},\\
v_{\rm net} &= \frac{1}{\sum_i\mathcal{I}_i}\sum_i\mathcal{V}_i,\\
\langle|m|\rangle &= \frac{1}{\sum_i\mathcal{I}_i}\sum_i\left(\mathcal{Q}_i^2+\mathcal{U}_i^2\right)^{1/2},\\
\langle|v|\rangle &= \frac{1}{\sum_i\mathcal{I}_i} \sum_i\left(\left|\frac{\mathcal{V}_i}{\mathcal{I}_i}\right|\mathcal{I}_i\right).
\end{align}
Note that $\langle|m|\rangle$ and $\langle|v|\rangle$ are sensitive to image resolution, i.e. the restoring beam size, while $|m|_{\rm net}$ and $v_{\rm net}$ remain unaffected by convolution. 

Another useful parameter for quantifying polarization structures and comparing polarimetric images is the complex coefficient of the azimuthal mode decomposition of $\mathcal{P}$ given by \citep{Palumbo2020} 
\begin{equation}\label{eq:beta}
\beta_m = \frac{1}{\textit{I}_{\rm ann}}\int_{\rho_{\rm min}}^{\rho_{\rm max}}\int_{0}^{2\pi}\mathcal{P}(\rho, \varphi)e^{-im\varphi} \rho \,d\varphi \,d\rho,
\end{equation}
where ($\rho$, $\varphi$) represent the polar coordinates on the image plane, {\it $I_{\rm ann}$} is the Stokes $\mathcal{I}$ ﬂux density contained within the annulus between the minimum and maximum radii $\rho_{\rm min,\,max}$. In this paper, $\rho_{\rm min}$ is set to zero, while $\rho_{\rm max}$ is set to $45\,\muas$ to focus on the compact core emission. Note that we defined the EVPA helicity of the ring's polarization pattern as the sign of $\angle \beta_2$ following conventions from \citet{Palumbo2020} and \citet{Chael23}. In addition to $|m|_{\rm net}$, $|v|_{\rm net}$, $\langle|m|\rangle$, and $\langle|v|\rangle$, the amplitude and phase of the second coefficient, $|\beta_2|$ and $\angle \beta_2$ are useful parameters to score the GRMHD accretion models against the EHT data. This parameter is useful for distinguishing between accretion states \citep{Palumbo2020} and will be used in a companion paper focusing on the theoretical interpretation of our results (Chael et al., in prep.).

\subsection{Calibration and post-processing}
Two pipelines are used for the post-correlation signal stabilization and data reduction \citep{2022Janssen}:  \texttt{rPICARD} described in \citet{2019Janssenproc, Janssen2019} and \texttt{EHT-HOPS} described in \citet{Blackburn2019}. The updated data processing steps needed to address unique data issues encountered in the 2021 observations are described in \autoref{appendix:data}.

\subsection{Polarized imaging}
In this study seven different polarized imaging algorithms from three different frameworks were used. Here, we present a concise overview of the methods (see also \autoref{tab:methods} for a summary) and provide a more complete description of the methods in \autoref{appendix:imaging-methods}.

For the two CLEAN-based pipelines, the first involved automated imaging with \difmap{} and polarimetric calibration procedures using \gpcal{} (see Appendix \ref{section:difmap_gpcal} for more details) and collectively will be denoted as \gpcal{}. The second method involved another CLEAN procedure with \difmap{} but using different hyperparameter settings and instrumental polarization calibration using \lpcal{} (see Appendix \ref{section:difmap_lpcal} for more details) and collectively will be denoted as \lpcal{}. Both methods achieved the final total intensity images through iterative CLEAN and self-calibration. The primary difference between these methods lies in their polarized imaging approach: \gpcal{} employed an automated parameter survey to produce a set of images, as seen in previous EHT imaging studies of M87 \citep{M87p4, M87_2018p1}, and selected the representative image based on closure chi-squares. Similarly, \lpcal{} involved imaging involving a small survey over different hyperparameters where the representative image was chosen by minimizing the closure chi-square. However, the specific hyperparameter survey and the overall imaging procedure differed from \gpcal. This allowed us to test the sensitivity of the CLEAN reconstructions to different assumptions.

For leakage calibration, \gpcal{} derived initial leakage solutions using the `similarity approximation', which assumes that the linear polarization structure is proportional to the total intensity structure within each sub-model. The solutions were then refined through iterative linear polarization imaging, leakage solution estimation, and correction. \lpcal{} used the standard CALIB and LPCAL AIPS tasks to estimate the leakages and apply the corrections to the data, and then used \difmap to make the final $\mathcal{I}$, $\mathcal{Q}$, and $\mathcal{U}$ maps. While both \gpcal and \lpcal used CLEAN to obtain final polarized images, their differing assumptions in deriving polarimetric leakages introduce a measure of uncertainty into the CLEAN polarimetric reconstructions. Finally, for both the \gpcal and \lpcal pipelines, the fiducial images were created by taking the final set of clean components and blurring them with a $20\,\muas$ Gaussian beam similar to the nominal resolution of the EHT array (see Fig. \ref{fig:polarmetric_cvg}).

The RML framework minimizes a weighted sum of multiple objectives, data fidelity functionals (loss functionals, $\chi^2$), and regularization terms, $R$:
\begin{equation}
    \hat{I} \in \mathrm{argmin}_I \sum_d \alpha_d \chi^2_d(I)+\sum_x \beta_x R_x(I). \label{eq:rml}
\end{equation}
The common minimization of data terms and regularization terms ensures a solution that matches observed data and is favourable with respect to the hand-crafted regularization terms. This framework has been realized in two different methods: \ehtim \citep{Chael2016, Chael2018} and \doghit \citep{Mueller2022,Mueller2023b}. \ehtim has been used in all previous EHT studies on \m87 \citep{M87p4, M87p7, M87p9, M87_2018p1} and Sgr\,A* \citep{SgrA.p3, SgrA.p7}, as well as for imaging various AGN sources \citep{Kim2020A&A...640A..69K,Janssen2021NatAs...5.1017J,Issaoun2022ApJ...934..145I,Jorstad2023ApJ...943..170J}. \doghit has also been applied to Sgr\,A* \citep{SgrA.p7} and lower frequency \m87 data \citep{kim_imaging_2025}. This paper is the first time \moead was applied to one of the primary EHT targets. The three RML approaches differ in the regularizer terms used, the optimality concept applied, and the minimization procedure, but use similar calibration procedures. For each RML pipeline, the representative image is the optimal image according to its loss function. For more information, see Appendix \ref{app:rml}.

To select the regularizer hyperparameters for \ehtim, we used a combination that was consistent with the top set results from \citet{M87p4, M87_2018p1} for total intensity and \citet{M87p7} for linear polarization. In Appendix \ref{app:rml} we verify that this combination performs well on synthetic data. However, unlike previous EHT publications \citep{M87p4, M87_2018p1}, no attempt was made to create the `top set' of hyperparameters within the \ehtim pipeline. A reduced parameter search has been performed as spot check instead, as described in Appendix \ref{app:rml}. Given that there are seven different imaging algorithms, including two Bayesian frameworks, there was sufficient diversity in our imaging algorithms to explore model uncertainty. Furthermore, we found that the uncertainty we reported from the 2017 image reconstructions (see Sect. \ref{sec:results}) was consistent with those previously inferred using the `top set' approach in \citet{M87p4}.

Finally, two Bayesian polarized imaging pipelines are used: \themis and \comrade. Both methods jointly solve for all four Stokes parameters and instrumental terms, such as gains and leakage corrections. Both methods used a rasterized image in all four Stokes parameters, but their priors differed to test the robustness of \m87's image. For both methods, every station in the array assumed gains that vary independently for each scan and frequency band. Leakages were assumed to be constant for each observing day, but could differ across frequency bands. Finally, the gain ratios were handled differently in \comrade and \themis. For each site in the array, \comrade fit an independent gain ratio for each scan and frequency band, while \themis enforced that the gain ratios were unity. For both Bayesian methods, the representative image was given by the mean image computed from their respective posterior samples.

\begin{figure*}[!t]
    \centering
    \includegraphics[height=5.79cm]{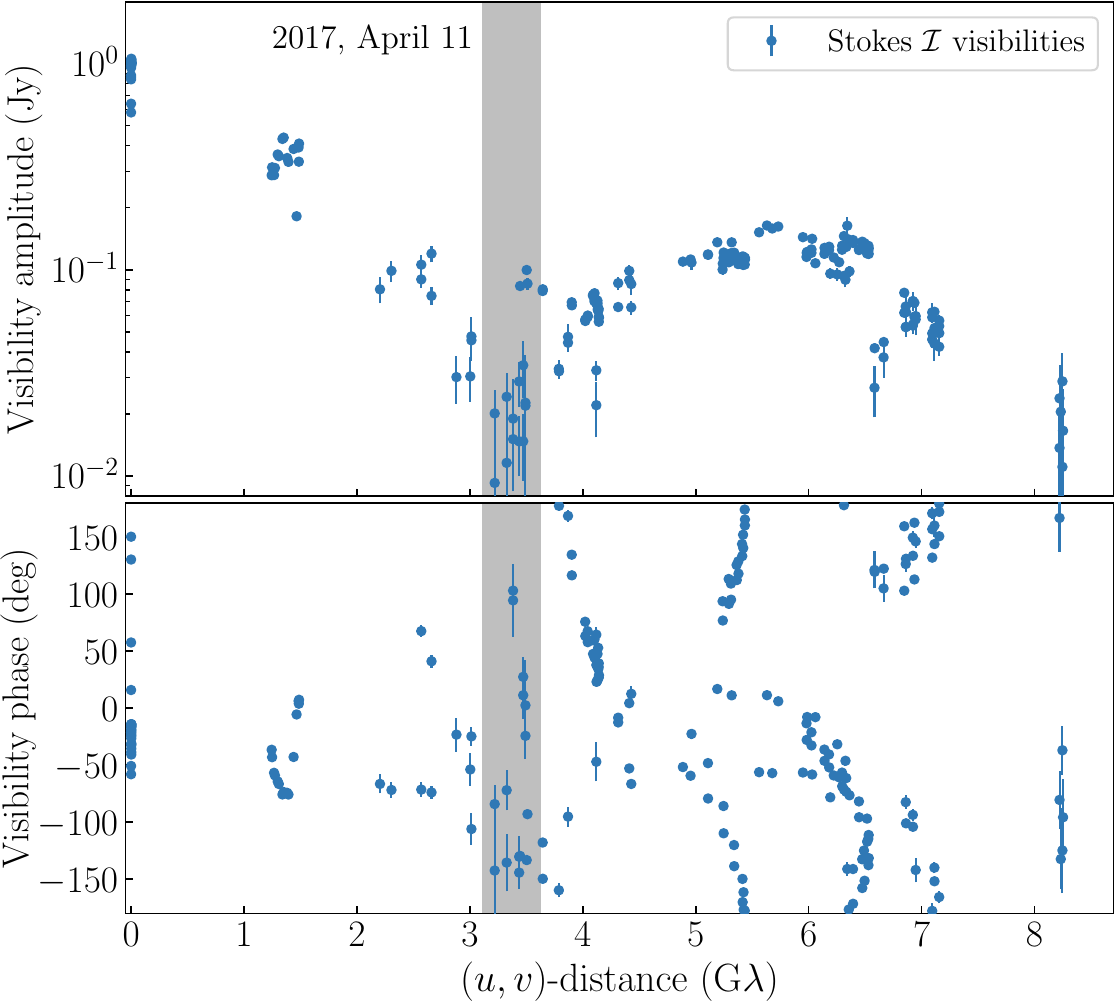} \hspace{-0.205cm}
    \includegraphics[height=5.79cm]{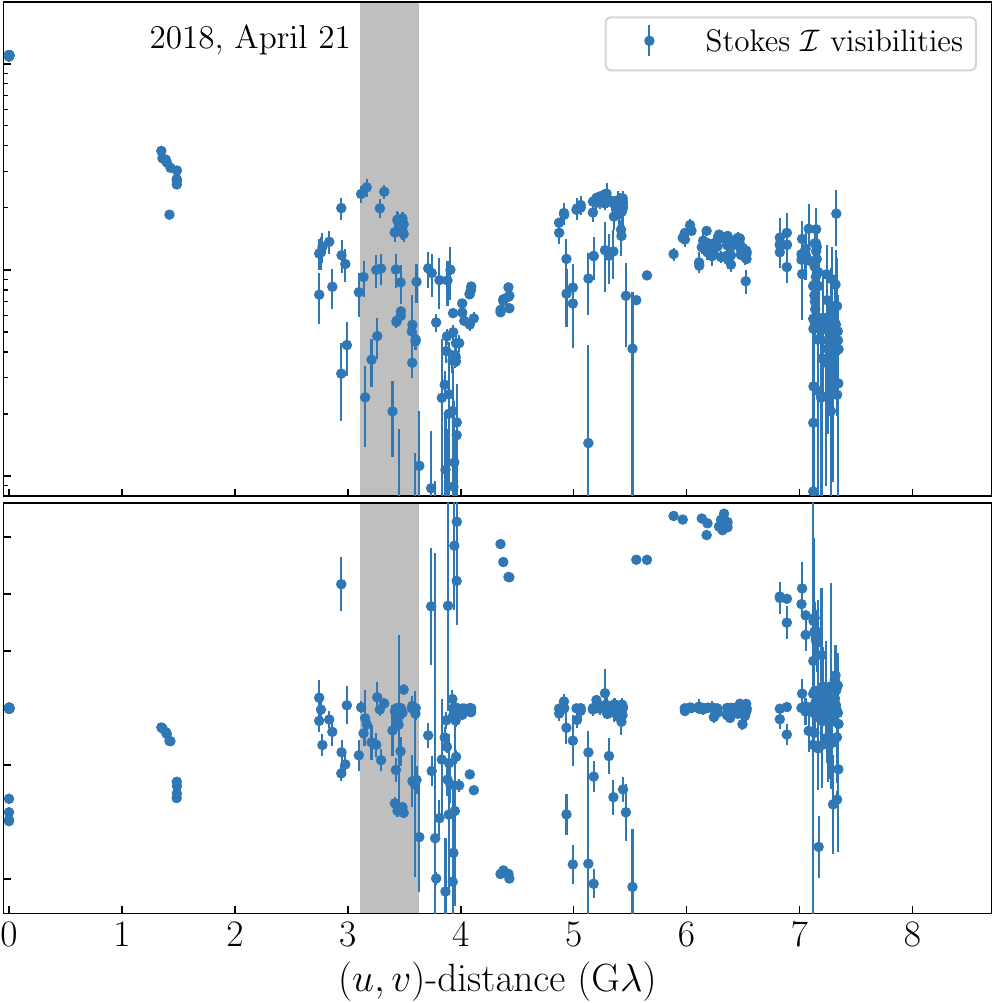} \hspace{-0.215cm}
    \includegraphics[height=5.79cm]{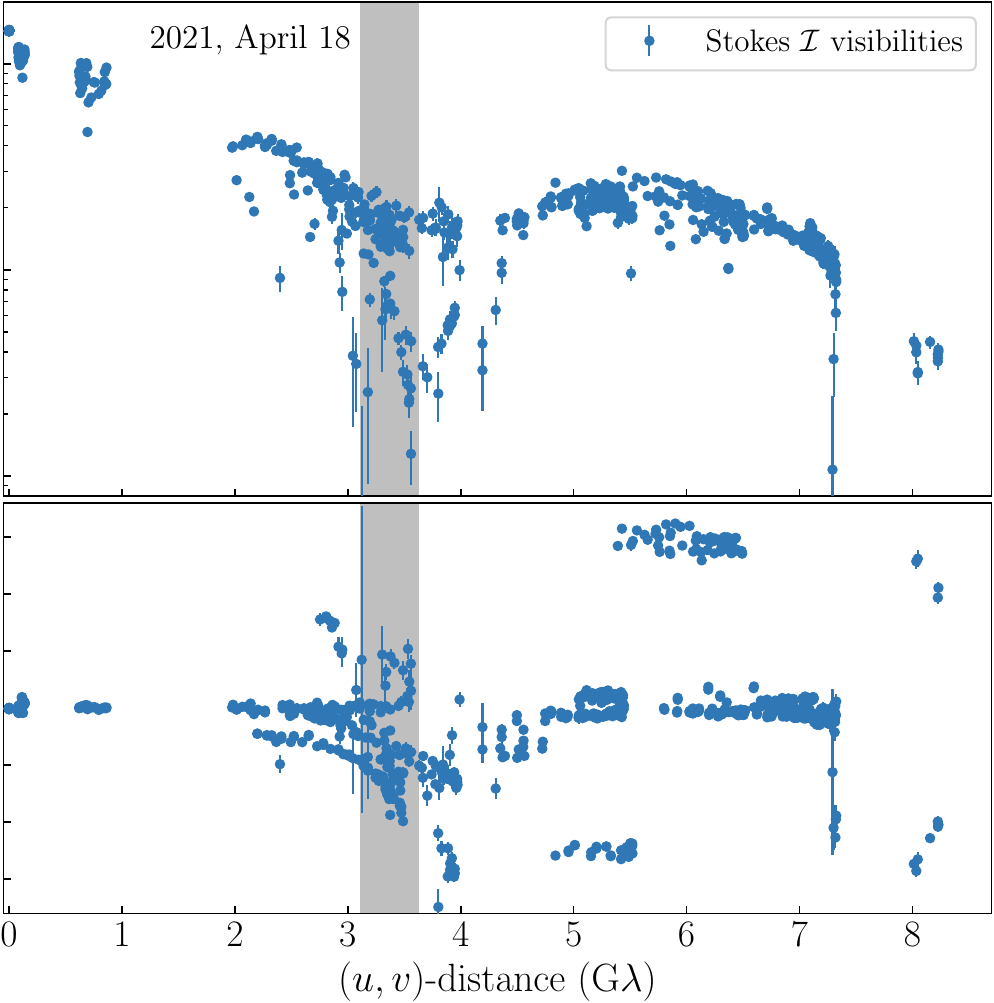}
    \caption{Band 3 (227.1\,GHz) \m87 total intensity amplitude and phase data measured in 2017, 2018, and 2021. The calibration applied is the pipeline-based signal stabilization and a priori flux density calibration without image-based self-calibration. The 2017 and 2018 data were produced by \texttt{EHT-HOPS}, and the 2021 data by \texttt{rPICARD}. The grey bands around $3.4\,\mathrm{G\lambda}$ indicate the approximate location of the first visibility minima.
    }
    \label{fig:radplot1318}
\end{figure*}

 \subsection{Feature extraction}

 To evaluate the images, we focused on each polarized imaging algorithm's ability to accurately measure the key image structural parameters and integrated polarization quantities. Quantifying the reliability of these quantities was critical since they are used to score the theoretical simulations in Chael et al. (in prep.) as well as the overall image quality across different sets of synthetic data.

For the Stokes $\mathcal{I}$ images, we extracted the following parameters: the zero-width or $\delta$ ring equivalent diameter, $\hat{d}$, (defined below), brightness asymmetry, $A$, brightness position angle, $\eta$, and compact ring flux density, $F_{\rm com}$. We used the template matching algorithm from \citet{vida} to extract these parameters. The radial profile of the template was a Gaussian distribution whose diameter, $d_0$, is defined as the diameter of the peak brightness. To harmonize the measured ring size across different pipelines, we converted $d_0$ to the $\delta$ ring equivalent diameter:
\begin{equation}
    \hat{d} = d_0 + \frac{1}{4 \ln(2)}\frac{w^2}{d_0},
\end{equation}
which is derived in \citet[][Appendix G]{M87p4}. This equation removes the approximate effect of blurring a $\delta$ ring with a Gaussian beam of size $w$, which we defined as the full width at half maximum (FWHM) of the Gaussian ring. 

The template azimuthal brightness profile was assumed to be \citep{vida},
\begin{equation}\label{eq:azi_ring_temp}
    S(\phi) = 1 - 2\sum_{k=1}^4 A_k \cos\left[k(\phi - \eta_k)\right].
\end{equation}
We restricted $A_k \in [0.0, 0.5]$ to limit the amount of negative brightness in the image. Following \citet{M87p4}, the ring brightness asymmetry, $A$, is defined as $A_1$, and the ring brightness position angle, $\eta$, is defined as $\eta_1$. 

To determine the optimal template, we optimized the cross-correlation coefficient 
\begin{equation}\label{eq:ls}
    \rho_\mathcal{I}(\bm{I},\bm{T}) = \frac{\avg{\bm{I} - \langle I \rangle , \bm{T} - \langle T \rangle}}
                        {\norm{\bm{I}-\langle I\rangle}\,\norm{\bm{T}-\langle T\rangle)}},
\end{equation}
where $\bm{T}$ is the template image and $\bm{I}$ is the image reconstruction. Given the optimal template, the compact ring flux density, $F_{\rm com}$, is defined as the total flux density within a $90\,\muas$ diameter disk, whose centre matches the fitted ring centre.

Finally, since each polarized imaging algorithm has a different image centre, field of view, and intrinsic resolution, we first normalized the image reconstructions across each method based on their total intensity images. To create a uniform image resolution, we selected a reference image. For the synthetic data tests described below, we used the ground-truth image blurred with a $20\,\muas$ FWHM Gaussian beam as the reference. This choice matched the conventions in \citet{M87p4, M87p7}. For the \m87 reconstructions, we used \lpcal as our reference image. Note that the two CLEAN pipelines gave consistent results. However, we chose \lpcal because it is an established pipeline in VLBI. Given the reference image, the reconstructions for each method are blurred to maximize their total intensity cross-correlation, Eq. \ref{eq:ls}. Note that this blurring may differ from the intrinsic resolution of the linear polarization maps. Given these harmonized images, we used the template matching procedure described above to estimate the total intensity parameters, $\{\hat{d},\; w,\; A,\; \eta\}$, and compute the centre of the ring. We then calculated $\mnet, \mavg, \beta_{1,2}$ and $F_\mathrm{com}$ by integrating radially about the fitted ring centre to a radius of $45\,\muas$ and azimuthally over $2\pi$ radians. Finally, to estimate the global polarization fidelity of the polarization reconstructions, we also computed the linear polarization cross-correlation between a linear polarization map $\bm{\mathcal{P}}$ and a reference map $\bm{\mathcal{P}}_0$ following \citet{M87p7},
\begin{equation}\label{eq:polcc}
    \rho_\mathcal{P}(\bm{\mathcal{P}},\bm{\mathcal{P}}_0) = \frac{{\rm Re}\left[\avg{\bm{\mathcal{P}}, \bm{\mathcal{P}}_0}\right]}
                                   {\norm{\bm{\mathcal{P}}} \;  \norm{\bm{\mathcal{P}}_0}},
\end{equation}
where $\langle \cdot, \cdot \rangle$ is the complex inner product, and $\norm{\cdot}$ is the complex norm. If $\bm{\mathcal{P}}$ and $\bm{\mathcal{P}}_0$ are co-linear, then $\rho_{\mathcal{P}}$ is unity, and if they are not, then $\rho_\mathcal{P} < 1$ by the Cauchy-Schwarz inequality.

\begin{figure*}[!t]
    \centering
    \includegraphics[width=\textwidth]{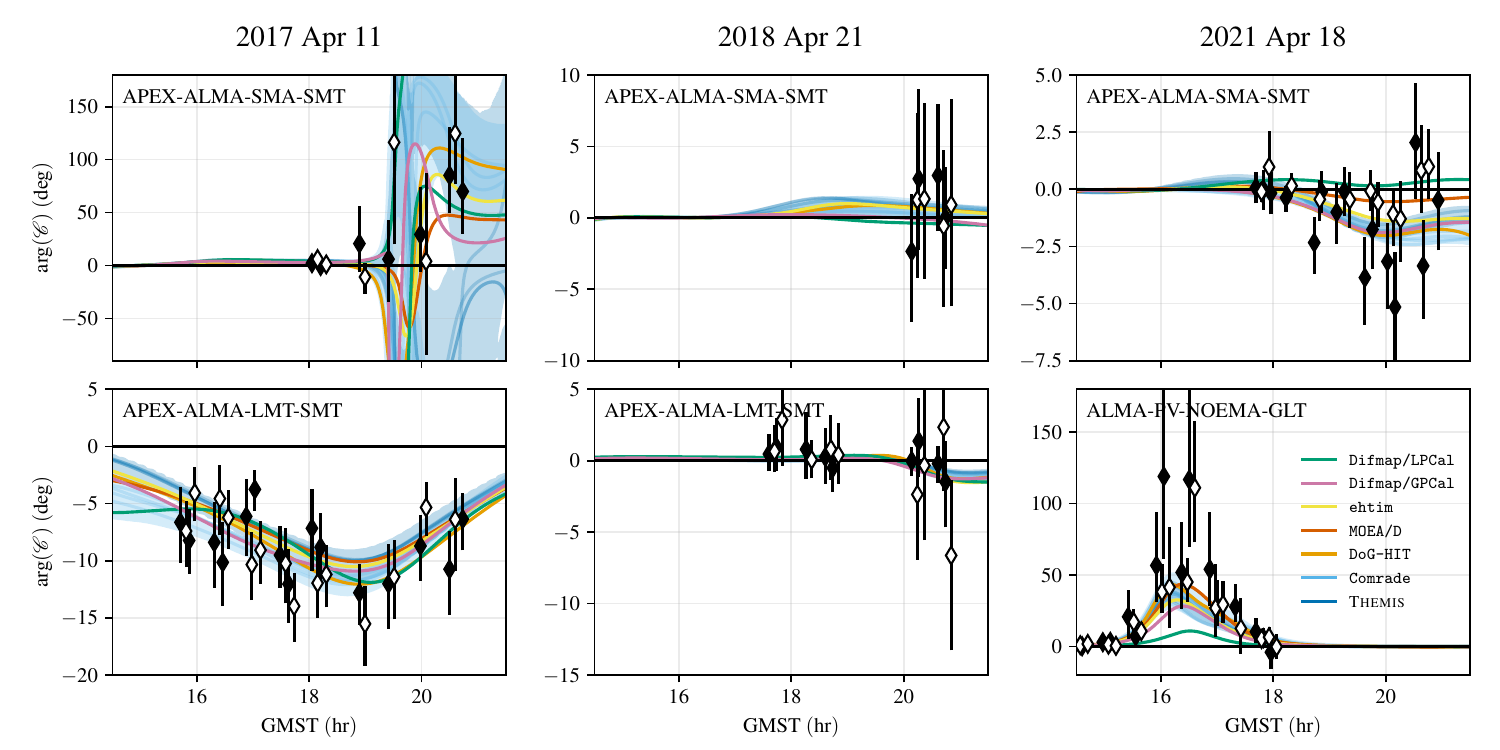}
    \caption{Comparisons of the conjugate closure trace phases across the three observing campaigns for two quadrangles. Deviations from $\arg(\mathscr{C})=0$ indicate the presence of significant polarization structure, independent of leakage and gain calibration. Closed and open points show low- and high-band data, respectively, offset in time for clarity.  The model predictions from the low-band reconstructions described in Sect. \ref{sec:imaging} are shown for reference (for the Bayesian methods, \themis and \comrade, the 95-percentile range is shown along with five draws from the posterior). Note the variation in the vertical axis ranges between panels.  
    Because LMT did not observe during the 2021 campaign, we show the ALMA-PV-NOEMA-GLT quadrangle in 2021, for which clear deviations from zero are apparent.}
    \label{fig:cctp}
\end{figure*}

\section{Data properties}\label{sec:data_properties}
Figure \ref{fig:polarmetric_cvg} shows the $(u,v)$ coverage for EHT observations in 2017, 2018, and 2021. The ticks denote the interferometric EVPA $\breve{\chi}$ and the linear polarization fraction $|\breve{m}|$ as defined in Eq. \ref{eqn:evpa_polfraction} after correcting for polarization leakage and amplitude noise bias. The 2018 data show some points with high $|\breve{m}|$, which are all related to the low signal-to-noise ratio data at baselines to the GLT. These points should be interpreted with care, as the GLT had a lower aperture efficiency in 2018 than in 2021. In particular, when considering only common regions in the $(u, v)$ space across the years, 2017 typically displays a higher $|\breve{m}|$.

The \m87 total intensity visibility amplitudes and phases as a function of $(u, v)$ distance are shown in Fig. \ref{fig:radplot1318}. The April 18, 2021, data have the best $(u, v)$ coverage of any EHT observation to date. Additionally, we achieved a high calibration quality with very coherent phases and few outliers in amplitudes that resulted from residual gain errors. The data show the characteristic secondary peak beyond a deep amplitude minimum of a ring-like structure at $\sim$$3.4\,\mathrm{G\lambda}$. Finally, the addition of the NOEMA station implies that the secondary visibility null located at $\approx8.3\,\mathrm{G\lambda}$ is probed by two baselines, PV-Hawaii and NOEMA-Hawaii, forming the first non-trivial closure triangle at such resolution scales. 

Compared to 2017 and 2018, the absence of LMT in the 2021 array implies that information from intermediate spatial scales from the $\sim$$1.5\,\mathrm{G\lambda}$ to $\sim$$2\,\mathrm{G\lambda}$ ($\sim$$100\,\muas$) LMT-SMT baseline are poorly constrained. At the same time, the addition of NOEMA provides a new $\sim$$700 \,\mathrm{M\lambda}$ ($\sim$$350\muas$) baseline to PV, and KP-SMT provides a new short $\sim$$100 \,\mathrm{M\lambda}$ ($\sim$$2\,{\rm mas}$) baseline. 
As in previous years, the ${\sim 3.5\,\mathrm{G\lambda}}$ GLT-PV baseline probes the source asymmetry; the addition of NOEMA allows us to constrain the asymmetry with a new high-fidelity triangle (GLT-PV-NOEMA).

Previous EHT observations of \m87 have shown significant polarization structures. Since polarization signals are particularly prone to calibration errors, we inspected the conjugate closure trace phases $\arg(\mathscr{C})$ defined in Eq. \ref{eqn:closure trace}. The conjugate closure trace quadrangle ALMA-APEX-SMA-SMT is present in all three years, which we show in the top panel of Fig. \ref{fig:cctp}. The bottom left and middle panels show the conjugate closure trace for the quadrangle ALMA-APEX-LMT-SMT for 2017 and 2018, while the bottom right panel shows the quadrangle ALMA-PV-NOEMA-GLT for 2021. Significant deviations from zero are visible in 2017 and 2021, but not in 2018. This result indicated the presence of significant and changing polarization structures throughout the years and provides calibration-independent evidence that \m87 was de-polarized in 2018.

\begin{figure*}[!t]
    \centering
    \includegraphics[width=\linewidth]{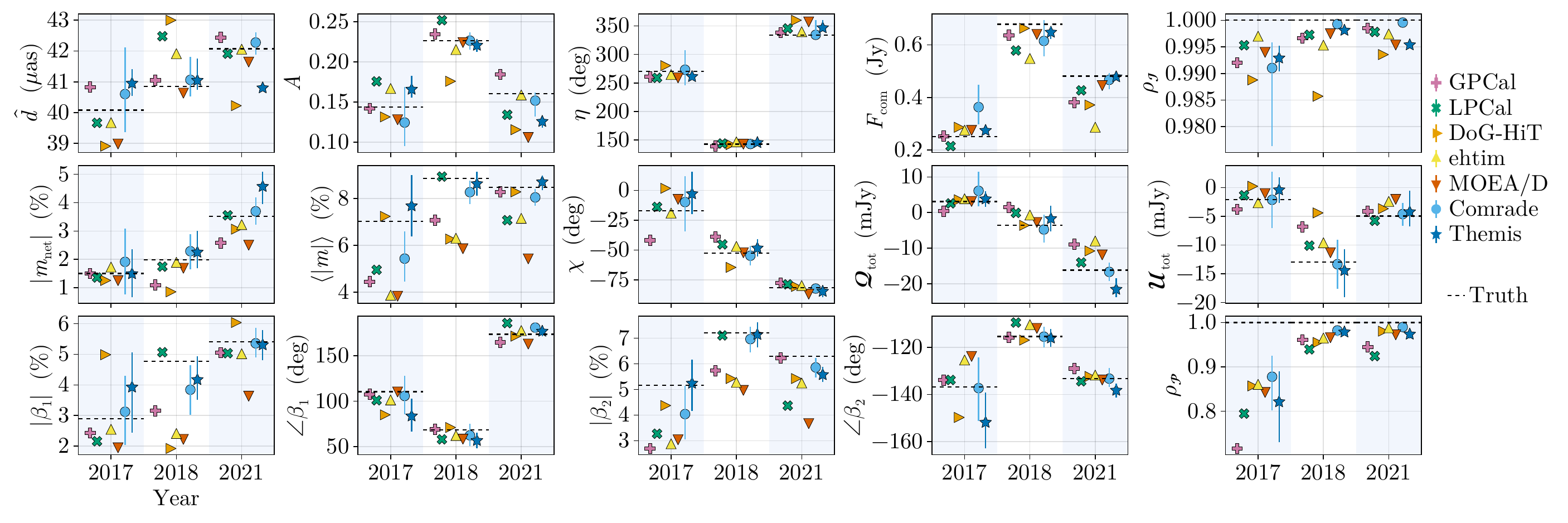}
    \caption{Comparison of extracted parameters for the blinded synthetic data test. All methods are blurred to match the ground truth GRMHD, blurred to $20\,\muas$ before the parameters are estimated. The markers show the results for each method. For the two Bayesian methods, \comrade and \themis, the markers denote the median, and the error bars the 95\% credible interval. Since the synthetic data only consider a single frequency, the non-Bayesian methods only produce a single image. The dashed black line shows the ground truth, estimated from the true on-sky image blurred with a $20\,\muas$ Gaussian beam.}
    \label{fig:stage3_params_res}
\end{figure*}

\section{Image validation}\label{sec:synthetic}

We conducted two separate synthetic polarized imaging and calibration tests to validate our imaging methods. These tests ensured that the features we extracted from \m87 were not significantly biased. Furthermore, the two tests were designed to investigate different potential systematics in the data, source model, and imaging methodologies. 

The first set was developed to test the robustness of the different polarized imaging methods to accretion turbulence and small-scale structure below the intrinsic EHT resolution. A significant component was ensuring that the image reconstructions were of high quality and that the parameter estimates for the total intensity and polarized quantities were accurately determined. The ground truth images and the data generation process were blinded for all imaging teams to prevent human biases from affecting the image reconstructions. However, an issue was identified during evaluation of the \moead results after the data had been un-blinded. This issue significantly impacted the \moead results and was not due to any assumptions in the imaging pipeline but to a bug in one of its polarized regularizers. After fixing this bug, the \moead results were rerun, which is what is shown below. As a result, the \moead results are not considered blinded. Note that this problem did not impact the six other imaging methods.

The second set of tests focused on checking whether the polarized image reconstructions could be biased by over-resolved polarized emission from \m87's extended jet. For this test, we created two versions of synthetic data that shared the same core component but differed in their polarization properties within the extended jet component. Note that these tests were not blinded during the data generation process. In the main text of this paper, we present results from the blinded synthetic data tests. Results from the extended emission test are presented in Appendix \ref{app:synth_geom}.

\subsection{Data generation}

We generated synthetic data for each source model with the software \ehtim using the $(u,v)$ coverage from the April 11, 2017, April 21, 2018, and April 18, 2021, observations. These synthetic data were generated for band 3 for all three years. The complex visibilities were generated by sampling the Fourier transform of each ground truth image with the observation's $(u,v)$ coverage. The sampled complex visibilities were then corrupted with $(i)$ time-stable polarimetric leakage terms, $(ii)$ station-based, time-variable gains in the visibility phase and amplitude, $(iii)$ station-based, time-variable complex gain ratios of the right-left circular polarization gains, and $(iv)$ baseline-based thermal noise estimated from the corresponding EHT observations.

This generation process is identical to that of \cite{M87p9} for the 2017 synthetic data. For 2018 and 2021, we changed the generation process to accommodate the changes in data properties throughout the three years of observations. These changes included adding or removing stations present during each year's observations, modifying station mount types, and replicating the significant JCMT polarimetric leakage observed during 2018 (Appendix \ref{appendix:data}).

For the blinded test source models, three snapshots from a polarized MAD GRMHD simulation with $i = 17\degr{}$, $a=-0.5$, $R_\text{low} = 10$, and $R_\text{high} = 40$ were used as the ground-truth images. This GRMHD simulation was chosen because it passed the EHT 2017 and 2018 multi-year theoretical constraints from \citet{M87_2018p2}. For each year of the synthetic data, we used a different random snapshot. For more details, see Appendix \ref{appendix:synth}.

\subsection{Synthetic data results}

The blinded parameter estimation results are shown in Fig. \ref{fig:stage3_params_res}. We found that for every year, the total intensity cross-correlation with the truth is $>0.975$ for all methods, demonstrating their high quality. Each method recovered the true $\delta$ ring diameter to within $2\,\muas$, the brightness asymmetry to within $\sim 0.05$, and the ring position angle estimate, $\eta$, to within $5-10\degr{}$. For the compact flux density, we found a larger spread across methods. This is not unexpected due to the lack of intermediate baselines in the EHT array, making compact flux constraints sensitive to the image priors and gain solutions as well as priors, as discussed in more detail below.

For the linear polarization reconstructions, the polarized cross-correlation improved dramatically from 2017 to 2021. In 2017, $0.7 < \rho_\mathcal{P} <  0.9$, while in 2018 and 2021, $\rho_\mathcal{P} > 0.9$ for all methods. This demonstrated the improved polarized imaging capabilities of the 2018 and 2021 EHT arrays. The image reconstructions and more details can be found in Appendix \ref{app:synth_blinded} and Fig. \ref{fig:stage3GRMHDsyntheticdata}.

Analysing the derived image-averaged polarimetric quantities in Fig. \ref{fig:stage3_params_res}, we found that the true $\mnet$ is contained within the spread across methods each year. Furthermore, \comrade's posterior contained the truth within its 95\% contours every year, while the estimates of \lpcal and \ehtim are within 0.25\% of the truth each year. Similarly to $\mnet$, the image-integrated EVPA $\chi$ is also recovered each year, and all methods recovered the truth to within 10\degr{}. 

For $\mavg$, we found a slightly more complicated result. Unlike $\mnet$ where the reconstructions tended to be distributed around the truth, we found that $\mavg$ tends to be biased low for all methods except \themis. This bias reflects the difficulty of measuring $\mavg$ due to its sensitivity to the linearly polarized resolution, and field of view of the reconstructions \citep[see Appendix G of][for a related discussion]{M87p9}. Although the images were blurred to match their resolution, this was based on total intensity, which may differ from the resolution of the polarized maps. The magnitude of this bias tended to decrease from 2017 to 2021, and in 2021, the results of four of the seven methods were close ($<0.5\%)$ to the true value. Therefore, after combining the results across all methods, our estimates of $\mavg$ recovered the true value. 

Analysing the azimuthal structure of the ring reconstruction, the phases of the first two $\beta$ modes are recovered each year. Furthermore, the improved coverage in 2018 and 2021, compared to 2017, increased the precision and accuracy of the measurements of $\angle\beta_2$. That is, we found a $10\degr{}$ dispersion around the truth for all methods. The amplitudes of the first two $\beta$ modes are also recovered, although the spread across methods is more pronounced. Specifically, for $|\beta_1|$, we found that the estimates from \doghit and \moead exhibited significant deviations from the true values. Furthermore, we found that, similar to $\mavg$, some methods tended to be biased towards lower values, specifically \ehtim and \moead for $|\beta_{1,2}|$ and \doghit for $|\beta_{2}|$ across all years. Similarly, \comrade was biased low for $|\beta_1|$ in 2018 and $|\beta_2|$ in 2017, although the truth is contained within the 99\% credible interval in both cases. Since $\mavg$, and $|\beta_i|$ are roughly proportional, the bias in the $\beta$ mode amplitudes is likely of a similar origin.

In summary, even if individual methods were sometimes biased, the combined estimates across all methods contained the truth. Therefore, to estimate the parameters of \m87, we combined the estimates from all methods. The estimates are combined by concatenating the results across all methods, weighted by the inverse number of samples produced. Specifically, for the non-Bayesian methods, the inverse weights are equal to two since each method produced an estimate for the band 3 and band 4 data. For the Bayesian methods, the inverse weight is given by the number of posterior samples from each method and frequency band. Given this set of samples, we then computed the 95\% percentile range to estimate the combined uncertainty.

\begin{figure*}[!t]
    \centering
    \includegraphics[width=\linewidth]{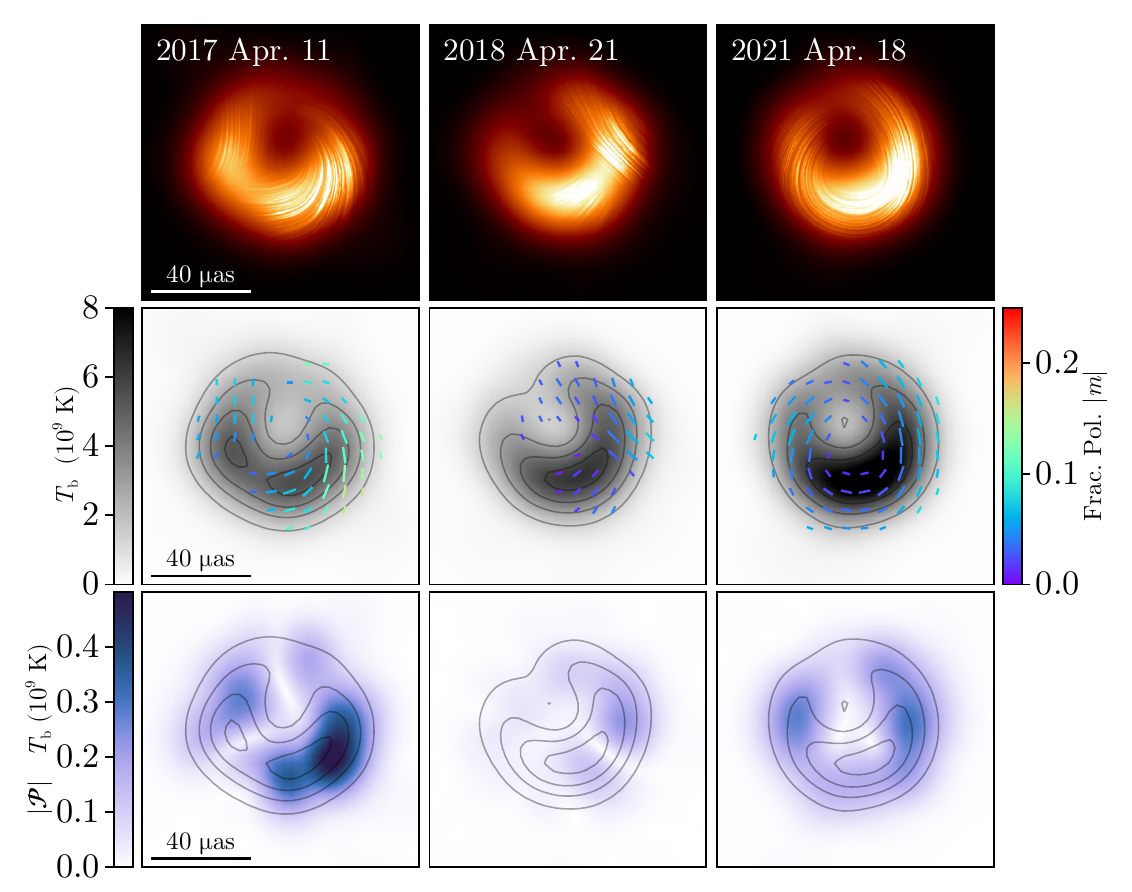}
    \caption{Fiducial images for 2017 (band 3), 2018 (bands 3 and 4), and 2021 (bands 3 and 4). The images were produced by averaging the reconstructions over the methods described in Sect. \ref{sec:imaging}. Each method's image reconstruction has been blurred to match the resolution of the representative \lpcal image. Top row: Polarization `field lines' overlaid on the total intensity image. Second row: Total intensity image in grey scale with the contours showing the 22.5\%, 45\%, 67.5\%, and 90\% peak brightness levels, overlaid with polarization ticks. The polarization ticks indicate the EVPA, the tick length is proportional to the linear polarization intensity, and their colour indicates the linear polarization fraction. Polarization ticks are only shown in regions where the total intensity is $>10\%$ of the maximum brightness and the linear polarization brightness is $>10\%$ of the peak linear polarized brightness. Bottom row: Total linear polarized brightness, $|\mathcal{P}|$.}
    \label{fig:fiducial}
\end{figure*}

In Appendix \ref{appendix:synth} and Fig. \ref{fig:synthetic_stage3_2021_D-term}, we also compare the leakage recovered from each method to the true value. In general, we found that the Bayesian methods recovered the leakage for every station to within 1\% for the blinded synthetic data. For the non-Bayesian methods, we observed more scatter, especially for GLT in 2018. This can be understood in light of the relatively small parallactic angle coverage of GLT for \m87 ($\sim 15\degr{}$) compared to other EHT sites ($50$ -- $100\degr{}$) over a whole observation track. As a result, the Bayesian methods reported a relatively large leakage uncertainty for GLT compared to the other sites, but the true value was still recovered. However, the non-Bayesian method leakage estimates are more prone to local minima since they report a single value rather than characterizing the parameter space, making them more susceptible to biases. Regardless of these discrepancies, the different leakage estimates for GLT did not impact our results.

Finally, for the non-blinded extended emission tests, we found similar results. Namely, each method's image reconstructions were not significantly impacted by the presence of extended emission. For more detailed information, see Appendix \ref{app:synth_geom}.

\section{Results} \label{sec:results}

Figure \ref{fig:fiducial} presents the total intensity and linear polarization maps of  \m87 in 2017, 2018, and 2021. Unless stated otherwise, the 2017 results were obtained from the \texttt{EHT-HOPS} band 3, April 11 data from \citet{M87p7}. Band 3 was chosen for consistency with \citet{M87p7}. The 2018 results were obtained from the HOPS reduction of bands 3 and 4 on April 21, 2018, in \citet{M87_2018p1}. The 2021 results from bands 3 and 4 on April 18 were produced using the \texttt{rPICARD} reduction. \texttt{rPICARD} was chosen in 2021 due to its ability to handle NOEMA's phase jumps (see Appendix \ref{appendix:data} and \citealt{vonFellenberg2025}). 

\begin{figure*}[!t]
    \centering
    \includegraphics[width=1.0\linewidth]{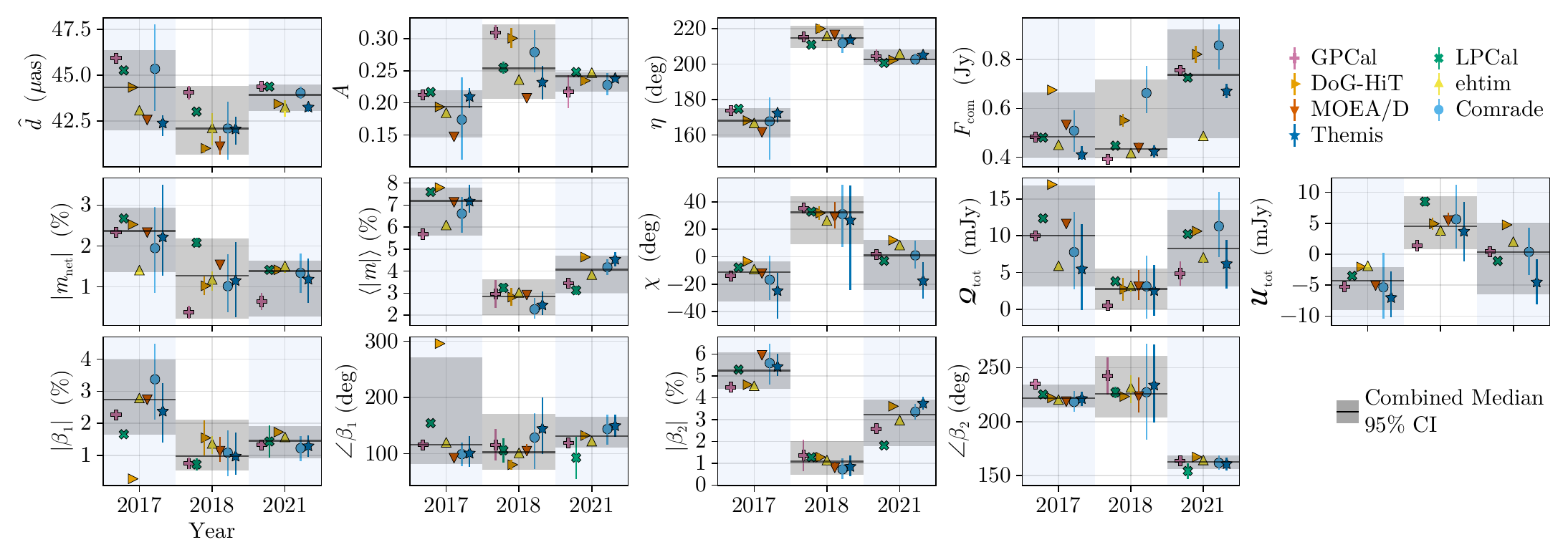}
    \caption{Extracted parameters of \m87 across the three observations and averaged over band 3 and band 4. Each panel shows the results from April 11, 2017, April 21, 2018, and April 18, 2021. For the non-Bayesian methods, the error bars show the spread between the high- and low-band estimates and are not a measure of the statistical uncertainty of the image reconstruction. The error bars for the Bayesian methods show the 95\% credible interval around the median and measure the statistical uncertainty of the reconstructions due to thermal noise, instrumental effects, incomplete coverage, and frequency dependence. The solid black  line is the median, and the grey band is the 95-percentile range from all image reconstructions, with each method weighted inversely to the number of images they produced. Note that each image reconstruction has been blurred to match the resolution of the \lpcal reconstruction before the parameters were estimated.}
    \label{fig:m87_params_uniform}
\end{figure*}

For each year, the images shown in Fig. \ref{fig:fiducial} are produced by averaging all the imaging methods, blurred to a common resolution. For individual results, we refer the reader to Appendix \ref{appendix:individual_results}. Generally, the individual methods blurred to a common resolution are consistent with the averaged results, especially for the 2021 data, where the improved array strongly constrains the ring emission. 

 The parameter and feature extraction results in \autoref{table:m87_results} show the EHT constraints averaged over all methods for each year. As mentioned above, the ranges are computed by averaging the results from each method, weighted equally, and calculating the $95\%$ percentile range.
 
For 2017, the values in this work are consistent with the original values reported in \citet{M87p7}. However, the uncertainty of our estimates is smaller than the original values \citet{M87p7}. This reduction is expected since \citet{M87p7} analysed all four days of observations, while this paper only considered April 11, 2017. Therefore, we used the reported polarization values from \citet{M87p7} and ring orientation from \citet{M87p4, M87p6} in \autoref{table:m87_results} for a conservative estimate of \m87's appearance in 2017. Note that the quantitative measurements of \m87's polarized emission using the methods in this paper are described below in Sect. \ref{sec:PolResults}.

Finally, note that in addition to estimating the linear polarization structure, every method also estimated the leakage for each station. Unlike \citet{M87p7}, we did not attempt multisource fitting to constrain the leakages for the EHT array. Instead, each imaging method estimated the polarization leakage for each site as a byproduct of polarimetric imaging. In general, we found consistent $D$-terms between imaging methods in 2018 and 2021, except for GLT in 2018. The measured GLT leakage displayed considerable scatter among the non-Bayesian methods. However, this scatter was also observed in the synthetic data tests and is explained by GLT's poor parallactic angle coverage and their leakage reconstruction approach as we describe in Sect. \ref{sec:synthetic}.

\subsection{Stokes I results} \label{sec:stokesIresults}

Figure \ref{fig:fiducial} demonstrates that \m87 consistently shows a central depression -- the black hole shadow -- across all observations. The ring around this central depression is consistently asymmetric in brightness, always appearing brighter in the south. This aligns with expectations from simulations of accretion physics, given the location of the large-scale jet observed at lower frequencies. The size of the ring remains stable over the years and is consistent with previous results \citep{M87p4,M87_2018p1}. 

Specifically, Fig. \ref{fig:m87_params_uniform} shows that the diameter of the ring is consistent between methods, although 2017 has the most scatter due to its sparser coverage. After averaging over methods, we found a diameter estimate of $[42.0\,\muas,\, 46.4\,\muas]$ for 2017, $[40.7\,\muas,\, 44.4\,\muas]$ for 2018, and $[43.1\,\muas,\, 44.4\,\muas]$ for 2021. Therefore, as predicted by our theoretical understanding of accretion physics, the diameter of the ring is stable across all three observing epochs, confirming the findings in \citet{M87_2018p1}. Note that in \citet{M87_2018p1} two estimates of \m87's diameter were provided. Our imaging results are consistent with their method averaged estimate of $\sim 43.3^{+1.5}_{-3.1}\,\muas$. This similarity demonstrates the robustness of the EHT ring diameter estimates, given that our results utilized different imaging and analysis pipelines.

Unlike the size of the ring, we found that its azimuthal structure changed from 2017 to 2018/2021. As first noted in \citet{M87_2018p1} in 2018, the brightness peak changed from 2017 to 2018, moving anti-clockwise by $\sim 45\degr{}$. From Fig. \ref{fig:fiducial} this shift appears to be due to the bright region in the eastern part of the ring disappearing. In 2021, we found that \m87's brightness profile is remarkably similar to the 2018 image. Quantitatively, calculating the cross-correlation between the fiducial 2017 and 2018 images gives $0.948$, while the cross-correlation between 2018 and 2021 gives $0.992$. This value is similar to the average cross-correlation between all pairs of imaging methods in 2021.

Analysing the total intensity ring properties in Fig. \ref{fig:m87_params_uniform} and \autoref{table:m87_results}, we found that the ring brightness asymmetry, $A$, is consistent in 2018 and 2021 and marginally discrepant in 2017. The difference in azimuthal brightness distribution is seen most clearly in the evolution of the position angle $\eta$ from 2017 to 2018/2021. The measured position angle was $[166\degr{}, 175\degr{}]$ in 2017, while in 2018 and 2021, $\eta$ was between $200\degr{}$ and $218\degr{}$ over the two years. The stability of $\eta$ from 2018 to 2021 is notable, as it aligns with the expected brightness maximum predicted by theoretical simulations, given the location of the low-frequency jet \citep{M87p5, M87_2018p1, M87_2018p2}.

Finally, we measured significant disagreements in the compact flux density in the 2021 images. \ehtim measured a compact flux density of around 0.5\,Jy, similar to the range found in 2017 and 2018, while the other methods found a compact flux between 0.7\,Jy and 0.95\,Jy. A similar disagreement in compact flux was reported in \citet{M87_2018p1}, where a detailed analysis revealed that the compact flux measured by EHT was very prior-dependent, producing values ranging from 0.3 to 1.1\,Jy. A significant factor in this uncertainty, beyond residual calibration issues, is the lack of intermediate baselines in the 2021 EHT array. For the 2017 and 2018 EHT array, the LMT-SMT baselines directly probed the emission on  $\sim$$100\,\muas$ scales. In 2021, the EHT gained a new intermediate baseline, NOEMA-PV (sensitive to $\sim$$300\,\muas$ scales) and a short baseline KP-SMT (sensitive to $\sim$2-3\,mas scales). Unfortunately, the LMT did not participate in the 2021 campaign, limiting our ability to directly constrain the flux on scales $\lesssim 100\,\muas$\footnote{Note that in the 2022 EHT campaign, the LMT rejoined the array}. Therefore, constraining the absolute emission on $\lesssim 100\,\muas$ scales strongly depends on the image and gain priors. Taking this uncertainty into account, we report a conservative range for the compact flux of \m87 in 2021 of $F_{\rm com} = [0.5\,\mathrm{Jy}\,,\,0.9\,\mathrm{Jy}]$. This constraint is similar to that reported in \citet{M87_2018p1} for \m87 in 2018. Note, the 2021 total flux constraint could be improved in the future by utilizing multifrequency information, for example, studies of the core \m87 at 86\,GHz similar to the analysis in \citet{M87_2018p1}.

\subsection{Polarization results} \label{sec:PolResults}

Unlike total intensity, where significant evolution in the ring parameters was only observed for the PA, \m87's linear polarization emission appeared distinct each year. Figure \ref{fig:fiducial} shows how the linear polarization emission of \m87 differs yearly in the EVPA pattern and the total linear polarization brightness. In 2017, a peak linear polarization fraction of $\sim 15\%$ was found near the brightest region of the ring. Similarly, the absolute linear polarized brightness peaked in the south-western region of the ring. In contrast, in 2018 and 2021, the total intensity brightness maximum is de-polarized with a measured linear polarization fraction of $\lesssim 5\%$. Moreover, in 2018, the ring is almost entirely de-polarized, except for a single region in the western part of the image that has a linear polarization fraction of $\sim$5 -- 10\%. While the ring is more polarized in 2021, it is still measured to be less than the previous 2017 estimates, and the peak polarization fraction of the ring never exceeds $10\%$ after blurring all methods to a common resolution of $20\,\muas$. 

\begin{table}[!t]
    \caption{Parameter constraints from 2017 -- 2021 (68\% credible interval).}
    \adjustbox{max width=\columnwidth}{
    \begin{tabular}{lccc}
    \hline\hline
    Parameter & 2017 & 2018 & 2021 \\
    \hline
    \rule{0pt}{2.5ex} 
      $\hat{d}\;(\muas$)     &  $[42.0 , 46.4]$   & $[40.7 , 44.4]$  &  $[43.1 , 44.5]$   \\[1mm]
      $A$                     & $[0.15 , 0.22]$   & $[0.21 , 0.32]$    & $[0.22 , 0.25]$    \\[1mm]
      $\eta$ (deg)            & $[150 , 200]^{\dagger}$     & $[209 , 222]$      &  $[200 , 208]$     \\[1mm]
      $F_{\rm com}$ (Jy)      & $[0.40 , 0.67]$   & $[0.40 , 0.72]$    & $[0.48 , 0.92]$     \\[1mm]
      $\mnet$  (\%)           & $[1.0 , 3.7]^\dagger$    & $[0.2 , 2.2]$     &  $[0.3 , 1.6]$    \\[1mm]
      $\mavg$ (\%)            & $[5.7 , 10.7]^{\dagger}$    & $[2.0 , 3.6]$     & $[3.0 , 4.7]$     \\[1mm]
      $\chi$ (deg)            & $[-33 , -3]$  & $[9 , 44]$   & $[-24 , 12]$ \\[1mm]
      $|\beta_1|$ (\%)        & $[1.6 , 4.0]$    & $[0.5 , 2.1]$       & $[0.9 , 1.9]$    \\[1mm]
      $\angle \beta_1$ (deg)  & $[82 , 271]$     & $[71 , 171]$  & $[111 , 165]$       \\[1mm]
      $|\beta_2|$ (\%)        & $[4.0 , 7.0]^{\dagger}$   & $[0.5 , 2.0]$  & $[1.8 , 3.9]$        \\[1mm]
      $\angle \beta_2$ (deg)  & $[-163 , -127]^{\dagger}$   & $[-156 , -99]$  & $[156 , 168]$       \\[1mm]
      \hline
    \end{tabular}}
    \tablefoot{$^\dagger$Using results from \citet{M87p7}}
    \label{table:m87_results}
\end{table}

 Examining the image-integrated non-structural parameters in Fig. \ref{fig:m87_params_uniform} and \autoref{table:m87_results}, a similar pattern is found. In 2017, using the methods described in this paper, we measured the image resolved fractional polarization to be $\mavg \in [5.7\%, 7.8\%]$ on April 11, 2017, consistent with the $5.7 - 10.7\%$ found in \citet{M87p7} after averaging over the four 2017 observations. This is significantly higher than the values we found for \m87 2018 ($\mavg\in [2.0\%, 3.6\%]$) and 2021 ($\mavg \in [3.0\%, 4.7\%]$). This result quantitatively demonstrates the relatively low polarization state of \m87 in 2018 and 2021. The unresolved fractional linear polarization in 2017, $\mnet$, while typically higher $(1.0\% - 3.7\%)$, is consistent with the values found in 2018 ($0.2\% - 2.1\%)$ and 2021 $(0.3\% - 1.6\%)$. For the image-integrated EVPA, $\chi$, we measured significant variability every year, where $\chi \in [-33\degr{}, -3\degr{}]$ in 2017, $[9\degr{}, 44\degr{}]$ in 2018, and $[-24\degr{}, 12\degr{}]$ in 2021, reflecting the changing EVPA pattern in the image reconstructions. 

Analysing the properties of the polarized ring, we found differences between the behaviour of $\beta_1$ and $\beta_2$ from 2017 to 2021. Interestingly, $\angle\beta_1$ was stable from 2017 to 2021, with $\angle\beta_1$ ranging from 71\degr{} to 171\degr{} for all years and methods, except for \doghit in 2017. Similarly, $|\beta_1|$ is consistent every year, although most methods found a larger $\beta_1$ in 2017 than in the other years. The exception to this is \doghit. However, \doghit struggled to recover $|\beta_1|$ in the blinded synthetic data test (see Fig. \ref{fig:stage3_params_res}); it was discrepant from the truth by $2-3\%$ in an absolute sense in 2017 and 2018.

Unlike $\beta_1$, $\beta_2$ evolves substantially from 2017 to 2021 in both amplitude and phase. Using the methods in this paper, we found $|\beta_2| \in [4.5\%, 5.5\%]$ on April 11, 2017, \$\ consistent with the values in \autoref{table:m87_results} taken from \citet{M87p7}, which are averaged over the four observations in 2017. Similarly to $\mavg$, $|\beta_2|$ decreases significantly in 2018 to $0.5\% - 2.0\%$, but recovers slightly in 2021 to $1.8\%-3.9\%$. The observed de-polarization is also evident from the polarization calibration-insensitive closure traces, as noted above.

Unexpectedly, we found that $\angle\beta_2$ evolved substantially from 2017 to 2021. Using the methods in this paper, we found that $\angle\beta_2 \in [-141\degr{}, -128\degr{}]$ on April 11, 2017, which is consistent with the values reported in \citet{M87p7}, with $\angle \beta_2 \approx [-163\degr{}, -127\degr{}]$ averaged over all four days of observations in 2017. Although very little $\beta_2$ is measured in 2018, we found a similar value to that of 2017, although with greater uncertainty, $[-156\degr{}, -99\degr{}]$. However, this consistency was broken in 2021. That is, we found that $\angle\beta_2$ rotated by about $60\degr{}$ and flips sign, signaling a change in the polarization helicity. The most apparent visual changes in the EVPA patterns between 2017 and 2021 are in the dimmer northern half of the ring. While the EVPA patterns look more similar in the brighter south-west region, there is still an apparent shift in EVPA in the SW; since the overall image $\beta_2$ is intensity weighted Eq. \ref{eq:beta}, changes in the polarization structure in the brighter south-western region in fact account for most of the overall $\approx 60\,\degr{}$ shift in $\angle \beta_2$ across the image. These results motivate further studies into the polarization structure in different parts of the emission ring and their relation to changes in the underlying magnetic field morphology or in the Faraday rotation along different lines of sight.

Note that from the synthetic data tests, we do not believe that the significant changes in $\angle\beta_2$ or the other polarimetric properties from 2017 to 2021 are due to changes in the coverage of \m87. The synthetic data results in Fig. \ref{fig:stage3_params_res} demonstrate that all algorithms recover the EVPA pattern to a high degree of certainty. Specifically, $\angle\beta_2$ is consistently one of the more robust quantities we measured in the synthetic data tests. Its values in 2021 are consistent on both April 13 and 18 (Fig. \ref{fig:app:April13_18_comp_params}) and for both reduction pipelines \texttt{rPICARD} and \texttt{EHT-HOPS} (Fig. \ref{fig:hops_v_casa_reco}). Additionally, while the image-integrated polarization quantities $\mavg$, $ |\beta_1| $, and $ |\beta_2|$ are biased low for some methods in the synthetic data tests, this bias decreases for the Bayesian methods in later years. Therefore, the fact that 2017 had a higher overall polarization is likely not a result of beam de-polarization or other instrumental effects. We discuss the interpretation of these changes in Sect. \ref{sec:discussion}.

Finally, we mention that, while most polarized imaging algorithms include circular polarization maps, we do not report them in this paper. In Appendix \ref{appendix:data} and Fig. \ref{fig:cplrdiff}, we inspect the right and left circular polarization closure phases differences in 2017 -- 2021 and found weak signals in the 2018 and 2021 observations. However, we found significant discrepancies in the circular polarization maps across methods due to residual instrumental systematics, such as right-left gain ratios. Like in past works \citep{M87p9}, we could not robustly recover horizon-scale circular polarization structure for this reason.

\subsection{Constraints on the extended non-ring emission} \label{sec:jet}

\begin{figure}[t]
    \centering
    \includegraphics[width=0.985\columnwidth]{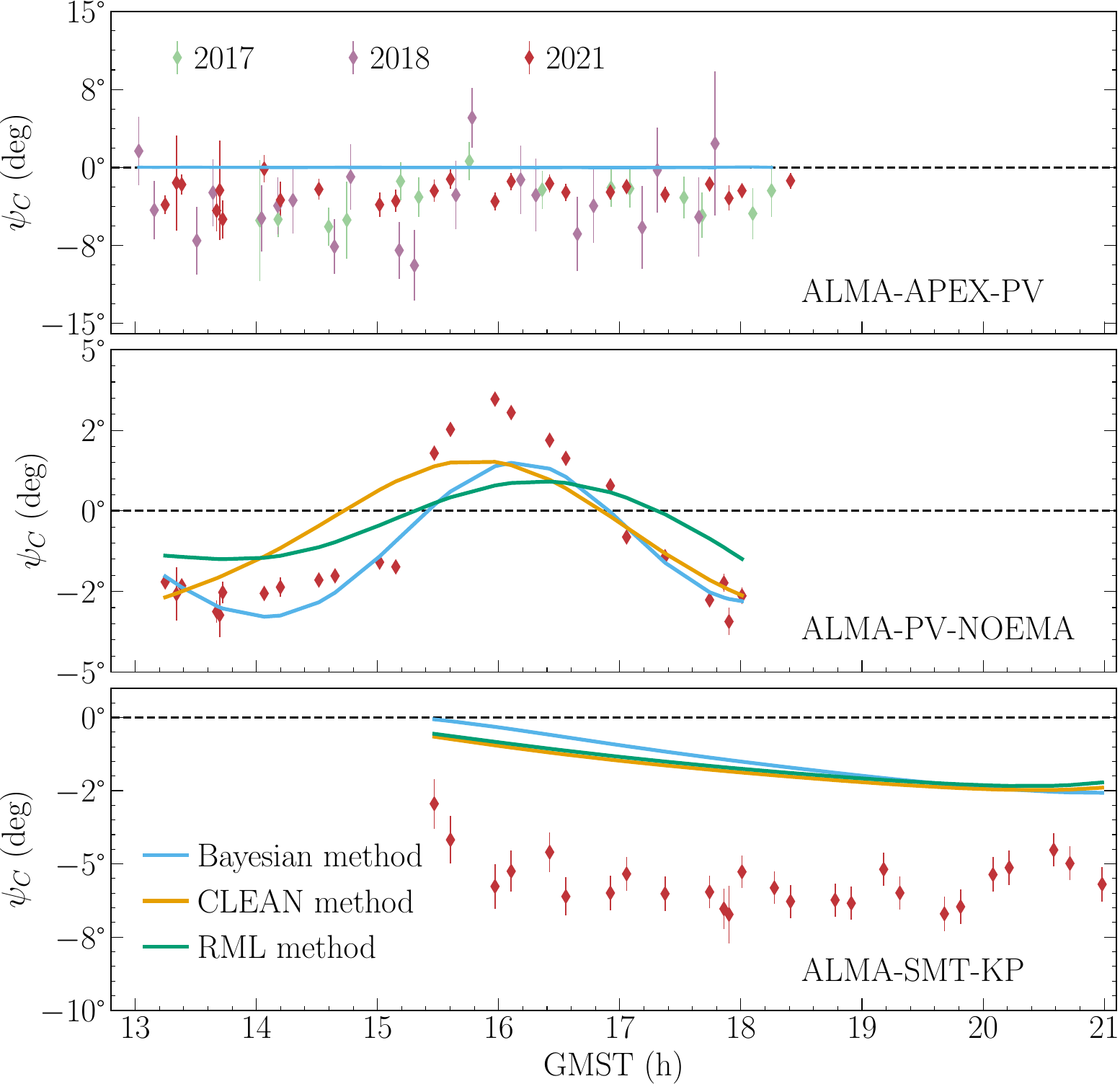}
    \caption{Closure phases (band 4) on the three closure triangles that probe extended non-ring emission trivial closures -- to AA-AX-PV (top), ALMA-PV-NOEMA (middle), and ALMA-SMT-KP (bottom) -- with fits from the final image reconstructions for the types of imaging algorithms -- Bayesian (\comrade), CLEAN (\gpcal), and the RML method (\ehtim). Other methods yield qualitatively similar fits where available. Strategies to fit the extended non-ring emission to the 2021 closure phases are presented in \citet{Georgiev2025}) and \citet{saurabh2025}}
    \label{fig:cp_bump} 
\end{figure}

 \m87 has a parsec-scale jet that is consistently detected at longer radio wavelengths over multiple magnitudes of spatial scales \citep{Kim2018M87, Lu2023, Cui2023, Walker2018, Lister2018, Nikonov2023}, including the inner few hundred microarcseconds with 86\,GHz GMVA observations \citep{Kim2018M87, Lu2023, kim_imaging_2025}. At 230\,GHz, ALMA observations revealed the jet is pointed $288\degr{}$ east of north on scales of $\sim\,100\,\mathrm{mas}$ \citep{Goddi2021}. EHT observations in all three years found missing flux between the intra-site and the next shortest baselines (see e.g. the discussion in \citealt{M87_2018p1} for details). Before 2021, the EHT lacked baselines $\gtrsim$\,200\,$\mu$as, limiting sensitivity to structures at jet-launching scales \cite[e.g.][]{Roelofs2020, Roelofs2023, Broderick2022, M87_2018p1, M87p4}. In 2021, the inclusion of KP and NOEMA added baselines beyond previous capabilities: NOEMA–PV with a baseline length of $\sim$\,1100\,km probes scales of approximately $\sim$\,340\,$\muas$, and KP–SMT ($\sim$\,100\,km) probes spatial scales $\sim$\,2700\,$\muas$. To analyse the characteristics of the emission on these scales, we inspected three sets of closure phases: ALMA-APEX closures, ALMA-SMT-KP closures, and ALMA-PV-NOEMA, which we show in Fig. \ref{fig:cp_bump}.

 We found that on ALMA-APEX triangles, the closure phases displayed a systematic negative offset of a few degrees. This offset is present in all years, but it is most notable in 2021. Focusing on the 2021 observation, the ALMA-SMT-KP triangle displayed a consistent offset of $-6\degr{}$ to $-7\degr{}$. Furthermore, the ALMA-PV-NOEMA triangle displayed a $\sim$$5\degr{}$ bump near 16 GMST. Interestingly, none of the 2021 reconstructions from the inner $100\,\muas$ are consistent with these closure phases when compared to the data under the assumption of small $\lesssim 1\%$ systematic errors for these triangles.

Unfortunately, even in 2021, we found that directly imaging the structure of \m87 on these scales with EHT data is still highly uncertain and model-dependent due to the sparse intermediate baseline coverage. A companion paper, \citet{saurabh2025}, explores the effects of including a Gaussian component with core $\lesssim 100\,\muas$ ring image on the ALMA-SMT-KP and ALMA-PV-NOEMA closure phases. They reported a preference for a faint ($\leq 100$\,mJy) component located $\sim300\,\muas$ to the west of the ring. However, the details of the locations are likely dependent on the morphology chosen for the extended non-ring emission.

While the addition of the Gaussian component improves the closure phase fits on the ALMA-SMT-KP and ALMA-PV-NOEMA triangles, the offset in the ALMA-APEX triangles still exists. If this offset is not due to instrumental effects, it implies the existence of non-trivial emission on scales $\gtrsim 300\,\muas$. This conclusion of extended emission is also supported by the measured compact flux of the ring and additional Gaussian component, which only accounts for 40--70\% of the measured flux on the ALMA-APEX and JCMT-SMA baselines for 2017, 2018, and 2021. To constrain the emission on larger scales, \citet{Georgiev2025} analysed the offset in the ALMA-APEX closure phases. They found that the offset can be explained by extended emission whose centroid is located about 1\,mas north-west of the ring. Encouragingly, the inferred centroid direction is consistent with the extended jet observed by ALMA \citep{Goddi2021}. This larger-scale emission does not impact the smaller-scale ALMA-KP-SMT or ALMA-PV-NOEMA triangles. Note that residual issues in the EHT data could be caused by systematic non-closing errors \citep[see e.g.][for a list]{M87p3} that may impact the recovered properties of this extended $\sim1$\,mas emission. We refer to \citet{Georgiev2025} for a thorough accounting of these systematics.

In conclusion, with the 2021 EHT array and the supporting analyses in \citet{saurabh2025}. and \citet{Georgiev2025} we found that emission on at least three spatial scales is necessary to fit \m87 data: a ring-like structure for scales $\gtrsim 1\, {\rm G}\lambda$, larger diffuse emission located about a few hundred microarcseconds from the core to match the KP-SMT and PV-NOEMA baselines, and an even larger component on $\sim 1$ mas scales to explain the offsets on ALMA-APEX triangles. We anticipate that the exact nature of this emission will be imageable in newer EHT observations, such as the 2022 array, which regained the LMT.

\begin{table*}
    \begin{center}
        \caption{\m87 large-scale quantities measured by ALMA and horizon-scale quantities measured by the EHT over the years in the 226 -- 230\,GHz band.}
    \begin{tabular}{l c c c c c}
    \hline \hline
         {Year} &  ${F_\mathrm{ALMA}}$ & ${\mathrm{RM}_\mathrm{ALMA}}$ & {Stokes} ${\mathcal{I}}$ {PA} ($\eta$) &  ${\langle|m|\rangle}$ & ${\angle \beta_2}$ \\
         \quad{month, day} & {(Jy)} & ${(10^5\,\mathrm{rad}\,\mathrm{m}^{-2})}$ & {(deg)} & {(\%)} & {(deg)} \\
         \hline
         2017&   &   &   &    &  \\
         \quad April 5& $1.27 \pm 0.13$ & $1.2 \pm 0.3$ & $[150,200]$ & \multirow{2}{*}{$[5.7, 10.7]$} & \multirow{2}{*}{$[-163, -127]$}\\
         \quad April 11& $1.30 \pm 0.13$ & $-0.3 \pm 0.2$ & $[150,183]$ & &\\
         \hline
         2018\textsuperscript{\textdagger} &   &   &   &    &  \\
         \quad April 21& $1.05\pm0.11$ & $-4.3 \pm 0.3$ & $[209, 222]$ & $[2.0, 3.6]$ & $[-156, -99]$\\
         \hline
         2021   &   &   &    &  \\
         \quad April 13 & $1.31 \pm 0.13$ & $-3.7 \pm 0.4$ & $[192 , 208]$ & $[2.8 , 3.9]$ & $[155 , 165]$\\
         \quad April 18 & $1.41\pm0.14$ & $-1.6 \pm 0.3$ & $[200, 208]$ & $[3.0, 4.7]$ & $[156, 168]$ \\
         \hline
    \end{tabular}
    \tablefoot{From ALMA we quote the flux density ($F$) and rotation measure (RM). These ALMA measurements were obtained using a calibration procedure similar to that described in \citet{Goddi2019}, but with additional refinements. These include T$_{\rm sys}$-based opacity corrections for the amplitude scale and residual $X-Y$ delay corrections for polarization calibration (see Carlos et al. in prep. for details). As a result, the derived $F_\mathrm{ALMA}$ and $\mathrm{RM}_\mathrm{ALMA}$ values for the 2017 and 2018 datasets differ slightly from those reported in \citet{Goddi2021}, although they remain consistent within $1 \sigma$ uncertainties. \textsuperscript{\textdagger}In 2018, the source underwent a very high-energy $\gamma$-ray flare between April 18 and 24 \citepalias{EHTMWL2018}.}
    \label{tab:acrossyearsresults}
   \end{center}
\end{table*}

\section{Discussion}\label{sec:discussion}

\subsection{Total intensity}

The general consistency between the images of \m87 reconstructed from the 2017, 2018, and 2021 observations is as expected from theoretical models and GRMHD simulations. All datasets produce a ring-like structure of $\sim 43\,\muas$ diameter, despite differences in the EHT array configuration and the resulting $(u,v)$ coverage. Nonetheless, well-constrained differences exist between the 2017 images and the subsequent ones; namely, an offset in the total intensity position angle and a higher source polarization in 2017. It is interesting to consider whether \m87 was in a special state in 2017, characterized by the aforementioned differences. 

In unresolved data, the 2018 observations show the most differences with other years, with a 3-day flare in very-high energy gamma-ray emission. Notably, there is no clear correlation between the EHT image or millimetre-flux variability and the very high-energy emission from \m87, which showed a $\sim 3$ day flare in 2018 (M87 MWL2018). The similarity between the 2018 and 2021 horizon-scale images suggests that this flare originated downstream in the jet, as supported by simple, single-zone modelling. However, we detected no accompanying changes in the flux of the nearest knot (HST-1) or outer jets. On the other hand, more complex scenarios are also consistent with our observations, such as magnetic reconnection in a current sheet near the black hole. Reconnection could produce the high-energy emission without altering the millimetre properties (see 
\citealt{Hakobyan2023};  \citetalias{EHTMWL2018} for more discussion of these and other possibilities). 

\subsubsection{Ring diameter}
With this work, we have further established the persistence of a ring-like emission feature on horizon scales in \m87. The stability of this feature further supports its interpretation as a black hole shadow from an optically thin accretion disk surrounding a Kerr black hole. After averaging the diameters from 2017, 2018, and 2021, we measured an average diameter of $43.9 \pm 0.6 \muas$. Our work corroborates the findings from \citet{Wielgus2020} that measured consistent ring diameters between the 2017 EHT measurements and the `proto-EHT' observations from 2009 to 2013. Additionally, given the stability of the ring, in a future work we will use the ring diameter measurements from the 2017, 2018, and 2021 observations to further improve the statistical error on the EHT's mass-to-distance ratio ($M/D$) estimate of \m87.

We did not compare the EHT spectral index measurements to ALMA measurements for two reasons. First, we only considered the upper two EHT sidebands in this publication, and second, ALMA's measurements of \m87's spectral index have significant contributions from regions $\gg 200\,\muas$. In future work, including new EHT observations with better $(u,v)$ coverage, we will investigate the spectral evolution of \m87 by computing spectral index maps between the 212\,--\,216 and 226\,--\,230\,GHz receiver sidebands. 

Finally, comparing our results to the 86\,GHz ring diameter measurements in \citet{Lu2023} and \citet{kim_imaging_2025}, the stability of the 230\,GHz ring supports the view that the larger diameter measured at 86\,GHz is due to synchrotron opacity effects. Moreover, future EHT analyses of the 260\,GHz data recorded in 2024 and the 345\,GHz data recorded in 2021, 2024, and 2025 will improve our understanding of the spectral nature of \m87's ring emission.

\subsubsection{Ring position angle}
The position angle shift of the ring's brightness peak between 2017 and 2018 is discussed in detail in \citet{M87_2018p2}.
As predicted in \citet{M87p5}, $\eta$ is expected to have a mean value around $\sim$$203\degr{}$ -- $209\degr{}$. This is about a $90\degr{}$ offset from the black hole spin axis, which, in turn, we assume to be closely aligned with the $288\degr{} \pm 10\degr{}$ large-scale jet direction \citep{Walker2018, Cui2023}. As the 2018 and 2021 measurements agree with this expectation and show a very similar source structure overall, it is imperative to understand the evolution and variability of the ring $\eta$ from further EHT observing campaigns in 2022 and onwards. This variability, most likely related to the turbulent character of the flow, may be an informative observable for comparing theoretical models of \m87 with observations \citep{Wielgus2020,M87_2018p2}. Comparing images from 2017, 2018, and 2021, it seems likely that \m87's ring-like structure is not as variable as inferred in \citet{Wielgus2020}, which may have been biased by the simple geometric model considered.

\subsection{Polarization}

The horizon-scale polarization measured by the EHT highly constrains numerical models of \m87, \citep{M87p8,M87p9} and Sgr A* \citep{SgrA.p8}. Past EHT papers have extensively compared EHT images to model images generated from GRMHD simulations, ray-traced with general relativistic radiative transfer codes. While the total intensity image from the EHT's 2017 observations only weakly constrained GRMHD models \citep{M87p5}, the addition of linear polarization constraints -- particularly $\mavg$ and $\angle \beta_2$ -- largely ruled out weak-magnetic field simulations in favour of MAD models as the preferred description of \m87's accretion flow \citep{M87p8}. These results are consistent with the upper limits on the circular polarization fraction reported in \citet{M87p9}. 

The qualitative differences in \m87's polarized image in 2017, 2018, and 2021 naturally raise two questions: are MAD models still preferred in all three years of EHT observations? Can existing GRMHD simulations explain the observed changes in the polarized image from year to year?

We defer a comprehensive scoring analysis of GRMHD images to future work and focus on the potential implications of two major differences in the polarized observations from 2018 and 2021 compared to 2017: the ${\approx}50\%$ lower beam-scale polarization fraction $\mavg$ in both 2018 and 2021, and the ${\approx}-60\degr{}$ shift in $\angle\beta_2$ from 2017 to 2021, resulting in a change in the sign of $\angle\beta_2$ from negative to positive. 

\subsubsection{Changes in $\mavg$}

Although both 2018 and 2021 EHT images of \m87 are significantly more de-polarized than the 2017 image ($\mavg\approx8\%$ in 2017 vs ${\approx}3-4\%$ in 2018 and 2021), images from both years are still likely to be more consistent with MAD GRMHD models than weakly magnetized `standard and normal evolution' (SANE) models. Although the observed changes in fractional polarization are significant, most SANE simulations are even more de-polarized than the 2018 and 2021 results would indicate.  
In particular, most SANE simulations have both $\mavg<2.5\%$ and $|\beta_2|<2.5\%$ (\citealt{M87p8}, Fig. 7, 9). SANE disks are significantly more de-polarized than MAD disks, partly due to increased plasma turbulence but largely due to significantly higher Faraday depths through the disk from the higher density plasma needed (compared to MAD models) to produce the observed total flux density \citep{Moscibrodzka2017}. SANE models are also more likely to overproduce circular polarization and violate the EHT's upper limits on the resolved circular polarization fraction $\langle|v|\rangle$ \citep{Ricarte21,M87p9}.

While SANE models are naturally de-polarized, most MAD models naturally produce more linear polarization than is observed in \m87 ($\mavg\gtrsim10\,$\% after convolution with a $20\,\muas$ Gaussian). De-polarizing MAD models requires relatively cold electrons, with the EHT model scoring tentatively preferring an ion-to-electron temperature ratio $R=T_{\rm i}/T_{\rm e}\gtrsim 40$ to explain the low fractional linear polarization observed in 2017. Producing the decreased fractional linear polarization seen in 2018 and 2021 with MAD GRMHD models will require even colder electrons (larger $R$). While ideal GRMHD simulations like those used in \citet{M87p5, M87p8, M87p9} require $R$ to be assigned manually in post-processing, radiative two-temperature simulations can predict $R$ by evolving separate ion and electron entropies self-consistently after adopting a sub-grid model for how the different species are heated. Recent surveys of two-temperature simulations found relatively low values of $R\approx1-10$ in the EHT emission region of GRMHD simulations  \citep{Moscibrodzka25,Chael25}, 
with correspondingly large $\langle|m|\rangle\approx20-40\%$. The fact that EHT observations suggest that cold thermal electrons in the inner $5r_{\rm g}$ are necessary to produce the observed low fractional polarization poses an intriguing tension with existing two-temperature models. This tension demonstrates that EHT observations can directly constrain plasma heating mechanisms near the black hole.

\subsubsection{Changes in $\angle\beta_2$}
In the absence of Faraday rotation, the sign of $\angle\beta_2$ for face-on systems like \m87 encodes the direction of electromagnetic energy flux; if magnetic fields are presumed to co-rotate with the plasma clockwise on the sky \cite{M87p5}, the observed sign of $\angle\beta_2<0\degr{}$ in 2017 and 2018 is consistent with outward electromagnetic energy flux in analytic Blandford-Znajek monopole \citep{BZ77} and GRMHD simulations \citep{Chael23}. The observed $\angle\beta_2\in[161\degr{},166\degr{}]$ in 2021 is not immediately consistent with either the measured value of $\angle\beta_2$ in 2017 and 2018 or this theoretical expectation.

The ${\approx}-60\degr{}$ shift in $\angle\beta_2$ from 2018 to 2021 could be caused by several different changes in the near-horizon emission region, including: (1) a change in the underlying magnetic field structure; (2) a change in the degree of Faraday rotation along the line of sight; and (3) evolving contributions from different emission regions along the line of sight (e.g. in the disk or jet); or some combination of all three.

As reported in \autoref{tab:acrossyearsresults}, the measured Faraday rotation measure from unresolved ALMA observations, $\mathrm{RM}_{\rm ALMA}$, is largely consistent in magnitude across the three years of EHT observations reported here, though the rotation measure is known to vary in sign and magnitude on short timescales (e.g. changing sign from positive to negative between April 5 and 11, 2017). However, the unresolved ALMA observations are significantly affected by large-scale polarized emission from M87's jet that does not contribute to EHT observations of the ${\sim}40\,\muas$ scale core structure. \citet{Goddi2021} used a two-component Faraday screen model, constrained by ALMA and EHT observations, to attempt to disentangle the $\mathrm{RM}$ in the core and the extended jet; they found that the magnitude of the variable core $\mathrm{RM}$ in this model can significantly exceed the value measured by ALMA. 

A preliminary application of the two-component model to the 2017, 2018, and 2021 observations suggests an increase in the core Faraday rotation in 2021 sufficient to provide consistent de-rotated $\angle\beta_2$ values across all three years. However, comparing the de-rotated $\angle\beta_2$ values with simulation images requires careful accounting of both the potentially large uncertainties in the core rotation measure from the two-component screen model \citep{Goddi2021} and the intrinsic internal rotation measure produced in GRMHD simulations \citep{Ricarte20}. Therefore, we leave the detailed quantitative analysis to a future paper.

If changes in \m87's Faraday screen are insufficient to explain the observed changes in $\angle\beta_2$, 
the likely alternative explanation is that the EHT is capturing changes in the emission region magnetic field or shifts in the location of the emission region(s) between 2017, 2018, and 2021.
MAD GRMHD simulations have intrinsically variable $\angle\beta_2$ values \citep{Palumbo2020}, but the distributions of $\angle\beta_2$ for MAD simulations with a fixed black hole typically do not change sign when applying standard assumptions (e.g. electron heating prescriptions) in post-processing \citep{M87p8, Chael23}. Further EHT monitoring of the polarization structure will indicate whether the observed variability in $\angle\beta_2$ is consistent with GRMHD variability, plasma propagation effects, or a large-scale change in the magnetic field topology between 2017 and 2021. 

In follow-up works, we will further constrain the core Faraday rotation measure with the \citet{Goddi2021} two-component model and directly compare de-rotated $\angle\beta_2$ in the 2018 and 2021 EHT observations to GRMHD simulations, including uncertainties in the modelled core rotation measure. We will present a comprehensive comparison of the polarized EHT images in 2017, 2018, and 2021 to several GRMHD simulation libraries and discuss whether all three years of EHT observations still support the conclusion that \m87 is in a MAD accretion state \citep{M87p8,M87p9}. Moreover, we will discuss whether or not the observed variability in $\angle\beta_2$ after Faraday de-rotation is consistent with the MAD picture that $\angle\beta_2$ tracks field lines with outward Poynting flux and is sensitive to black hole spin \citep{Palumbo2020,Chael23}.

\section{Summary and conclusions}\label{sec:summary}

We have presented the first multi-epoch, polarimetric imaging of \m87 on event-horizon scales using EHT observations from 2017, 2018, and 2021. This work includes the first results from the 2021 EHT observing campaign, which added the 12\,m Kitt Peak Telescope and NOEMA, substantially improving the baseline coverage.

From the measured visibility data, we observed the same overall radial visibility amplitude plot across 2017, 2018, and 2021, with the 2021 data being of very high quality. For all three years, we found a visibility amplitude null at $\sim 3.4\,{\rm G}\lambda$ and a second null at $\sim 8.3\,{\rm G}\lambda$, characteristic of a ring-like structure in the image. Analysing the polarized or cross-hand visibilities, we found that in 2018 and 2021, the data are de-polarized relative to 2017. Furthermore, we found that in 2018, there were no significant non-zero conjugate closure trace products, further evidence that \m87 was largely de-polarized. Similar to 2017, we found small offsets in the differences between the right and left circular polarization closure phases in 2021. However, structural maps of the measured circular polarization are still dominated by instrumental systematics. 

We employed seven distinct methods for polarimetric imaging and instrumental calibration, each validated via several synthetic data tests. When applied to EHT observations of \m87, each method produces images dominated by a \mbox{${\sim}\,42-46\,\muas$} ring for each year, with a brightness maximum in the south. Despite this broad consistency, the images show variation from year to year in terms of the azimuthal brightness distribution of the ring. The 2017 images exhibit the largest differences from the two other years. Notably, the ring appears identical in 2018 and 2021, even though \m87 had a recorded gamma-ray flare just prior to the 2018 EHT observation. Future EHT observations of \m87, including completed campaigns in 2022 and 2024 \citep[see e.g.][]{EHT_midrange}, will further elucidate the typical state of its accretion flow and its turbulence statistics.

Unlike the total intensity, we found that the polarized structure of \m87 changes dramatically every year. The ring observed in 2018 and 2021 is strongly de-polarized relative to 2017, with only a single region having a fractional linear polarization of $\sim$5 -- 10\%. Furthermore, the polarization pattern shows significant variations, with $\angle\beta_2$ rotating by $\sim$$60\degr{}$ and changing the overall EVPA helicity in 2021. These changes highlight the importance of the EHT multi-frequency observations at 345 GHz in 2021, 2024, and 2025 and at 260\,GHz in 2024 to separate the internal polarization structure from external Faraday effects. The implications of these results on our understanding of the accretion flow around \m87 will be analysed in a future publication and demonstrate the EHT's capabilities to constrain plasma physics around SMBHs.

To clarify the location and process driving the 2018 very high-energy $\gamma$-ray flaring, a full theoretical study that simultaneously takes into account  the changes in the EHT image flux and polarization, as well as the ring and inner jet PA between 2017-2021, is needed. However, such a study will be limited by the annual cadence of our observations, and the image quality in 2018 was overall worse than in 2017. The potential to link dynamics to particle acceleration in real time provides strong motivation for a high-cadence (every few days), longer-duration (over months) intensive monitoring campaign of \m87 \citep{EHT_midrange}, together with multi-wavelength coverage, including the high-energy facilities.  
If another $\gamma$-ray flare is detected, such a campaign would drastically improve our chances of definitively identifying the physics driving at least one mechanism for very high-energy flaring in AGNs.

Finally, the 2021 data provide the first structural hints of \m87's extended non-ring emission up to $1\,{\rm mas}$ thanks to the addition of the Kitt Peak and NOEMA telescopes. Using closure phases on triangles that include short EHT baselines, we demonstrate that the compact ring emission does not fully fit the EHT data. Specifically, all our imaging methods fail to fit the ALMA-PV-NOEMA and ALMA-SMT-KP closure phases when confined to $\sim100\times100\,\muas^2$, showing residuals of the order of a few degrees. However, given the limited EHT coverage on these scales, we could not robustly image the extended emission.
The measurements that probe the extended non-ring emission are best described by simple Gaussian components. A more involved modelling or imaging of the extended jet structure would overfit the data, particularly for pre-2021 EHT observations, when the important closure triangles were not present.
Future EHT observations with improved intermediate baseline coverage, for example the 2022 EHT array, may enable the direct imaging of these extended structures at 230\,GHz. These images would provide a direct view of the connection between the jet launch region and the black hole shadow \citep[see e.g.][]{EHT_midrange}.

\bibliographystyle{aa}
\bibliography{references}

\begin{appendix}
\onecolumn
\section{Data consistency and further checks}
\begin{figure}[!h]
    \centering
    \includegraphics[width=0.49\columnwidth]{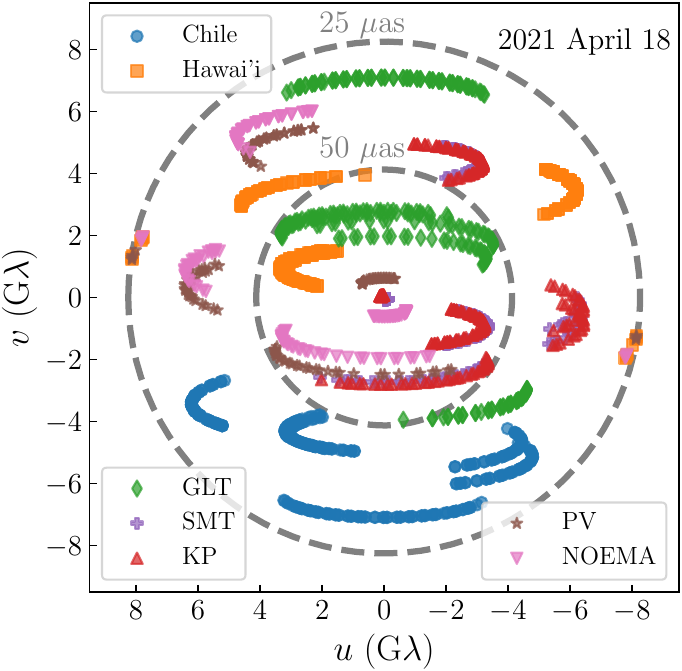}
    \includegraphics[width=0.49\columnwidth]{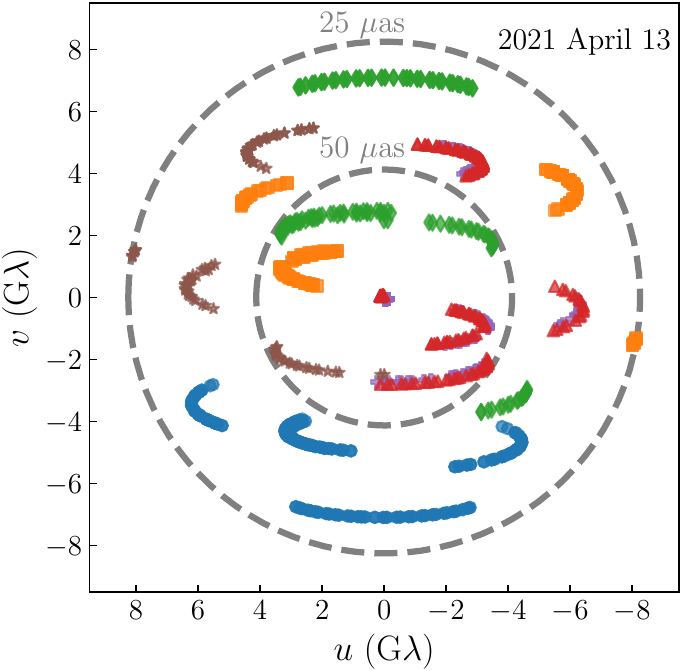}
    \caption{\m87 $(u, v)$ coverage of VLBI scans shown for the band 3 (227.1\,GHz) data on April 13 and 18, 2021. The co-located ALMA, APEX, JCMT, and SMA stations are described by the Chile and Hawaii points, respectively. The labelling is done by antenna, using conjugate baselines to display the other station.}
    \label{fig:uvcov}
\end{figure}

\subsection{Coverage and systematic errors}

Figure \ref{fig:uvcov} compares the \m87 $(u, v)$ coverage between the two 2021 observing days, April 13 and 18, analysed in this work. We focused on the April 18 data due to the substantially improved $(u,v)$ coverage provided by NOEMA.

\autoref{tab:syserr} shows an overview of our systematic errors from observations over the years. 2021 has the best $(u, v)$ coverage, allowing for more accurate calibration and overall the lowest systematics, as reflected by the data quality. Our sensitivity to non-zero trivial closure quantities is explored in \citet{Georgiev2025} and briefly discussed in Sect. \ref{sec:discussion}.

\begin{table*}[h]
\centering
\caption{Non-closing systematic uncertainties for \m87.}
\adjustbox{max width=\textwidth}{
\begin{tabular}{c|ccccccccccc}
\hline
\hline
 Test  & \multicolumn{3}{c}{2017 Apr 11} &~~&  \multicolumn{3}{c}{2018 Apr 21} &~~& \multicolumn{3}{c}{2021 Apr 18}  \\ \cline{2-4} \cline{6-8} \cline{10-12}
& $s$ & $s/\sigma_{\rm th}$ & $n$ && $s$ & $s/\sigma_{\rm th}$ & $n$ && $s$ & $s/\sigma_{\rm th}$ & $n$ \\
\hline
lo\,$-$\,hi closure trace phases         &  2.8$\degr{}$  &  0.5 &  220 &&  3.4$\degr{}$ &  0.6 &   99 &&  2.1$\degr{}$ &  0.5 & 4321 \\
trivial closure trace phases             &  0.3$\degr{}$  &  0.0 &   63 &&  0.0$\degr{}$ &  0.0 &   63 &&  0.9$\degr{}$ &  0.2 & 1844 \\
lo\,$-$\,hi log closure trace amplitudes &  5.4\%        &  0.5 &  220 &&  7.9\%       &  0.7 &   99 &&  3.9\% &  0.5 & 4321 \\
trivial log closure trace amplitudes     &  3.2\%        &  0.3 &  151 &&  7.1\%       &  0.7 &   63 &&  3.3\% &  0.5 & 1844 \\
\hline
\hline
\end{tabular}}
\tablefoot{Non-closing systematic uncertainties, $s$ (and in units of thermal noise, $s/\sigma_{\rm th}$), are estimated using various statistical tests for each year.  For reference, the number of closure quantities used for each test, $n$, is listed.}
\label{tab:syserr}
\end{table*}

\newpage
\subsection{Evidence of circular polarization}

\begin{figure}[h!]
    \centering
    \includegraphics[width=\textwidth]{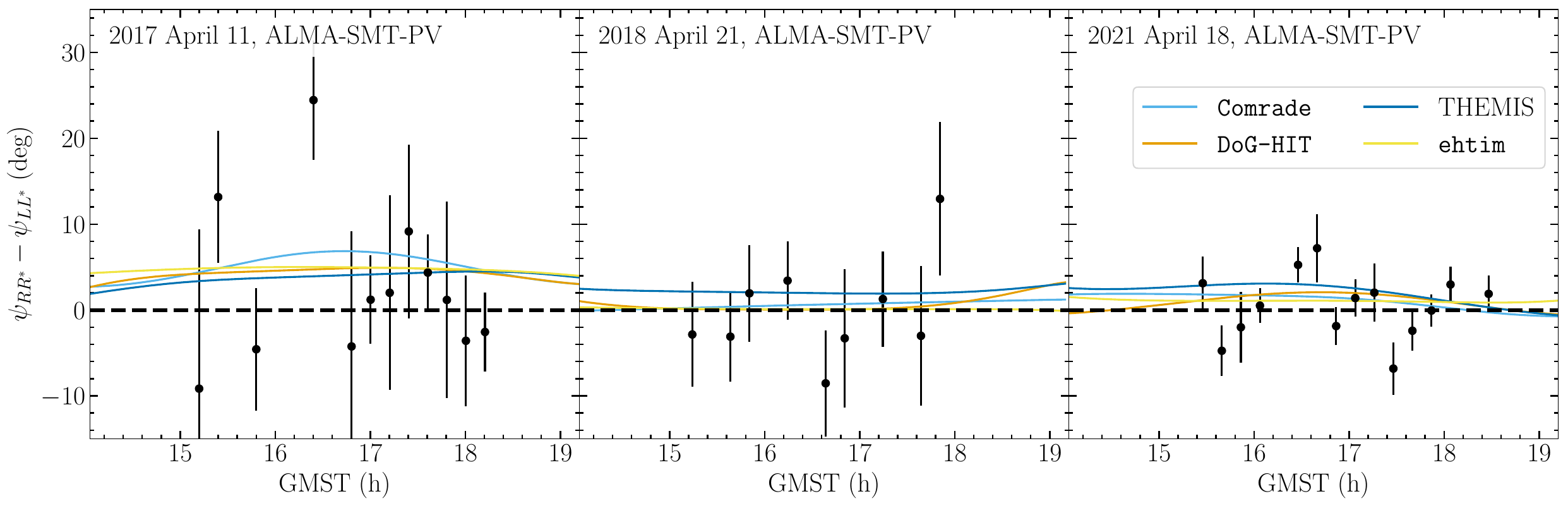}
    \includegraphics[width=\textwidth]{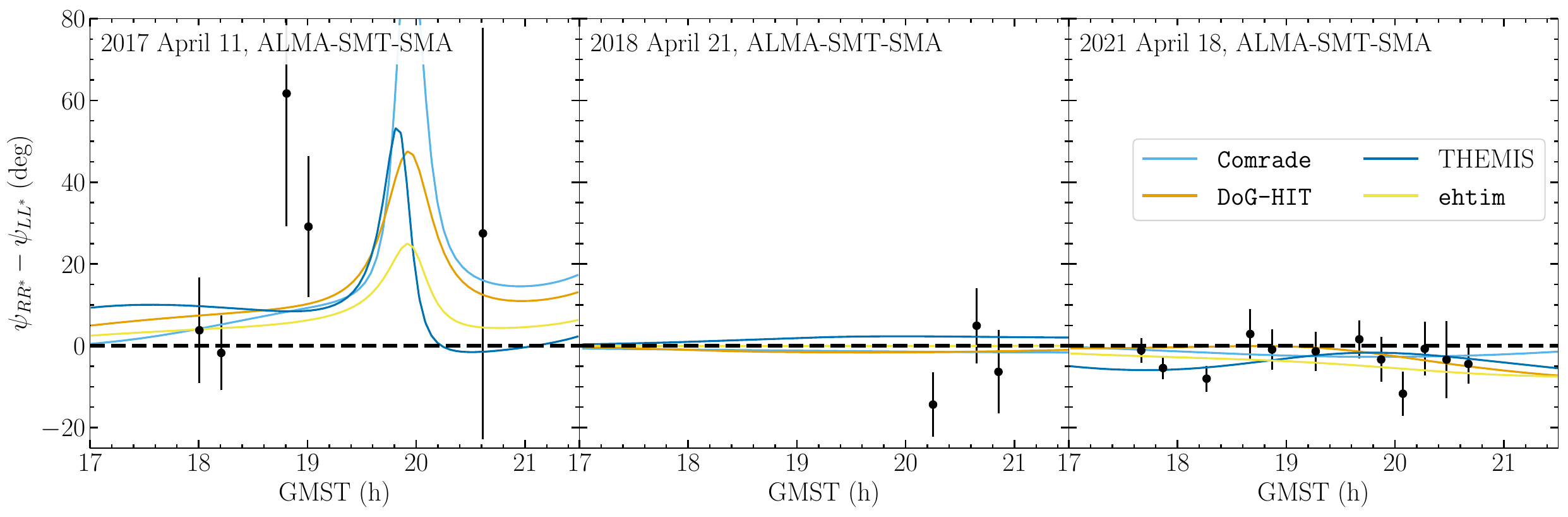}
    \caption{$RR^* - LL^*$ closure phase differences that describe source circular polarization structure shown for two triangles in 2017, 2018, and 2021. The data are averaged over 12-minute-long segments.}
    \label{fig:cplrdiff}
\end{figure}

Figure \ref{fig:cplrdiff} shows the RR-LL closure phase differences, which are a robust estimate of the existence of circular polarization. Overall, smaller uncertainties in the 2021 data lead to small but visible non-zero Stokes~$\mathcal{V}$ signals on some closure triangles. Our polarized imaging tools pick up these biases, but residual gain ratio uncertainties are too significant to reconstruct a reliable circular polarization map from the data.
\FloatBarrier

\twocolumn
\subsection{2021 HOPS comparison}
\begin{figure}[h!]
    \centering
    \includegraphics[width=\linewidth]{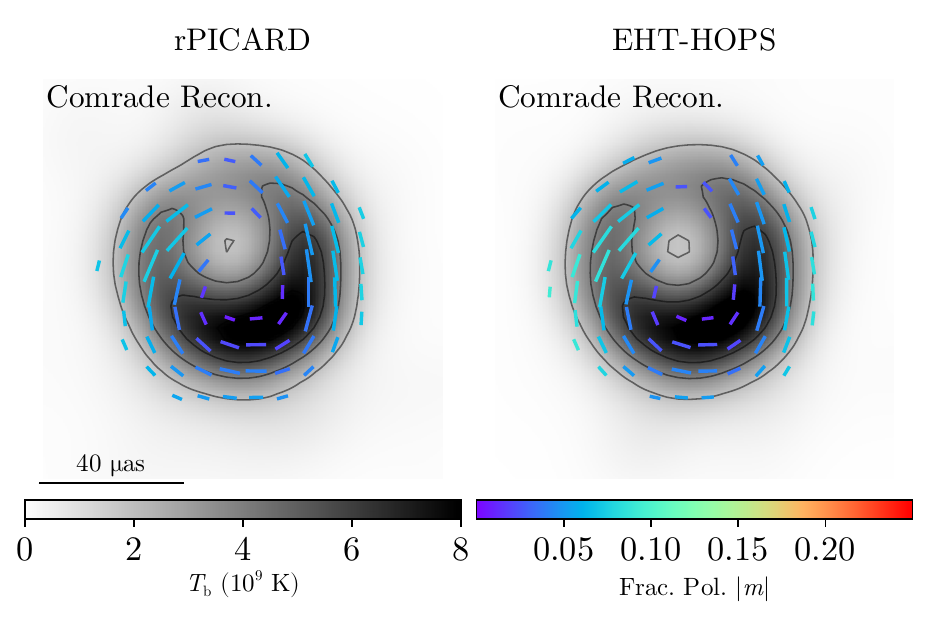}
    \caption{Image reconstructions averaged over bands 3 and 4 using the \comrade pipeline and the April 18, 2021, reduction from the standard \texttt{rPICARD} pipeline used in this paper (left) and the \texttt{EHT-HOPS} pipeline (right). Both images have been normalized to have the same compact flux due to residual issues with the EHT-HOPS calibration of NOEMA.}
    \label{fig:hops_v_casa_reco}
\end{figure}

In the main text, we only considered the \texttt{rPICARD} reduction for the 2021 data. The primary reason was that \texttt{HOPS} cannot currently correct for the NOEMA phase jumps noted in Appendix \ref{app:noema}. To remove the effects of the phase jumps, the \texttt{EHT-HOPS} pipeline instead selects the largest part of the data within each sub-band that exhibits a stable phase and flags the rest at the cost of sensitivity on NOEMA baselines. 

To test whether the NOEMA differences and other differences between the \texttt{EHT-HOPS} and \texttt{rPICARD} pipelines could impact the results \citep[see][for an explanation]{M87p3}, we reimaged the \m87 April 18, 2021, HOPS reduction using the \comrade pipeline and compared it with the results presented in the main text. Figure \ref{fig:hops_v_casa_reco} shows the image reconstructions from the \texttt{rPICARD} and \texttt{EHT-HOPS} pipelines. Likely due to residual issues in the HOPS NOEMA calibration, we found a large amplitude offset in the NOEMA gain. Due to this offset, we measured a smaller compact flux for the HOPS data. To fix this, we renormalized the total flux to match the total flux from the fiducial 2021 image. After this adjustment, we found very similar image reconstructions between the \texttt{rPICARD} and \texttt{EHT-HOPS} pipelines. This result demonstrates the robustness of our results to different calibration pipelines.

\subsection{Cross-check of the 2021 polarized imaging results 
from April 13}\label{appendix:april13_2021}
\begin{figure}[h!]
    \centering
    \includegraphics[width=\columnwidth]{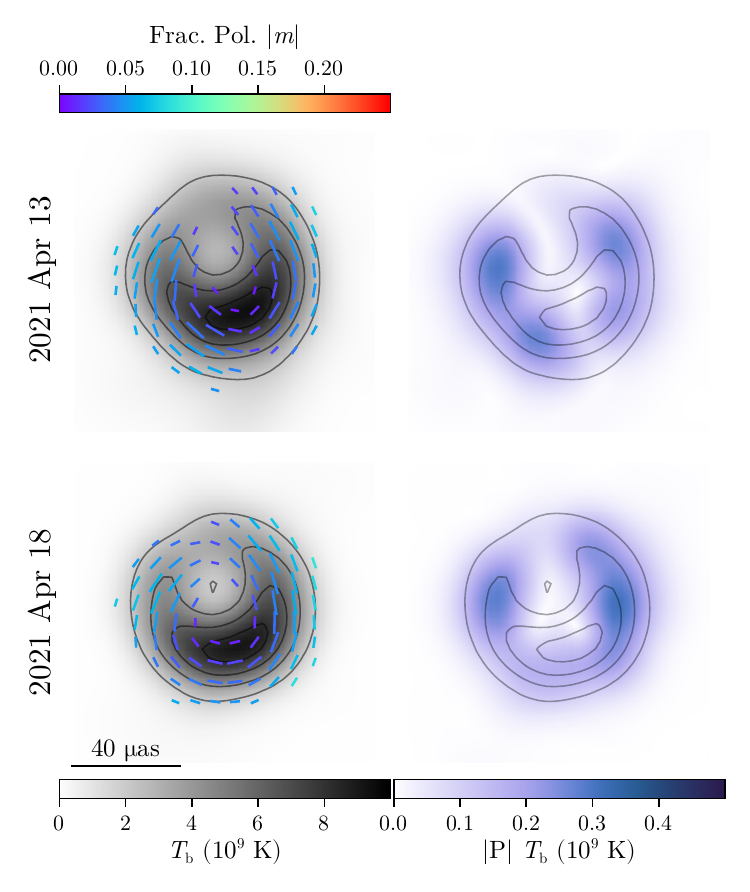}
    \caption{Fiducial images from April 13, 2021, and April 18, 2021. The April 13 image is constructed in the same manner as Fig. \ref{fig:fiducial}.}
    \label{fig:app:April13_18_comp}
\end{figure}

\begin{figure*}[h!]
    \centering
    \includegraphics[width=\textwidth]{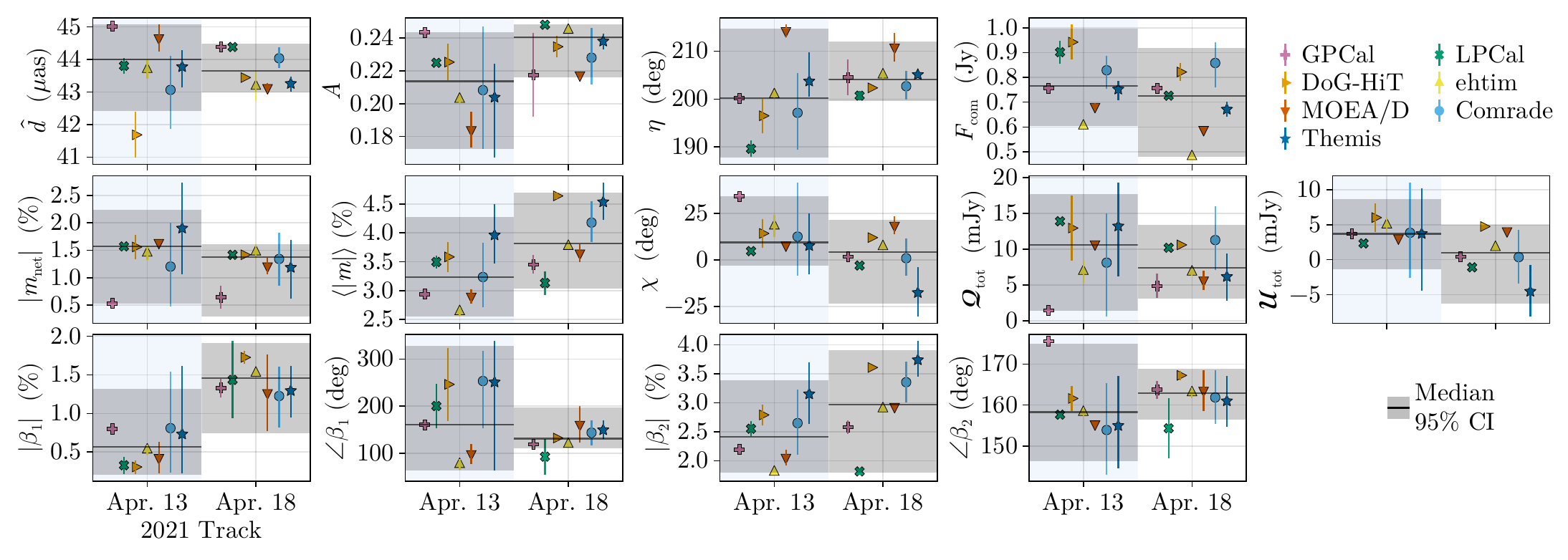}
    \caption{Comparison of imaging parameter estimation results from April 13 and 18, 2021. The conventions follow Fig. \ref{fig:m87_params_uniform}.}
    \label{fig:app:April13_18_comp_params}
\end{figure*}

In the main text, we focus on the results from April 18, 2021. This day was selected because it was the only track in 2021 where the entire 2021 EHT array observed \m87 for a significant period. The next best day is April 13, 2021, in terms of coverage; however, NOEMA did not participate due to bad weather. As a result, April 13, 2021, lacks any baselines on the scales of $\sim 100-500\,\muas$, weakening the constraining power of the EHT on intermediate scales. Given the issues with NOEMA in the 2021 observation described in Appendix \ref{appendix:data}, April 13 provides an additional check that NOEMA is not dramatically altering the conclusions of April 18. Additionally, since the \m87 observations are 5 days apart, this allowed us to test for any significant evolution in the source, which is expected to be minimal given the dynamical timescale of \m87 \citep{M87p4, M87p5, Georgiev}. 

Figure \ref{fig:app:April13_18_comp} shows the fiducial image reconstruction produced by averaging over the 7 methods in Sect. \ref{sec:imaging} on April 13, 2021  (top row) and April 18, 2021 (bottom row). From Fig. \ref{fig:app:April13_18_comp}, we see that the two images display a very similar EVPA pattern and that the overall fractional polarization appears qualitatively similar. Both images indicate that the ring is de-polarized in the brightest region, and the polarized emission is concentrated in the east-west portion of the ring. The similarity in the EVPA pattern is noteworthy given that the ALMA rotation measure differed by a factor of 2 (see \autoref{tab:acrossyearsresults}) between April 13 and April 18.

Looking at the images on April 13 and 18 quantitatively, Fig. \ref{fig:app:April13_18_comp_params} shows the estimated parameters for both days. After averaging over all methods, the grey bands indicated that the total intensity image parameters were consistent across the two days. We see a similar result for the linear polarization parameters, except for $|\beta_1|$, where there was a $>1\sigma$ difference in the method average, although a $|\beta_1| \approx 1\%$ is within the $2\sigma$ error bars. The measurement of $\angle\beta_1$ also showed some discrepancies between April 13 and 18, although again it is not significant. Finally, we also found that the leakage estimates on April 13 agree with those on April 18 across all methods and sites.

\section{Data issues and mitigation strategies}
\label{appendix:data}

\subsection{NOEMA instrumental phase calibration}\label{app:noema}
NOEMA experienced phase jumps at the edges of each of its 64\,MHz sub-bands. Each jump amounted to a multiple of $90\degr{}$. The phase jumps are mostly constant in time, but change in value when the NOEMA PolyFiX backend is reset during observations \citep[see][for a description of NOEMA's phasing system]{pietu2025}. During the April 18 observations, the backends were reset twice, leading to multiple changes in the instrumental phases. In addition, the EHT data are correlated using the $58\,\mathrm{MHz}$ sub-band \textit{outputbands} mode, which results in phase jumps within the final EHT sub-bands. The CASA \textsc{fringefit} task used by \texttt{rPICARD} cannot fit for phase jumps within sub-bands (spectral windows in the CASA Measurement Set). Thus, we calibrated these phase jumps in a dedicated calibration step. 

For the \texttt{rPICARD}-calibrated data, we extended our pipeline with a dedicated NOEMA instrumental phase calibration step. We re-grid the data virtually by defining visibility chunks in the native $64\,\mathrm{MHz}$ gridding used by the NOEMA backend. These are then fit separately for each scan and NOEMA frequency window. To increase robustness, we required that the determined phase jumps are multiples of $90\degr{}$. We also require that the phase jumps are consistent for more than $5$ scans, to ensure that only real changes due to the infrequent resets of the PolyFiX backend are accepted.

The current HOPS processing environment does not support non-linear phase bandpass correction within a 58 MHz sub-band at this moment\footnote{This capability will be introduced in HOPS-4 \citep{hoak_2022}}. To mitigate de-coherence from the uncorrected phase jump, the \texttt{EHT-HOPS} pipeline flags the smaller portion of each sub-band before or after the jump, which allows the remaining bandpass to be calibrated using standard corrections. This process removes $\sim$25\% of the total integrated bandwidth for NOEMA baselines, reducing overall sensitivity.

In late 2023, the origin of the phase jumps was identified in the
signal processing firmware of the PolyFiX correlator. A PolyFiX firmware update was implemented to remove these phase offsets in the NOEMA passband. The new firmware version was installed at the NOEMA observatory in January 2024 and has been successfully tested on-sky. Consequently, the dedicated phase calibration of the NOEMA visibilities will not be required for EHT data from 2024 onwards. Further details on the rPICARD calibration steps and the PolyFiX firmware update are
described in \citet{vonFellenberg2025}.

\subsection{JCMT leakage}
\label{app:jcmtD}

In the 2018 data, we infer that JCMT had an unusually large polarization leakage of $\sim\,30\,$\%.
We traced the likely origin to a $17\degr{}$ misalignment of the quarter-wave plate (QWP).
The QWP is aligned regularly using data from a baseline between the JCMT and a single SMA antenna. The optimal angle is found by nulling the signal, i.e. minimizing the cross-correlation signal for a calibrator target (3C\,84 was used in 2018). From this position, where JCMT and SMA record in opposite polarizations, the QWP is rotated back by $90\degr{}$ to align the JCMT polarization with the SMA. In 2018, this procedure was complicated by noisy data from poor weather conditions, making QWP alignment difficult. To remove the impact of the large JCMT leakage in 2018, all imaging methods flagged JCMT baselines.

\subsection{2021 specific calibration steps}
Three stations required dedicated calibration steps in 2021. 
PV experienced a change in instrumental R-L delay at 00:37 UT on April 18. To account for this, we allowed for dedicated, per-VLBI-scan instrumental delay calibration of PV. KP showed a small drift in the L-R phase over time, which was corrected through self-calibration during imaging for the \comrade pipeline.

On April 13, APEX erroneously executed observations with the activated `wobbler' for the first $\approx 2\,\mathrm{h}$, which resulted in the antenna pointing off-source for roughly $50\%$ of the time, leading to a correspondingly halved visibility amplitude, without affecting the phase stability. The resultant time-dependent amplitude gain error led to poor network calibration. 
We thus decided to calibrate the resulting APEX gain error before post-processing by fitting a Gaussian component to the data in \difmap. We then amplitude self-calibrate the APEX station, keeping all other stations fixed. The resulting corrected data are exported and used for further post-processing (i.e. network calibration).

\subsection{GLT system temperatures on April 13}
The GLT did not record system temperatures on April 13. Using the methods outlined in Section 5.2.4 of \citet{eht_memo_koay_2023-l1-01}, we instead used modelled $T_\mathrm{sys}$ values, which resulted in large GLT amplitude gain errors that need to be corrected through self-calibration.

\section{Details of the polarized imaging methods} \label{appendix:imaging-methods} 

\subsection{CLEAN methods}

\subsubsection{\difmap/\gpcal}
\label{section:difmap_gpcal}
We conducted imaging with CLEAN using \difmap{} \citep{Shepherd_1997,Shepherd_2011} based on the imaging pipeline utilized for the 2018 EHT images of \m87{} \citep{M87_2018p1}. In summary, the pipeline first searches for the best geometric model among a circular Gaussian component with 15\,\uas{} FWHM (representing an unresolved symmetric model), a uniform disk with sizes ranging from 56 to 84\,\uas{} in steps of 4\,\uas{}, and a uniform ring with sizes ranging from 36 to 68\,\uas{} (also in steps of 4\,\uas{}, without width). The best model was selected based on the closure phase normalized $\chi^2$, and phase-only self-calibration was conducted using the model. The pipeline surveys five parameters: the total assumed compact flux density, cleaning stopping condition, relative weight correction factor for ALMA in self-calibration, diameter of the CLEAN window, and the power-law scaling of the $(u,v)$ density weighting function. The same parameter ranges used for the 2018 \m87{} imaging were used. The final representative image was selected based on the lowest overall $\chi^2_{\rm CPh}$ and $\chi^2_{\rm LCA}$. We did not perform the top-set image selection procedure used in the 2017 and 2018 EHT imaging of \m87{}.

We utilized \gpcal \citep{Park2021, Park2023a, Park2023b}, a novel instrumental polarization calibration pipeline based on AIPS and \difmap, to correct the polarimetric leakages in the visibilities. The pipeline first derives the leakages of the Chile and Hawaii intra-site stations, ALMA, APEX, SMA, and JCMT, using the visibilities of the intra-site baselines (ALMA-APEX and SMA-JCMT) only. In the derivation, the source structure was assumed to be point-like. Then, the pipeline derives the leakages of the other stations by utilizing all the baselines except for the intra-site baselines and by fixing the leakages of the intra-site stations during the fitting procedure.

Linear polarization images were obtained with CLEAN using \difmap{} and the leakage-corrected visibilities, as done in the 2017 polarimetric imaging of \m87{} \citep{M87p7}. CLEAN was used for Stokes $\mathcal{Q}$ and $\mathcal{U}$ images using the CLEAN windows utilized for the Stokes $\mathcal{I}$ imaging, until the peak in the residual dirty image within the windows reached the RMS noise level of the dirty image.

\subsubsection{\difmap/\lpcal}
\label{section:difmap_lpcal}
For the determination of the instrumental polarization of the antennas (so-called $D$ terms), we also utilized the traditional AIPS-based task \lpcal \citep{Leppanen1995}, which has been used extensively in VLBI polarimetry during the past two decades. The total intensity image (Stokes I) was performed with \difmap{} in the same way as described in Appendix \ref{section:difmap_gpcal} but with different hyperparameters, for example the number of iterations and data flags. The edited and fully self-calibrated data were then imported to AIPS, where the AIPS task CALIB was used to remove the gain ratios between the parallel-hand polarizations ($RR^*$ and $LL^*$). We then applied the AIPS task CCEDT to split the $\delta$ components of the Stokes I image into up to 10 compact sub-regions of constant fractional polarization, which are used as input models for the AIPS task \lpcal. The antenna $D$-terms derived from \lpcal were then applied to the data using another round of amplitude and phase self-calibration with the
option DOPOL in the CALIB AIPS task. Finally, the data are exported from AIPS back to a local data file, from which the final linear polarization images are obtained in the same manner as described in the last paragraph of the previous section.

\subsection{Regularized maximum likelihood methods} \label{app:rml}

\begin{table*}[!t]
\caption{Data and regularization terms used for the RML techniques \ehtim, \doghit, and \moead.}\label{tab:objectives}
\vspace{-6mm}
\label{tab: rml}
\begin{center}
    \begin{tabular}{l l l l l l l}
    \hline
    \hline
        Data & Type & Obj. & Description & Ref. & 
weight & Method \\ \hline
        $\mathcal{I}$ & Data& $\chi^2_{\rm CPh}$ & fit quality to closure phases & 1& 10 & \ehtim, \doghit\\
         && $\chi^2_{\rm LCA}$ & log. of closure amplitudes & 1& 10 & \ehtim, \moead, \doghit\\
        && $\chi^2_{amp}$ & amplitudes & 1& 0.1 & \ehtim, \moead\\
        && $\chi^2_{vis}$ & complex visibilities & 1& 0 & \ehtim \\
        & Reg. & $R_{entr}$ & entropy & 1& 100 & \ehtim, \moead\\
        &  & $R_{tv}$ & total variation & 1& 1 & \ehtim, \moead\\
        &  & $R_{tv2}$ & total squared variation & 1& 1 & \ehtim, \moead\\
        &  & $R_{l1}$ & $l^1$-norm & 1& 0 & \ehtim, \texttt{MOEA/D}\\
        &  & $R_{flux}$ & total flux constraint & 1& 100 & \ehtim, \moead\\
        &  & $R_{l1w}$ & $l^1$-norm of wavelets & 2 & & \doghit \\
        \hline
        lin. pol. &  Data & $\chi^2_{pvis}$ & fit quality to LP visibility $\mathcal{P}$ & 3& 1 & \ehtim\\ 
        & & $\chi^2_{m}$ & visibility polarimetric ratio $\mathcal{P}/\mathcal{I}$ & 3 & 1 & \ehtim \\
        & Reg. & $R_{ptv}$& total variation of $\mathcal{P}=\mathcal{Q}+i\mathcal{U}$ & 3& 1 & \ehtim, \doghit\\
        &  & $R_{ms}$& entropy of $\mathcal{P}=\mathcal{Q}+i\mathcal{U}$ & 3& 0 & \ehtim, \moead\\
        &  & $R_{hw}$& HW-entropy of lin. pol. fraction $m$ & 3& 100& \ehtim\\
        &  & $R_{hw+cp}$& HW-entropy of total pol. fraction $m+v$ & 4 & & \moead\\ 
        &  & $\mathbf{1}_{ms}$ & constrained to Stokes $\mathcal{I}$ wavelets & 5 & & \doghit \\ \hline
        circ. pol. & Data & $\chi^2_{cvis}$ & fit quality to CP visibility $\mathcal{V}$ &6 & 0.1 & \ehtim, \moead\\
        & Reg. & $R_{l1v}$ & $l^1$-norm of circ. pol. & 6 & 0.1 & \ehtim \\
        &  & $R_{vtv}$ & total variation of circ. pol. & 6 & 10.0 & \ehtim \\
        &  & $R_{hw+cp}$& HW-entropy of total pol. fraction $m+v$ & 4& & \moead\\
        &  & $\mathbf{1}_{ms}$ & constrained to Stokes $\mathcal{I}$ wavelets & 5 & & \doghit \\ \hline
    \end{tabular}
\end{center}
\vspace{-6mm}
\tablefoot{We show the top-set weights for \ehtim in the sixth column. \moead surveys all weight combinations internally, \doghit performs constrained optimization. \\References. 1: \cite{M87p4}, 2: \cite{Mueller2022}, 3: \citep{Chael2016}, 4: Toscano et al. (in prep.), 5: \cite{Mueller2023b}, 6:  \cite{SgrA.p7}. }
\end{table*}

Regularized maximum likelihood techniques balance data fidelity and regularization assumptions on the image, such as smoothness, sparsity, or entropy. This is achieved by minimizing a weighted sum of terms measuring data fit quality, and terms measuring its feasibility:
\begin{align}
    \hat{{I}} \in \mathrm{argmin}_{I} \sum_d \alpha_d \chi^2_d({I})+\sum_x \beta_x R_x({I}). \tag{\ref{eq:rml} revisited}
\end{align}

Throughout this manuscript, three algorithms are applied that utilize this concept. These are \ehtim \citep{Chael2016, Chael2018}, \doghit \citep{Mueller2022,Mueller2023b}, and \moead \citep{Mueller2023c, Mus2024a}. While all these algorithms apply similar heuristics, they differ significantly in the regularization assumptions applied and in the minimization problem in Eq. \eqref{eq:rml} is solved. A comprehensive overview is presented in \autoref{tab:objectives}.

In particular, \ehtim has been used intensively and successfully in previous EHT data analyses of black hole shadows in total intensity \citep{M87p4, SgrA.p3, M87_2018p1} and polarization \citep{M87p7, SgrA.p7}. The robustness of and confidence in the \ehtim imaging stemmed partly from thorough and detailed parameter surveys and studies on synthetic data. We did not perform the full evaluation and parameter search in this manuscript. However, we performed spot checks with hyperparameter configurations from previous years and accompanied the RML imaging procedure with two alternative approaches: \doghit and \moead. \doghit is a lightweight compressive sensing approach that, due to its simple design, the small number of hyperparameters, and the data-driven selection of the regularization assumption (ie, the exact form of the sparsifying basis), is a robust and fast alternative well suited for the EHT. \moead replaces parameter surveys on synthetic data with a multi-objective exploration of the optimization landscape based on the data itself. 

\subsubsection{\doghit}
\doghit is a compressive sensing algorithm that models the image by wavelets \citep{Mueller2022}. In contrast to traditional compressive sensing approaches, the wavelets have been specifically designed to fit to the $(u,v)$ coverage \citep{Mueller2023a}. This is done to offer an ideal separation between covered and non-covered Fourier coefficients, highlighting the features introduced by the former, and suppressing the latter, motivated by the exceptional sparse coverage of the EHT. All in all, \doghit solves the optimization problem \citep{Mueller2022}:
\begin{align} \nonumber
    &\hat{\omega} \in \mathrm{argmin}_\omega \left\{ \chi^2_{\rm CPh} (\Psi \omega)+\chi^2_{\rm LCA}(\Psi \omega)+\norm{\omega}_{l^0}+R_{\rm flux}(\Psi \omega)\right\}, \\
    &\hat{\mathcal{I}} = \Psi \hat{\omega}.
\end{align}
Here, $\Psi$ denotes the wavelet dictionary, and $\omega$ the wavelet coefficients.

Due to its sparsifying approach, \doghit computes the multi-resolution support (the spatial scales and positions of all wavelet coefficients that are statistically significant to represent the features of the image) as a byproduct of the total intensity imaging. \citet{Mueller2023b} noted that the multi-resolution support provides a powerful prior information for the reconstruction of channelized datasets (e.g. spectral channels, polarization channels, or time-dynamic reconstructions). Linear polarization and circular polarization were recovered by constrained minimization, i.e. we minimized $\chi^2_{pvis}$ and $\chi^2_{\rm CPh}$, respectively, and only allowed coefficients in the multi-resolution support to vary. Since the form of the basis functions is rigidly set by the instrumental configuration and there is no manual selection, the number of free hyperparameters and the human bias in imaging are reduced.

\doghit handles total intensity, linear polarization, and circular polarization separately. First, we recover an image in total intensity by closure-only imaging. In a second step, we calibrate the gains and leakages, and recover linear polarization with the above-mentioned strategy of only changing the parameters in the multi-resolution support. Finally, we recover circular polarization using a gradient descent approach.

\subsubsection{\ehtim}
\ehtim directly implements the minimization problem Eq. \eqref{eq:rml}, with objective functionals summarized in \autoref{tab:objectives}. Reconstruction with \ehtim depends strongly on the regularization hyperparameters $\alpha_d, \beta_x$, which balance the data fidelity and regularization assumptions. These parameters have been selected by parameter surveys in previous studies \citep[e.g.][]{M87p4}, i.e. various parameter combinations have been tested against a variety of geometric models. \texttt{ehtim} can process visibilities in total intensity, linear, and circular polarization simultaneously, by adding all the objectives together. However, the combined parameter space would be very high-dimensional. Instead, we followed the approach taken in \citet{M87p7} and \citet{SgrA.p7}. We recovered total intensity first, then fixed the total intensity structure, and recovered linear and circular polarization afterwards.

Reflecting this splitting strategy, the parameter search has also been split into total intensity and polarization. A parameter survey has been recomputed in total intensity on the four original geometric models (double, disk, ring, and crescent) used to validate the first release of an image of the black hole shadow \citep{M87p4}. Additionally, the data and regularization terms for linear polarization and circular polarization have been surveyed for a number of crescent images with varying asymmetry, orientation of the EVPA pattern, and polarization fraction as a spot-check. The best parameter combination (in terms of cross-correlation to the ground truth) matches the one reported in \citet{M87p7} and has therefore been adapted for the three years for the sake of consistency. Our final weights are shown in \autoref{tab:objectives}. Ultimately, the validation for this configuration results from the validation described in Sect. \ref{sec:synthetic}.

The full polarized imaging and calibration pipeline consists first of imaging the source in total intensity. The reconstruction is iteratively blurred, and the reconstruction is redone to avoid getting trapped in local minima. The observation is scan-averaged, and imaging optimization alternates with self-calibration steps iteratively. During the initial self-calibration steps, we assume that the gains for the right- and left-handed polarization feeds are identical. As a second step, we fix the total intensity image and then recover the linear polarization image. The linear polarization map is recovered in a loop, alternating between image reconstruction and leakage estimation. Finally, we self-calibrate the data again, allowing for different gains in the two polarization feeds, and solve for the circular polarization image.

\subsubsection{\moead}
\moead is a widely used optimization framework for multi-objective optimization originally proposed by \citet{Zhang2008} and \citet{Li2009}. Recently, it has been transferred to VLBI imaging \citep{Mueller2023c}, and expanded to include polarization \citep{Mus2024a}. RML methods proceed by minimizing a weighted sum that balances data fidelity terms and regularizers against each other. \moead, aims to find the `best-compromise' solutions between the different regularizers. The notion of Pareto optimality identifies this: For a multi-objective vector of functionals $[\chi^2_1, \chi^2_2, ..., R_1, R_2, ...]$, a solution is called `Pareto optimal' if the further optimization along one of the functionals automatically has to worsen the scoring in another.

The set of all Pareto-optimal image structures is known as the Pareto front. \moead approximates this front. Compared to \ehtim, the solution to the optimization problem defined by \moead is not a single image, but a space of images (the Pareto front), representing the best-compromise solutions among all objectives. Minimization is achieved through genetic evolution, where every genome in every generation represents a single image \citep{Mueller2023c}. The final population approximates the Pareto front.

It has been demonstrated both for total intensity \citep{Mueller2023c} and polarization \citep{Mueller2024b} that the Pareto front of the calibration-independent structure separates into multiple disjoint clusters of solutions, interpreted as local minima of a potentially multimodal optimization landscape. In this work, we report the image closest to the utopian ideal point following the strategy outlined in \citet{Mueller2023c, Mus2024b}, and leave the detailed analysis of the multi-modality of the problem to a subsequent work.

\moead has been implemented in the \texttt{MrBeam} \citep{Mueller2022, Mueller2023c, Mus2024b} software package, which calls \ehtim-subroutines for the calibration and evaluation of the data terms. The full imaging and calibration procedure is therefore similar to the one adapted for \texttt{ehtim}, although not equivalent due to the difference in the surveying strategy (i.e. no survey on synthetic data are needed), the different concept of optimality, and the optimization procedure. The data fidelity and penalty terms are summarized in \autoref{tab:objectives}.  

Due to this proximity, \moead's imaging pipeline resembles the \texttt{ehtim} strategy. We first recovered the total intensity image by fitting only closure quantities. To help \moead converge, we derived a starting point for the population using an un-regularized run of \texttt{ehtim}. Afterwards, we selected a representative cluster of solutions (proximity to the utopian) and performed self-calibration and leakage calibration. Then, we ran \moead again to recover the linear and circular polarization images. In contrast to \ehtim, we solved for linear polarization and circular polarization at the same time.

\subsection{Bayesian methods}\label{appendix:bayes_imaging}
This section describes the two Bayesian polarized imaging algorithms, \themis and \comrade, used for image reconstruction, residual calibration, and leakage corrections. Both Bayesian codes aim to simultaneously solve for all four Stokes parameters in the image and calibration and leakage terms through forward modelling of the measurement process. 

Both \themis and \comrade utilize a Cartesian grid of fluxes, where for each pixel, the Stokes vector is parameterized using the Poincar\'{e} representation,
\begin{equation}
    \begin{pmatrix}
        \mathcal{I}_{ij} \\
        \mathcal{Q}_{ij} \\
        \mathcal{U}_{ij} \\
        \mathcal{V}_{ij}
    \end{pmatrix}
       =
    \mathcal{I}_{ij}
    \begin{pmatrix}
        1 \\
        p_{ij} \hat{\bm{p}}_{ij,x}\\
        p_{ij} \hat{\bm{p}}_{ij,y}\\
        p_{ij} \hat{\bm{p}}_{ij,z},
    \end{pmatrix}
,\end{equation}
where $p_{ij}$ is the total polarization fraction and $\hat{\bm{p}}$ is a unit vector in $\mathbb{R}^3$, which are all included as parameters in the forward model. Each pixel is modelled as a point source and then convolved with a kernel to create a continuous image function.  The ideal visibilities are calculated using Eq. \ref{eq:rvis} based on the Stokes Images produced from each method. Site-based instrumental corruptions are applied using a RIME formalism, utilizing the Jones matrices for feed rotation, leakage, and residual gains from Eq. \ref{eq:jones_decomp}. The model visibilities are then compared to the data using a complex Gaussian likelihood, which is calculated as the product of all individual likelihoods for all parallel and cross-hand products. We now describe the individual details for each method.

\subsubsection{\comrade}

\comrade \citep{Tiede2022} is a Bayesian polarized imaging and calibration software suite written in the Julia programming language \citep{bezanson2017julia}. Here, we briefly describe the polarized imaging process and the priors used; a complete explanation can be found in \citet{Tiede2022} and Tiede et al. (in prep.).

To prepare the data, \comrade first coherently averaged the data over observation scans using the \texttt{Pyehtim.jl} \texttt{scan\_average} function and then added 2\% systematic noise to model residual calibration errors, such as loss of coherence due to time and frequency averaging. Additionally, due to the over-resolved flux on short baselines, \comrade flagged the ALMA-APEX, JCMT-SMA, and KP-SMT baselines, and fit the total flux of the compact emission.

\comrade's image model is aligned with the equatorial coordinate axes and uses a $64\times64$ raster with a field of view of $200\,\muas$. For the total intensity image raster $\mathcal{I}$, we use a first-order Gaussian Markov random field (GMRF) prior on the log-ratio transformed pixel fluxes $r_{ij}$, which are related to the total intensity fluxes $\mathcal{I}_{ij}$ by
\begin{equation}
    \mathcal{I}_{ij} = F_0 \frac{e^{r_{ij}}}{\sum_{ij}e^{r_{ij}}},
\end{equation}
where $F_0$ is the total flux of the raster component. For the GMRF, the variance and correlation length of the random field are included as hyperparameters. For more information, see Tiede et al. (in prep.). The GMRF acts as multiplicative fluctuations in some mean image structures, which we assume to be given by a $60\,\muas$ Gaussian blob. However, during testing, we found that the size of the Gaussian did not appreciably change the resulting images. 

The total fractional polarization amplitude, $p_{ij}$, is given by
\begin{equation}
    p_{ij} = \left[1 + \exp(\bar{\ell} + \sigma_{p} \ell_{ij})\right]^{-1},
\end{equation}
where $\bar{\ell}$ is the average logistic total fractional polarization, $\sigma_p$ is the variance of the logistic fractional polarization, and $\ell_{ij}$ are the scaled fluctuations of the total fractional polarization, whose prior is another independent GMRF. This corresponds to modelling the total fractional polarization in \textit{logit} space, with some constant mean fractional polarization and some variance. This parameterization ensures that $0 \leq p_{ij} \leq 1$ for all values. For the total polarization direction on the Poincar\'{e} sphere, for each unit vector $\hat{p}_{ij}$, we used the uniform distribution on the sphere by modelling each direction with an independent unit normal distribution and then dividing by the total length to ensure that the vector is normalized.

Gain and gain ratios are allowed to vary every scan, whereas the $D$-terms are fixed to constant values over each track. For each station, we use a Gaussian prior on the log-gain amplitudes with a mean of 0 and a standard deviation of $0.2$, except for GLT, which uses a standard deviation of $1.0$ to model the large-gain-amplitude fluctuations. For the gain phases and gain phase ratios, first lock the gain phase and gain phase ratio for ALMA to zero to model the a priori EVPA calibration included in the PolConvert procedure. If ALMA is absent in the scan, we alphabetically select a different reference site and set only the right-hand gain phase to zero while keeping the gain ratio phase as a model parameter. For the gain phases, we use a zero-mean von Mises prior with concentration parameter $\pi^{-2}$ that essentially creates a uniform prior in the interval $[-\pi, \pi]$. For the gain ratio phases, we use a zero-mean von Mises prior with a concentration parameter $0.5^2$ to model the a priori calibration that the \texttt{rPICARD} pipeline does to remove gain phase ratio. 

During imaging, \comrade simultaneously models all four Stokes parameters and the instrumental terms. Since \comrade is a Bayesian imaging algorithm, it estimates the uncertainty for all model parameters, including the image and instrumental terms. Its output is a set of samples approximately drawn from the posterior using Markov chain Monte Carlo (MCMC). Specifically, \comrade uses the No-U-Turn Sampler \citep[NUTS;][]{hoffman_no-u-turn_2011} in the Julia sampling package \texttt{AdvancedHMC.jl}, utilizing the automatic differentiation package \texttt{Enzyme.jl} \citep{Enzyme1, Enzyme2} to compute gradients. To initialize the chain, we ran ten rounds of optimization to find the approximate posterior maximum. From this location, we then ran NUTS for 10,000 adaptation steps to tune the sampler, followed by 10,000 sampling steps. From these 10,000 samples, we randomly selected 500 images for further analysis.

 \subsubsection{\themis}
 \label{sec:Themis}
\themis provides a Bayesian framework for modelling and parameter estimation from EHT data \citep{Broderick2020}. Imaging with \themis employs a model consisting of a rectilinear set of control points spanned via a bicubic spline to represent the intensity map, the polarization fraction map, EVPA map, and the Stokes $V$ fraction \citep{Themaging:2020,M87p7}. The orientation of the raster and the field of view are free parameters, which allow the raster to expand, contract, and rotate to maximize the model's freedom to fit the data. The resolution of the raster is typically determined by maximizing the Bayesian evidence on the raster dimension; this is small due to the limited number of EHT resolution elements across \m87. To facilitate cross-epoch comparisons, here we adopted a 5x5 raster based on the polarized imaging study of the 2017 observations \citep{M87p7}. Finally, a large-scale asymmetric Gaussian was included to capture potential discrepancies between the intra-site and VLBI baselines.

\themis reconstructions are fit directly to each band's scan-averaged complex visibilities ($RR^*$, $LL^*$, $RL^*$, $LR^*$) independently. Site-specific leakages and complex gains are fit simultaneously with image generation as described in \citet{M87p7}. \themis assumes that the right and left-hand gains are equal; a 3\% systematic error is added to partially mitigate non-unity gain ratios.

The result of the \themis fits is an approximate posterior composed of a set of images used for Bayesian uncertainty quantification. \themis provides a number of posterior sampling methods, for which the most common output is a MCMC chain that supports subsequent Bayesian interpretation. To ensure efficient sampling of the posterior, we use the differential even-odd parallel tempering scheme, with each tempering level explored via the Hamiltonian Monte Carlo NUTS algorithm implemented by the \texttt{Stan} package \citep{DEO:2019,Stan:2017}. This sampler has been shown to effectively capture multimodal posteriors \citep[see e.g.][]{M87p7}. Chain convergence is assessed by visual inspection of parameter traces and quantitative chain statistics, including integrated autocorrelation time, split-$\hat{R}$, and parameter rank distributions \citep{Vehtari}, and typically requires $\sim10^5$ MCMC steps. The number of tempering levels is chosen to ensure efficient communication between the highest- and lowest-temperature levels, typically set at 80 due to the complicated nature of the model.

\section{Synthetic data validation}\label{appendix:synth}
\subsection{Models}
Here, we describe the models used to generate the synthetic data. For the blinded GRMHD synthetic data, a KHARMA MAD simulation \citep{M87_2018p2} with inclination $i = 17\degr{}$, spin $a=-0.5$, $R_\text{low} = 10$, and $R_\text{high} =40$ was used. For each epoch, 2017, 2018, and 2021, a random snapshot was chosen with parameters that did not necessarily match the properties of the measured data for each epoch. The selected GRMHD snapshots are shown in the first column of Fig. \ref{fig:stage3GRMHDsyntheticdata}.

To test the impact of over-resolved polarized flux on ring reconstructions, we constructed a pair of geometric models consisting of an m-ring and an extended jet component. The jet consists of three polarized, elliptical, Gaussian components --- this is the same tri-Gaussian jet model used in the synthetic data generation of \citep{M87p9}, only polarized. The amount of polarized flux in the jet was constructed to roughly match the upper and lower levels of polarization observed on the short baseline ALMA-APEX, JCMT-SMA, and KP-SMT.

\subsection{Blinded GRMHD results: Individual methods}\label{app:synth_blinded}

\begin{figure*}[h!]
    \centering
    \includegraphics[width=\linewidth]{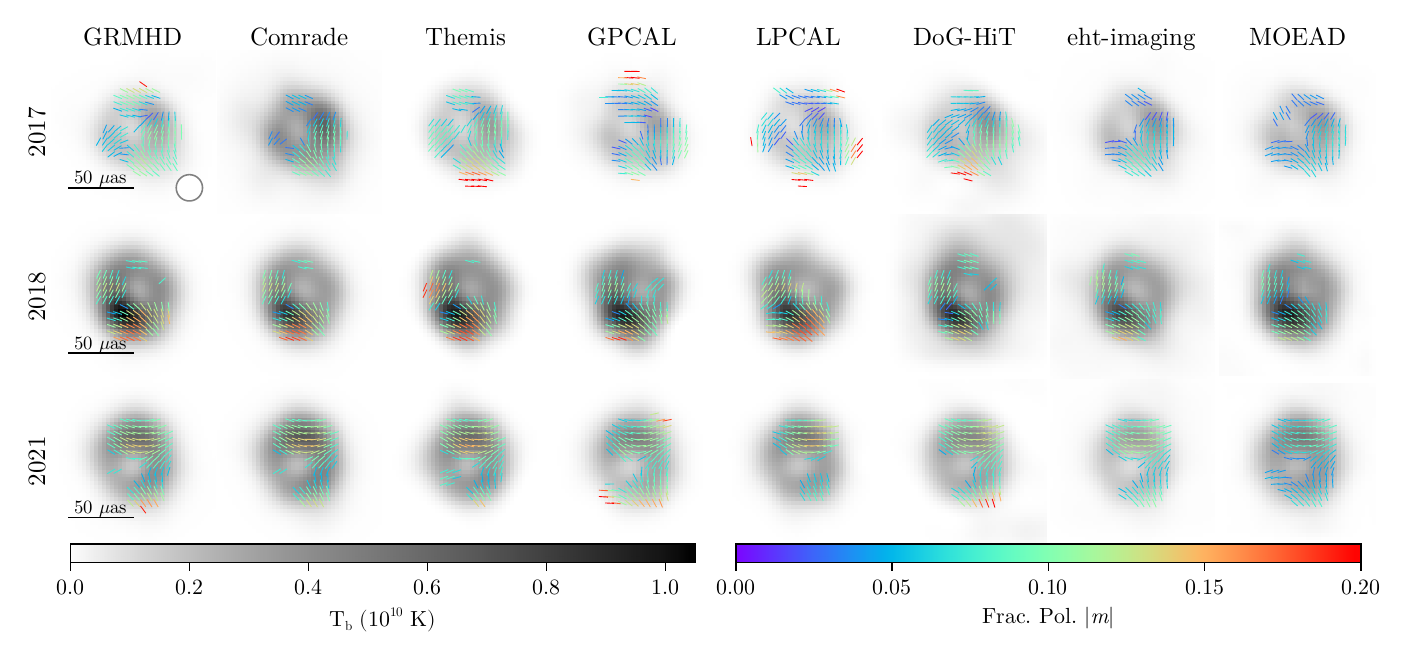}
    \caption{Polarized imaging results from the blinded GRMHD synthetic data test. The leftmost column shows the ground truth GRMHD image blurred by a 20\,$\muas$ Gaussian beam; the other columns show the imaging results from each method blurred to an equivalent resolution. From top to bottom, the rows correspond to the 2017, 2018, and 2021 results.}
    \label{fig:stage3GRMHDsyntheticdata}
\end{figure*}

\begin{figure*}[h!]
    \centering
    \includegraphics[width=0.98\linewidth]{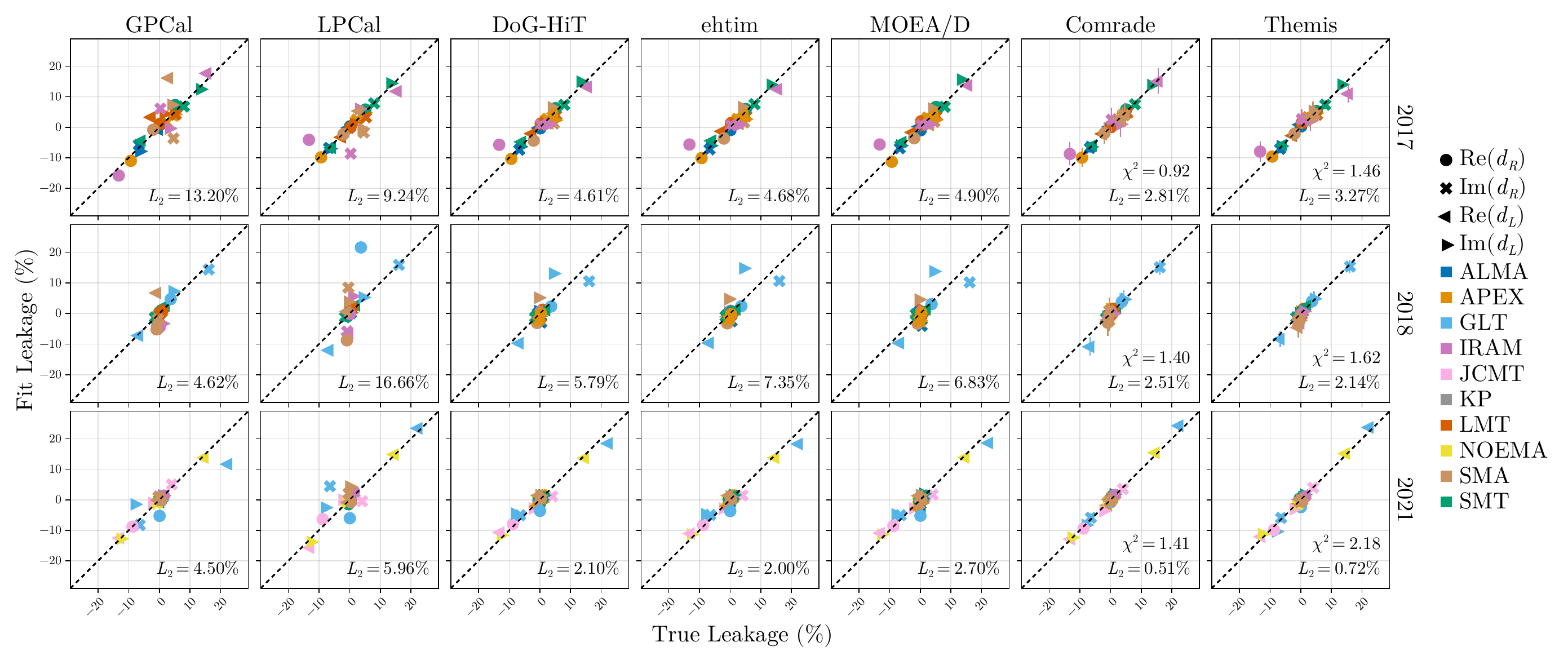}
    \caption{D-term recovery results of blinded GRMHD synthetic data from each polarized imaging method (columns) and year (rows). The x-axis of each plot is the true leakage, and the y-axis is the measured leakage. Perfect recovery means that all points should live on the $y=x$ line. The total $L_2$ difference between the ground truth and recovered leakage averaged over all stations is shown in the bottom right of each plot. For the Bayesian methods, we also show a $\chi^2$ value computed as the square difference between the true value and the recovered leakage divided by the variance of the posterior estimate. For the Bayesian methods, the points denote the posterior median leakage, and error bars show the 95\% credible interval about the median.}\label{fig:synthetic_stage3_2021_D-term}
\end{figure*}

The blinded synthetic data image reconstructions for each method are shown in Fig. \ref{fig:stage3GRMHDsyntheticdata} with image reconstruction blurred to match the resolution of the truth convolved with a $20\,\muas$ circular Gaussian beam. For 2017, we see the most diversity across the methods as expected due to the overall poorer coverage of the 2017 data. This is similar to the diversity between methods found in \citet{M87p7}. In 2018 and 2021, we found that once blurred to the same resolution, the image reconstructions show more consistency between methods, especially in the brighter regions of the ring.

Figure \ref{fig:synthetic_stage3_2021_D-term} shows the recovered leakage compared to the true values used when creating the synthetic data. The rows show the recovered leakage for each epoch of synthetic data. Generally, the two Bayesian methods produce the best leakage estimates each year and provide reliable estimates of the true leakage to within a few percent for all years. 

For the non-Bayesian methods, there is more scatter in the estimated leakage, but the error in the leakage estimates improves each year due to the increased number of stations in the array, making it easier to solve for leakage. In 2018 and 2021, we found that GLT's leakage estimates tended to have the largest discrepancies from the truth. This error is expected given the relatively poor parallactic angle coverage $(< 15\degr{})$ of the site. Encouragingly, the Bayesian methods can still constrain GLT's leakage, with the posterior for \comrade and \themis containing the truth within their 95\% credible interval for each station.

\subsection{Impact of polarized extended emission on the core reconstructions}\label{app:synth_geom}
\begin{figure}[h!]
    \centering
    \includegraphics[width=\linewidth]{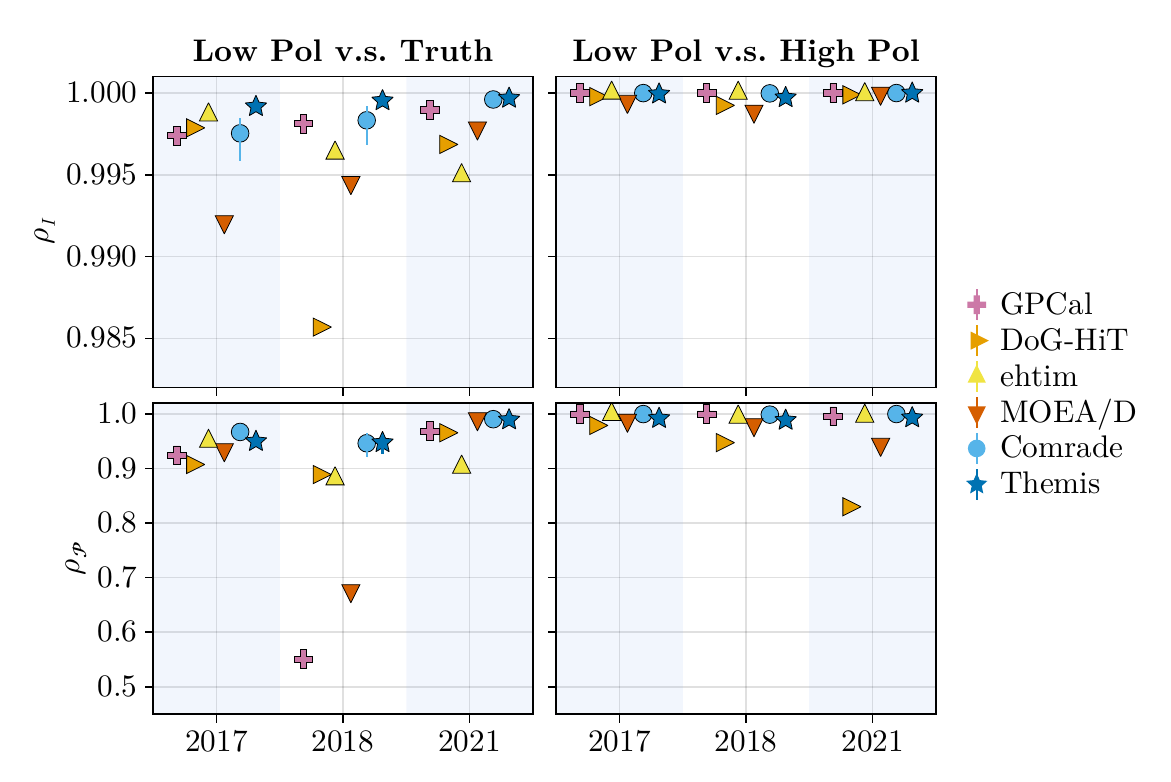}
    \caption{Cross-correlation between the low polarized and high polarized extended jet geometric tests. The ring model for the two tests is identical, meaning that the cross-correlation in the core region would be equal to unity for perfect reconstructions. Left column: Total intensity (top row) and linear polarization cross-correlation (bottom row) between the low-polarized jet reconstructions. Right column: Same quantities but with the cross-correlation computed between the low and high polarized jet reconstructions. }
    \label{fig:stage2}
\end{figure}

The impact of polarized extended emission for each imaging pipeline is shown in Fig. \ref{fig:stage2}. The overall reconstruction quality of the low-polarized jet synthetic data is shown in the left column of Fig. \ref{fig:stage2}. For total intensity (top row), we found that all methods produced reliable estimates of the total intensity structure across all epochs with $\rho_\mathcal{I} > 0.985$. For linear polarization, we found more mixed results. Both Bayesian methods \themis and \comrade performed well every year, finding a $\rho_\mathcal{P} \gtrsim 0.95$ each year. In addition, \ehtim performed well for the low-polarization jet reconstructions. \doghit performed well in 2017 and 2021, but its total intensity image got noticeably worse in 2018; however, the reconstruction was still of high quality overall. In 2018, \gpcal and \moead struggled to recover the true linear polarized structure but recovered in 2021, suggesting it may have been a convergence issue in the algorithms. 

To compare the reconstructions of the same compact features in the low- and high-polarized jet synthetic data, we then cross-correlated the reconstructions from each pipeline, which is shown in the right column of Fig. \ref{fig:stage2}. Note that we cross-correlated the mean images for the Bayesian methods to remove the impact of different thermal noise realizations on the image reconstructions. The total intensity cross correlation between the low and high polarized jet reconstructions was substantially closer to unity $(> 0.998)$ for all methods than the cross correlation between the low polarized reconstructions and the truth. A similar result was found for the linear polarization cross-correlation between the two synthetic datasets, albeit with a few exceptions. Namely, for 5/7 methods, we found a linear polarization cross-correlation of $> 0.98$, and this consistency is always greater than the consistency between the reconstruction and ground truth. The only exceptions to this are \doghit. For \doghit, the low-polarized jet reconstruction showed a polarized cross-correlation of almost unity (0.96). In contrast, the high-polarized jet reconstruction performed worse (0.88), suggesting that the method is weakly sensitive to extended emission. The \comrade, \themis, and \ehtim are not affected by different levels of extended emission when reconstructing the compact ring structure.

\section{Leakage estimates in 2018 and 2021}\label{label:leakage}
\begin{figure}[h!]
    \centering
    \includegraphics[width=\linewidth]{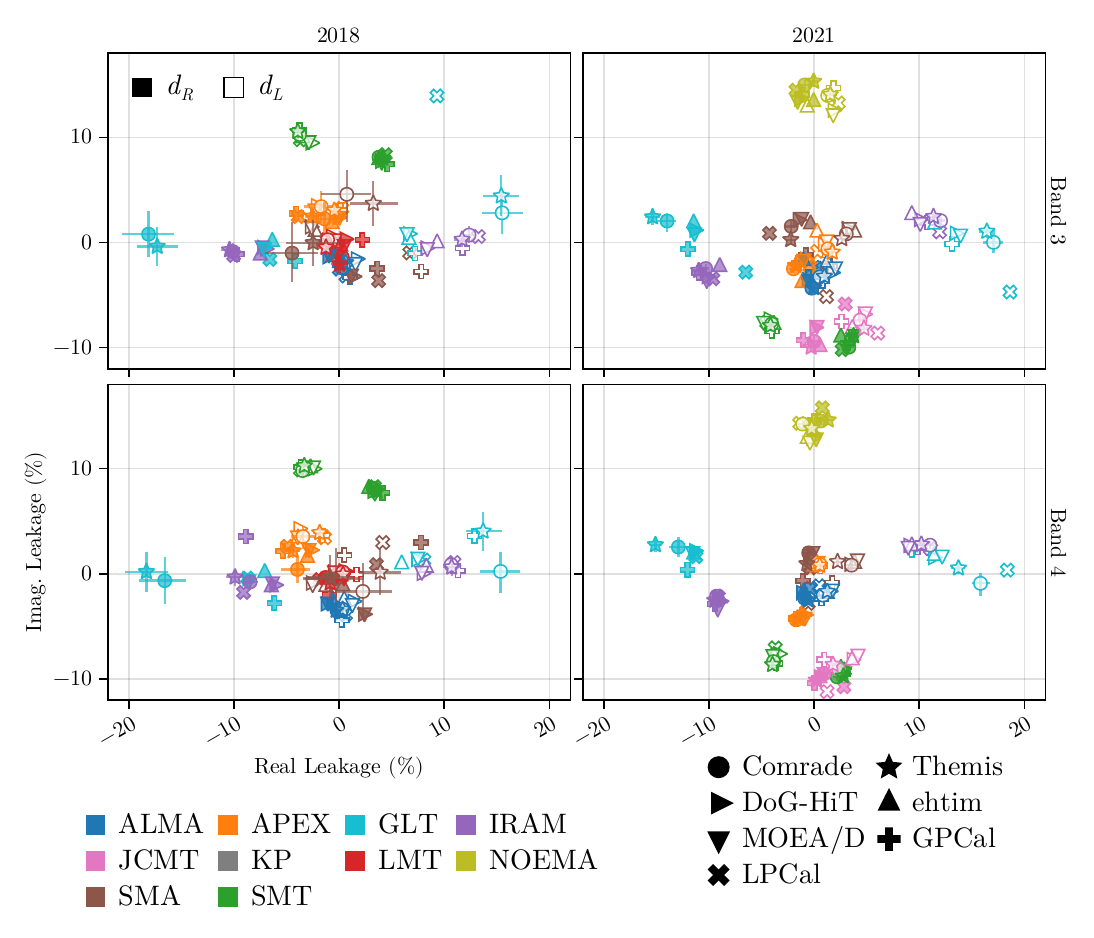}
    \caption{Leakage estimates across all methods for 2018 and 2021. In 2018, we flagged JCMT (Appendix~\ref{app:jcmtD}). For the Bayesian methods, the error bars show the 95\% credible intervals about the median leakage denoted by the marker.}
    \label{fig:m87_leakage}
\end{figure}

Individual leakage solutions are shown in Fig. \ref{fig:m87_leakage} for 2018 and 2021. For the 2017 leakage estimates, we refer the reader to \citet{M87p7} for a multisource analysis. Note that in general, the 2017 leakage estimates we found in this paper are consistent with those found in \citet{M87p7}. We found that the leakage estimates produced by each pipeline are consistent to within $5\%$, with a few notable exceptions. For the 2018 reconstructions, GLT leakage estimates show significant discrepancies between methods. GLT's leakage, being difficult to estimate, was also observed in the synthetic data tests, where non-Bayesian methods could not measure GLT's leakage to better than 10\%. The discrepancy in the estimated leakage of GLT is not surprising due to its limited parallactic angle coverage. Similarly, in 2018, SMA had limited parallactic angle coverage for \m87, with only $\sim$$5\degr{}$ evolution of the parallactic angle. As a result, the leakage estimate for SMA is highly uncertain from just the \m87 coverage. In 2021, we found that all leakage estimates, besides GLT, are consistent to within $\lesssim 3\%$. GLT still has the largest spread between methods due to its poor parallactic angle coverage, although the spread improved by a factor of $\sim 2$ compared to 2018

Finally, the uncertainty of GLT's leakage in 2018 and 2021, and SMA's in 2018, does not appear to dramatically alter the image reconstructions across pipelines (see Fig. \ref{fig:m87_all_images}) or the polarized parameter estimates (see Fig. \ref{fig:m87_params_uniform}). Although we expect leakage estimates for GLT and SMA to improve from multisource fitting in the future, it is unlikely that they will significantly alter the image reconstructions or change our quantitative results.

\onecolumn
\section{Individual method reconstructions of \m87}\label{appendix:individual_results}
\begin{figure}[h!]
    \centering
    \includegraphics[width=\textwidth]{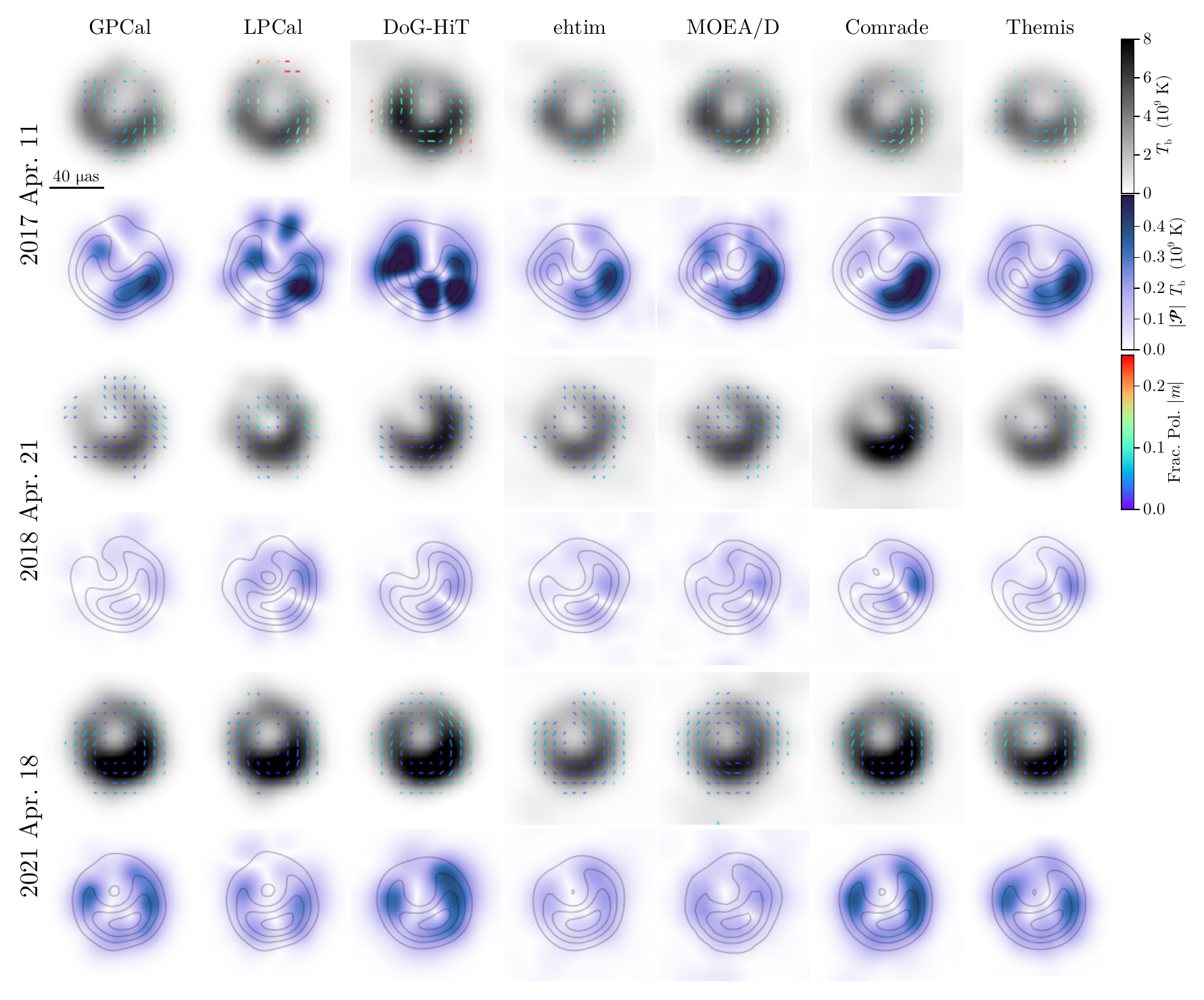}
    \caption{Reconstructions of \m87 across 2017, 2018, and 2021 from all methods considered. All methods have been blurred to match the resolution of the two CLEAN reconstructions \gpcal and \lpcal.}
    \label{fig:m87_all_images}
\end{figure}

Figure \ref{fig:m87_all_images} shows the results of each imaging method for the three epochs presented, blurred to a common resolution. In addition to the total flux differences, we found excellent agreement in the structure of \m87 every year. For 2017, all methods found that the south-west region of the ring is the most polarized region and has a similar total intensity azimuthal profile. However, for the same year, there are some differences in the EVPA pattern, especially in the eastern and northern parts of the images. This uncertainty was also found in \citet{M87p7}, and is a result of the sparseness of the 2017 coverage.

Remarkably, the uncertainty in the linear polarization and total intensity images in 2018, especially in 2021, has been significantly reduced. All methods found that the position angle of the ring brightness has shifted in 2018 and 2021 compared to 2017. Furthermore, all methods found that \m87 in 2018 is significantly de-polarized, where only the western part of the image has significant polarized emission across all methods. For the 2021 reconstructions, all methods found a very similar EVPA pattern, further demonstrating the robustness of the $\angle\beta_2$ rotation and the changes in the EVPA helicity of the ring. Furthermore, unlike 2017, the brightest part of the total intensity was de-polarized relative to other parts in 2021. 

\begin{figure}[h!]
    \centering
    \includegraphics[width=1.0\linewidth]{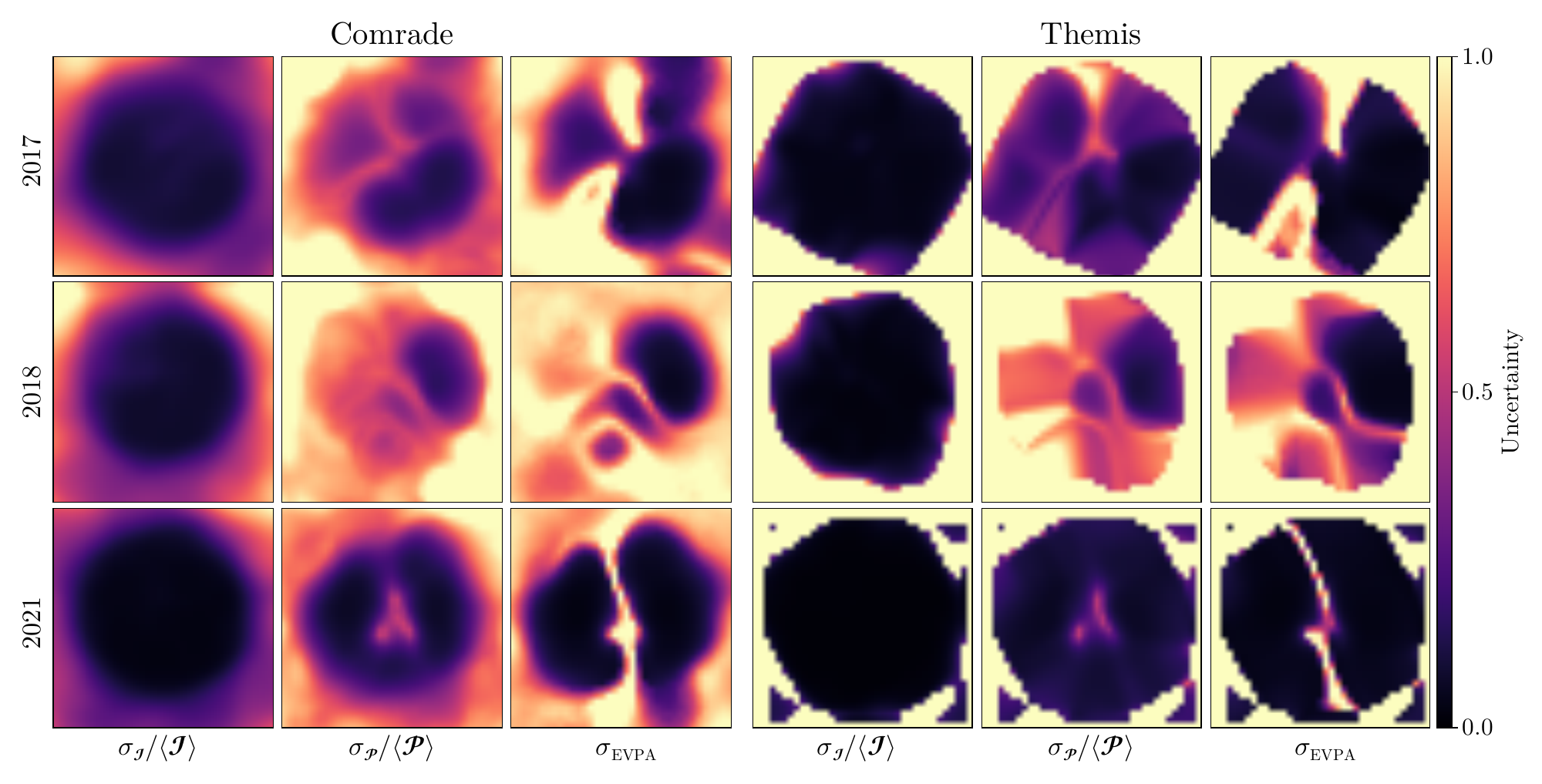}
    \caption{Uncertainty maps for the 2017 (top), 2018 (middle), and 2021 (bottom) \m87 reconstructions using the \comrade (left) and \themis (right) pipelines. Left columns: Fractional total intensity uncertainty. Middle columns: Fractional linear polarization uncertainty. Right columns: Absolute EVPA uncertainty, which is defined as the circular standard deviation in radians of each pixel. Any pixels that are outside the field of view of the raster are assigned an uncertainty of $>1$.}
    \label{fig:uncertainty}
\end{figure}
\FloatBarrier

Finally, Fig. \ref{fig:uncertainty} shows the uncertainty maps for the \comrade and \themis pipelines. We found that the central ring has a small fractional uncertainty over all three years, demonstrating its robustness. Furthermore, the regions of the ring that are polarized in 2017 (SW), 2018 (W), and 2021 (E and W) have a small fractional uncertainty in total linear polarization and overall uncertainty in EVPA.

In summary, all key results presented in Sect. \ref{sec:results} are seen in individual image reconstructions and do not depend on the particular choices of the imaging algorithms.

\twocolumn
\section{Acknowledgments}

\begin{acknowledgements}
The Event Horizon Telescope Collaboration thanks the following
organizations and programs: the Academia Sinica; the Academy
of Finland (projects 274477, 284495, 312496, 315721); the Agencia Nacional de Investigaci\'{o}n 
y Desarrollo (ANID), Chile via NCN$19\_058$ (TITANs), Fondecyt 1221421 and BASAL FB210003; the Alexander
von Humboldt Stiftung (including the Feodor Lynen Fellowship); an Alfred P. Sloan Research Fellowship;
Allegro, the European ALMA Regional Centre node in the Netherlands, the NL astronomy
research network NOVA and the astronomy institutes of the University of Amsterdam, Leiden University, and Radboud University;
the ALMA North America Development Fund; the Astrophysics and High Energy Physics programme by MCIN (with funding from European Union NextGenerationEU, PRTR-C17I1); the Black Hole Initiative, which is funded by grants from the John Templeton Foundation (60477, 61497, 62286) and the Gordon and Betty Moore Foundation (Grant GBMF-8273) - although the opinions expressed in this work are those of the author and do not necessarily reflect the views of these Foundations;  
the Brinson Foundation; the Canada Research Chairs (CRC) program; Chandra DD7-18089X and TM6-17006X; the China Scholarship
Council; the China Postdoctoral Science Foundation fellowships (2020M671266, 2022M712084); ANID through Fondecyt Postdoctorado (project 3250762); Conicyt through Fondecyt Postdoctorado (project 3220195); Consejo Nacional de Humanidades, Ciencia y Tecnología (CONAHCYT, Mexico, projects U0004-246083, U0004-259839, F0003-272050, M0037-279006, F0003-281692, 104497, 275201, 263356, CBF2023-2024-1102, 257435); the Colfuturo Scholarship; the Consejo Superior de Investigaciones 
Cient\'{i}ficas (grant 2019AEP112);
the Delaney Family via the Delaney Family John A.
Wheeler Chair at Perimeter Institute; Dirección General de Asuntos del Personal Académico-Universidad Nacional Autónoma de México (DGAPA-UNAM, projects IN112820 and IN108324); the Dutch Research Council (NWO) for the VICI award (grant 639.043.513), the grant OCENW.KLEIN.113, and the Dutch Black Hole Consortium (with project No. NWA 1292.19.202) of the research programme the National Science Agenda; the Dutch National Supercomputers, Cartesius and Snellius  (NWO grant 2021.013); 
the EACOA Fellowship awarded by the East Asia Core
Observatories Association, which consists of the Academia Sinica Institute of Astronomy and Astrophysics, the National Astronomical Observatory of Japan, Center for Astronomical Mega-Science,
Chinese Academy of Sciences, and the Korea Astronomy and Space Science Institute; 
the European Research Council (ERC) Synergy Grant ``BlackHoleCam: Imaging the Event Horizon of Black Holes'' (grant 610058) and Synergy Grant ``BlackHolistic:  Colour Movies of Black Holes:
Understanding Black Hole Astrophysics from the Event Horizon to Galactic Scales'' (grant 10107164); 
the European Union Horizon 2020
research and innovation programme under grant agreements
RadioNet (No. 730562), 
M2FINDERS (No. 101018682) and FunFiCO (No. 777740); the European Research Council for advanced grant ``JETSET: Launching, propagation and 
emission of relativistic jets from binary mergers and across mass scales'' (grant No. 884631); the European Horizon Europe staff exchange (SE) programme HORIZON-MSCA-2021-SE-01 grant NewFunFiCO (No. 10108625); the Horizon ERC Grants 2021 programme under grant agreement No. 101040021; the FAPESP (Funda\c{c}\~ao de Amparo \'a Pesquisa do Estado de S\~ao Paulo) under grant 2021/01183-8; the Fondes de Recherche Nature et Technologies (FRQNT); the Fondo CAS-ANID folio CAS220010; the Generalitat Valenciana (grants APOSTD/2018/177 and  ASFAE/2022/018) and
GenT Program (project CIDEGENT/2018/021); the Gordon and Betty Moore Foundation (GBMF-3561, GBMF-5278, GBMF-10423);   
the Institute for Advanced Study; the ICSC – Centro Nazionale di Ricerca in High Performance Computing, Big Data and Quantum Computing, funded by European Union – NextGenerationEU; the Istituto Nazionale di Fisica
Nucleare (INFN) sezione di Napoli, iniziative specifiche
TEONGRAV; 
the International Max Planck Research
School for Astronomy and Astrophysics at the
Universities of Bonn and Cologne; the Italian Ministry of University and Research (MUR)– Project CUP F53D23001260001, funded by the European Union – NextGenerationEU; 
Deutsche Forschungsgemeinschaft (DFG) research grant ``Jet physics on horizon scales and beyond'' (grant No. 443220636) and DFG research grant 443220636;
Joint Columbia/Flatiron Postdoctoral Fellowship (research at the Flatiron Institute is supported by the Simons Foundation); 
the Japan Ministry of Education, Culture, Sports, Science and Technology (MEXT; grant JPMXP1020200109);  
the Japan Society for the Promotion of Science (JSPS) Grant-in-Aid for JSPS
Research Fellowship (JP17J08829); the Joint Institute for Computational Fundamental Science, Japan; the Key Research
Program of Frontier Sciences, Chinese Academy of
Sciences (CAS, grants QYZDJ-SSW-SLH057, QYZDJSSW-SYS008, ZDBS-LY-SLH011); 
the Leverhulme Trust Early Career Research
Fellowship; the Max-Planck-Gesellschaft (MPG);
the Max Planck Partner Group of the MPG and the
CAS; the MEXT/JSPS KAKENHI (grants 18KK0090, JP21H01137,
JP18H03721, JP18K13594, 18K03709, JP19K14761, 18H01245, 25120007, 19H01943, 21H01137, 21H04488, 22H00157, 23K03453); the MICINN Research Projects PID2019-108995GB-C22, PID2022-140888NB-C22; the MIT International Science
and Technology Initiatives (MISTI) Funds; 
the Ministry of Science and Technology (MOST) of Taiwan (103-2119-M-001-010-MY2, 105-2112-M-001-025-MY3, 105-2119-M-001-042, 106-2112-M-001-011, 106-2119-M-001-013, 106-2119-M-001-027, 106-2923-M-001-005, 107-2119-M-001-017, 107-2119-M-001-020, 107-2119-M-001-041, 107-2119-M-110-005, 107-2923-M-001-009, 108-2112-M-001-048, 108-2112-M-001-051, 108-2923-M-001-002, 109-2112-M-001-025, 109-2124-M-001-005, 109-2923-M-001-001,  
110-2112-M-001-033, 110-2124-M-001-007 and 110-2923-M-001-001); the National Science and Technology Council (NSTC) of Taiwan
(111-2124-M-001-005, 112-2124-M-001-014,  112-2112-M-003-010-MY3, and 113-2124-M-001-008);
the Ministry of Education (MoE) of Taiwan Yushan Young Scholar Program;
the Physics Division, National Center for Theoretical Sciences of Taiwan;
the National Aeronautics and
Space Administration (NASA, Fermi Guest Investigator
grant 
80NSSC23K1508, NASA Astrophysics Theory Program grant 80NSSC20K0527, NASA NuSTAR award 
80NSSC20K0645); NASA Hubble Fellowship Program Einstein Fellowship;
NASA Hubble Fellowship 
grants HST-HF2-51431.001-A, HST-HF2-51482.001-A, HST-HF2-51539.001-A, HST-HF2-51552.001A awarded 
by the Space Telescope Science Institute, which is operated by the Association of Universities for 
Research in Astronomy, Inc., for NASA, under contract NAS5-26555; 
the National Institute of Natural Sciences (NINS) of Japan; the National
Key Research and Development Program of China
(grant 2016YFA0400704, 2017YFA0402703, 2016YFA0400702); the National Science and Technology Council (NSTC, grants NSTC 111-2112-M-001 -041, NSTC 111-2124-M-001-005, NSTC 112-2124-M-001-014); the US National
Science Foundation (NSF, grants AST-0096454,
AST-0352953, AST-0521233, AST-0705062, AST-0905844, AST-0922984, AST-1126433, OIA-1126433, AST-1140030,
DGE-1144085, AST-1207704, AST-1207730, AST-1207752, MRI-1228509, OPP-1248097, AST-1310896, AST-1440254, 
AST-1555365, AST-1614868, AST-1615796, AST-1715061, AST-1716327,  AST-1726637,  
OISE-1743747, AST-1743747, AST-1816420, AST-1935980, AST-1952099, AST-2034306,  AST-2205908, AST-2307887, AST-2407810); 
NSF Astronomy and Astrophysics Postdoctoral Fellowship (AST-1903847); 
the Natural Science Foundation of China (grants 11650110427, 10625314, 11721303, 11725312, 11873028, 11933007, 11991052, 11991053, 12192220, 12192223, 12273022, 12325302, 12303021); 
the Natural Sciences and Engineering Research Council of
Canada (NSERC);  
the National Research Foundation of Korea (the Global PhD Fellowship Grant: grants NRF-2015H1A2A1033752; the Korea Research Fellowship Program: NRF-2015H1D3A1066561; Brain Pool Program: RS-2024-00407499;  Basic Research Support Grant 2019R1F1A1059721, 2021R1A6A3A01086420, 2022R1C1C1005255, 2022R1F1A1075115); the POSCO Science Fellowship of the POSCO TJ Park Foundation; 
Netherlands Research School for Astronomy (NOVA) Virtual Institute of Accretion (VIA) postdoctoral fellowships; NOIRLab, which is managed by the Association of Universities for Research in Astronomy (AURA) under a cooperative agreement with the National Science Foundation; 
Onsala Space Observatory (OSO) national infrastructure, for the provisioning
of its facilities/observational support (OSO receives funding through the Swedish Research Council under grant 2017-00648);  the Perimeter Institute for Theoretical Physics (research at Perimeter Institute is supported by the Government of Canada through the Department of Innovation, Science and Economic Development and by the Province of Ontario through the Ministry of Research, Innovation and Science); the Portuguese Foundation for Science and Technology (FCT) grants (Individual CEEC program - 5th edition, \url{https://doi.org/10.54499/UIDB/04106/2020}, \url{https://doi.org/10.54499/UIDP/04106/2020}, PTDC/FIS-AST/3041/2020, CERN/FIS-PAR/0024/2021, 2022.04560.PTDC); the Princeton Gravity Initiative; the Spanish Ministerio de Ciencia, Innovaci\'{o}n  y Universidades (grants PID2022-140888NB-C21, PID2022-140888NB-C22, PID2023-147883NB-C21, RYC2023-042988-I); the Severo Ochoa grant CEX2021-001131-S funded by MICIU/AEI/10.13039/501100011033; The European Union’s Horizon Europe research and innovation program under grant agreement No. 101093934 (RADIOBLOCKS); The European Union “NextGenerationEU”, the Recovery, Transformation and Resilience Plan, the CUII of the Andalusian Regional Government and the Spanish CSIC through grant AST22\_00001\_Subproject\_10; ``la Caixa'' Foundation (ID 100010434) through fellowship codes LCF/BQ/DI22/11940027 and LCF/BQ/DI22/11940030; 
the University of Pretoria for financial aid in the provision of the new 
Cluster Server nodes and SuperMicro (USA) for a SEEDING GRANT approved toward these 
nodes in 2020; the Shanghai Municipality orientation program of basic research for international scientists (grant no. 22JC1410600); 
the Shanghai Pilot Program for Basic Research, Chinese Academy of Science, 
Shanghai Branch (JCYJ-SHFY-2021-013); the Simons Foundation (grant 00001470); the Spanish Ministry for Science and Innovation grant CEX2021-001131-S funded by MCIN/AEI/10.13039/501100011033; the Spinoza Prize SPI 78-409; the South African Research Chairs Initiative, through the 
South African Radio Astronomy Observatory (SARAO, grant ID 77948),  which is a facility of the National 
Research Foundation (NRF), an agency of the Department of Science and Innovation (DSI) of South Africa; the Swedish Research Council (VR); the Taplin Fellowship; the Toray Science Foundation; the UK Science and Technology Facilities Council (grant no. ST/X508329/1); the US Department of Energy (USDOE) through the Los Alamos National
Laboratory (operated by Triad National Security,
LLC, for the National Nuclear Security Administration
of the USDOE, contract 89233218CNA000001); and the YCAA Prize Postdoctoral Fellowship. This work was also supported by the National Research Foundation of Korea (NRF) grant funded by the Korea government(MSIT) (RS-2024-00449206). We acknowledge support from the Coordenação de Aperfeiçoamento de Pessoal de Nível Superior (CAPES) of Brazil through PROEX grant number 88887.845378/2023-00. We acknowledge financial support from Millenium Nucleus NCN23\_002 (TITANs) and Comité Mixto ESO-Chile.

We thank
the staff at the participating observatories, correlation
centers, and institutions for their enthusiastic support.
This paper makes use of the following ALMA data:
ADS/JAO.ALMA\#2017.1.00841.V and ADS/JAO.ALMA\#2019.1.01797.V.
ALMA is a partnership
of the European Southern Observatory (ESO;
Europe, representing its member states), NSF, and
National Institutes of Natural Sciences of Japan, together
with National Research Council (Canada), Ministry
of Science and Technology (MOST; Taiwan),
Academia Sinica Institute of Astronomy and Astrophysics
(ASIAA; Taiwan), and Korea Astronomy and
Space Science Institute (KASI; Republic of Korea), in
cooperation with the Republic of Chile. The Joint
ALMA Observatory is operated by ESO, Associated
Universities, Inc. (AUI)/NRAO, and the National Astronomical
Observatory of Japan (NAOJ). The NRAO
is a facility of the NSF operated under cooperative agreement
by AUI.
This research used resources of the Oak Ridge Leadership Computing Facility at the Oak Ridge National
Laboratory, which is supported by the Office of Science of the U.S. Department of Energy under contract
No. DE-AC05-00OR22725; the ASTROVIVES FEDER infrastructure, with project code IDIFEDER-2021-086; the computing cluster of Shanghai VLBI correlator supported by the Special Fund 
for Astronomy from the Ministry of Finance in China;  
We also thank the Center for Computational Astrophysics, National Astronomical Observatory of Japan. This work was supported by FAPESP (Fundacao de Amparo a Pesquisa do Estado de Sao Paulo) under grant 2021/01183-8.

APEX is a collaboration between the
Max-Planck-Institut f{\"u}r Radioastronomie (Germany),
ESO, and the Onsala Space Observatory (Sweden). The
SMA is a joint project between the SAO and ASIAA
and is funded by the Smithsonian Institution and the
Academia Sinica. The JCMT is operated by the East
Asian Observatory on behalf of the NAOJ, ASIAA, and
KASI, as well as the Ministry of Finance of China, Chinese
Academy of Sciences, and the National Key Research and Development
Program (No. 2017YFA0402700) of China
and Natural Science Foundation of China grant 11873028.
Additional funding support for the JCMT is provided by the Science
and Technologies Facility Council (UK) and participating
universities in the UK and Canada. 
The LMT is a project operated by the Instituto Nacional
de Astr\'{o}fisica, \'{O}ptica, y Electr\'{o}nica (Mexico) and the
University of Massachusetts at Amherst (USA).
The IRAM 30 m telescope on Pico Veleta, Spain and the NOEMA interferometer on Plateau de Bure,
France are operated by IRAM and supported by CNRS (Centre National de la Recherche Scientifique, France), MPG (Max-Planck-Gesellschaft, Germany), and IGN (Instituto Geográfico Nacional, Spain).
The SMT is operated by the Arizona
Radio Observatory, a part of the Steward Observatory
of the University of Arizona, with financial support of
operations from the State of Arizona and financial support
for instrumentation development from the NSF.
Support for SPT participation in the EHT is provided by the National Science Foundation through award OPP-1852617 
to the University of Chicago. Partial support is also 
provided by the Kavli Institute of Cosmological Physics at the University of Chicago. The SPT hydrogen maser was 
provided on loan from the GLT, courtesy of ASIAA.

This work used the
Extreme Science and Engineering Discovery Environment
(XSEDE), supported by NSF grant ACI-1548562,
and CyVerse, supported by NSF grants DBI-0735191,
DBI-1265383, and DBI-1743442. XSEDE Stampede2 resource
at TACC was allocated through TG-AST170024
and TG-AST080026N. XSEDE JetStream resource at
PTI and TACC was allocated through AST170028.
This research is part of the Frontera computing project at the Texas Advanced 
Computing Center through the Frontera Large-Scale Community Partnerships allocation
AST20023. Frontera is made possible by National Science Foundation award OAC-1818253.
This research was done using services provided by the OSG Consortium~\citep{osg07,osg09}, which is supported by the National Science Foundation award Nos. 2030508 and 1836650.
Additional work used ABACUS2.0, which is part of the eScience center at Southern Denmark University, and the Kultrun Astronomy Hybrid Cluster (projects Conicyt Programa de Astronomia Fondo Quimal QUIMAL170001, Conicyt PIA ACT172033, Fondecyt Iniciacion 11170268, Quimal 220002). 
Simulations were also performed on the SuperMUC cluster at the LRZ in Garching, 
on the LOEWE cluster in CSC in Frankfurt, on the HazelHen cluster at the HLRS in Stuttgart, 
and on the Pi2.0 and Siyuan Mark-I at Shanghai Jiao Tong University.
The computer resources of the Finnish IT Center for Science (CSC) and the Finnish Computing 
Competence Infrastructure (FCCI) project are acknowledged. This
research was enabled in part by support provided
by Compute Ontario (http://computeontario.ca), Calcul
Quebec (http://www.calculquebec.ca), and the Digital Research Alliance of Canada (https://alliancecan.ca/en).

The EHTC has
received generous donations of FPGA chips from Xilinx
Inc., under the Xilinx University Program. The EHTC
has benefited from technology shared under open-source
license by the Collaboration for Astronomy Signal Processing
and Electronics Research (CASPER). The EHT
project is grateful to T4Science and Microsemi for their
assistance with hydrogen masers. This research has
made use of NASA's Astrophysics Data System. We
gratefully acknowledge the support provided by the extended
staff of the ALMA, from the inception of
the ALMA Phasing Project through the observational
campaigns of 2017 and 2018. We would like to thank
A. Deller and W. Brisken for EHT-specific support with
the use of DiFX. We thank Martin Shepherd for the addition of extra features in the Difmap software 
that were used for the CLEAN imaging results presented in this paper.
We acknowledge the significance that
Maunakea, where the SMA and JCMT EHT stations
are located, has for the indigenous Hawaiian people.

\end{acknowledgements}

\end{appendix}

\end{document}

%% file: authorlist.tex
\author{
        The Event Horizon Telescope Collaboration
        \and
        Kazunori Akiyama\orcid{0000-0002-9475-4254}\inst{\ref{inst1},\ref{inst2},\ref{inst3}}\and
        Ezequiel Albentosa-Ruíz\orcid{0000-0002-7816-6401}\inst{\ref{inst4}}\and
        Antxon Alberdi\orcid{0000-0002-9371-1033}\inst{\ref{inst5}}\and
        Walter Alef\inst{\ref{inst6}}\and
        Juan Carlos Algaba\orcid{0000-0001-6993-1696}\inst{\ref{inst7}}\and
        Richard Anantua\orcid{0000-0003-3457-7660}\inst{\ref{inst8},\ref{inst9},\ref{inst3},\ref{inst10}}\and
        Keiichi Asada\orcid{0000-0001-6988-8763}\inst{\ref{inst11}}\and
        Rebecca Azulay\orcid{0000-0002-2200-5393}\inst{\ref{inst4},\ref{inst12},\ref{inst6}}\and
        Uwe Bach\orcid{0000-0002-7722-8412}\inst{\ref{inst6}}\and
        Anne-Kathrin Baczko\orcid{0000-0003-3090-3975}\inst{\ref{inst13},\ref{inst6}}\and
        David Ball\inst{\ref{inst14}}\and
        Mislav Baloković\orcid{0000-0003-0476-6647}\inst{\ref{inst15}}\and
        Bidisha Bandyopadhyay\orcid{0000-0002-2138-8564}\inst{\ref{inst16}}\and
        John Barrett\orcid{0000-0002-9290-0764}\inst{\ref{inst1}}\and
        Michi Bauböck\orcid{0000-0002-5518-2812}\inst{\ref{inst17}}\and
        Bradford A. Benson\orcid{0000-0002-5108-6823}\inst{\ref{inst18},\ref{inst19}}\and
        Dan Bintley\inst{\ref{inst20},\ref{inst21}}\and
        Lindy Blackburn\orcid{0000-0002-9030-642X}\inst{\ref{inst10},\ref{inst3}}\and
        Raymond Blundell\orcid{0000-0002-5929-5857}\inst{\ref{inst10}}\and
        Katherine L. Bouman\orcid{0000-0003-0077-4367}\inst{\ref{inst22}}\and
        Geoffrey C. Bower\orcid{0000-0003-4056-9982}\inst{\ref{inst20},\ref{inst21},\ref{inst23},\ref{inst24}}\and
        Michael Bremer\inst{\ref{inst25}}\and
        Roger Brissenden\orcid{0000-0002-2556-0894}\inst{\ref{inst10}}\and
        Silke Britzen\orcid{0000-0001-9240-6734}\inst{\ref{inst6}}\and
        Avery E. Broderick\orcid{0000-0002-3351-760X}\inst{\ref{inst26},\ref{inst27},\ref{inst28}}\and
        Dominique Broguiere\orcid{0000-0001-9151-6683}\inst{\ref{inst25}}\and
        Thomas Bronzwaer\orcid{0000-0003-1151-3971}\inst{\ref{inst29}}\and
        Sandra Bustamante\orcid{0000-0001-6169-1894}\inst{\ref{inst30}}\and
        Douglas F. Carlos\orcid{0000-0002-1340-7702}\inst{\ref{inst31}}\and
        John E. Carlstrom\orcid{0000-0002-2044-7665}\inst{\ref{inst32},\ref{inst19},\ref{inst33},\ref{inst34}}\and
        Andrew Chael\orcid{0000-0003-2966-6220}\inst{\ref{inst35}}\and
        Chi-kwan Chan\orcid{0000-0001-6337-6126}\inst{\ref{inst14},\ref{inst36},\ref{inst37}}\and
        Dominic O. Chang\orcid{0000-0001-9939-5257}\inst{\ref{inst10},\ref{inst3}}\and
        Erandi Chavez\orcid{0000-0003-4143-9717}\inst{\ref{inst10}}\and
        Koushik Chatterjee\orcid{0000-0002-2825-3590}\inst{\ref{inst38},\ref{inst3},\ref{inst10}}\and
        Shami Chatterjee\orcid{0000-0002-2878-1502}\inst{\ref{inst39}}\and
        Ming-Tang Chen\orcid{0000-0001-6573-3318}\inst{\ref{inst23}}\and
        Yongjun Chen (\cntext{陈永军})\orcid{0000-0001-5650-6770}\inst{\ref{inst40},\ref{inst41}}\and
        Xiaopeng Cheng\orcid{0000-0003-4407-9868}\inst{\ref{inst42}}\and
        Paul Chichura\orcid{0000-0002-5397-9035}\inst{\ref{inst33},\ref{inst32}}\and
        Ilje Cho\orcid{0000-0001-6083-7521}\inst{\ref{inst42},\ref{inst43},\ref{inst5}}\and
        Pierre Christian\orcid{0000-0001-6820-9941}\inst{\ref{inst44}}\and
        Nicholas S. Conroy\orcid{0000-0003-2886-2377}\inst{\ref{inst45},\ref{inst10}}\and
        John E. Conway\orcid{0000-0003-2448-9181}\inst{\ref{inst13}}\and
        Thomas M. Crawford\orcid{0000-0001-9000-5013}\inst{\ref{inst19},\ref{inst32}}\and
        Geoffrey B. Crew\orcid{0000-0002-2079-3189}\inst{\ref{inst1}}\and
        Alejandro Cruz-Osorio\orcid{0000-0002-3945-6342}\inst{\ref{inst46},\ref{inst47}}\and
        Yuzhu Cui (\cntext{崔玉竹})\orcid{0000-0001-6311-4345}\inst{\ref{inst48}}\and
        Brandon Curd\orcid{0000-0002-8650-0879}\inst{\ref{inst8},\ref{inst3},\ref{inst10}}\and
        Rohan Dahale\orcid{0000-0001-6982-9034}\inst{\ref{inst5}}\and
        Jordy Davelaar\orcid{0000-0002-2685-2434}\inst{\ref{inst49},\ref{inst50}}\and
        Mariafelicia De Laurentis\orcid{0000-0002-9945-682X}\inst{\ref{inst51},\ref{inst52}}\and
        Roger Deane\orcid{0000-0003-1027-5043}\inst{\ref{inst53},\ref{inst54},\ref{inst55}}\and
        Gregory Desvignes\orcid{0000-0003-3922-4055}\inst{\ref{inst6},\ref{inst56}}\and
        Jason Dexter\orcid{0000-0003-3903-0373}\inst{\ref{inst57}}\and
        Vedant Dhruv\orcid{0000-0001-6765-877X}\inst{\ref{inst17}}\and
        Indu K. Dihingia\orcid{0000-0002-4064-0446}\inst{\ref{inst58}}\and
        Sheperd S. Doeleman\orcid{0000-0002-9031-0904}\inst{\ref{inst10},\ref{inst3}}\and
        Sergio A. Dzib\orcid{0000-0001-6010-6200}\inst{\ref{inst6}}\and
        Ralph P. Eatough\orcid{0000-0001-6196-4135}\inst{\ref{inst59},\ref{inst6}}\and
        Razieh Emami\orcid{0000-0002-2791-5011}\inst{\ref{inst10}}\and
        Heino Falcke\orcid{0000-0002-2526-6724}\inst{\ref{inst29}}\and
        Joseph Farah\orcid{0000-0003-4914-5625}\inst{\ref{inst60},\ref{inst61}}\and
        Vincent L. Fish\orcid{0000-0002-7128-9345}\inst{\ref{inst1}}\and
        Edward Fomalont\orcid{0000-0002-9036-2747}\inst{\ref{inst62}}\and
        H. Alyson Ford\orcid{0000-0002-9797-0972}\inst{\ref{inst14}}\and
        Marianna Foschi\orcid{0000-0001-8147-4993}\inst{\ref{inst5}}\and
        Raquel Fraga-Encinas\orcid{0000-0002-5222-1361}\inst{\ref{inst29}}\and
        William T. Freeman\inst{\ref{inst63},\ref{inst64}}\and
        Per Friberg\orcid{0000-0002-8010-8454}\inst{\ref{inst20},\ref{inst21}}\and
        Christian M. Fromm\orcid{0000-0002-1827-1656}\inst{\ref{inst65},\ref{inst47},\ref{inst6}}\and
        Antonio Fuentes\orcid{0000-0002-8773-4933}\inst{\ref{inst5}}\and
        Peter Galison\orcid{0000-0002-6429-3872}\inst{\ref{inst3},\ref{inst66},\ref{inst67}}\and
        Charles F. Gammie\orcid{0000-0001-7451-8935}\inst{\ref{inst17},\ref{inst45},\ref{inst68}}\and
        Roberto García\orcid{0000-0002-6584-7443}\inst{\ref{inst25}}\and
        Olivier Gentaz\orcid{0000-0002-0115-4605}\inst{\ref{inst25}}\and
        Boris Georgiev\orcid{0000-0002-3586-6424}\inst{\ref{inst14}}\and
        Ciriaco Goddi\orcid{0000-0002-2542-7743}\inst{\ref{inst31},\ref{inst69},\ref{inst70},\ref{inst71}}\and
        Roman Gold\orcid{0000-0003-2492-1966}\inst{\ref{inst72},\ref{inst73},\ref{inst74}}\and
        Arturo I. Gómez-Ruiz\orcid{0000-0001-9395-1670}\inst{\ref{inst75},\ref{inst76}}\and
        José L. Gómez\orcid{0000-0003-4190-7613}\inst{\ref{inst5}}\and
        Minfeng Gu (\cntext{顾敏峰})\orcid{0000-0002-4455-6946}\inst{\ref{inst40},\ref{inst77}}\and
        Mark Gurwell\orcid{0000-0003-0685-3621}\inst{\ref{inst10}}\and
        Kazuhiro Hada\orcid{0000-0001-6906-772X}\inst{\ref{inst78},\ref{inst79}}\and
        Daryl Haggard\orcid{0000-0001-6803-2138}\inst{\ref{inst80},\ref{inst81}}\and
        Ronald Hesper\orcid{0000-0003-1918-6098}\inst{\ref{inst82}}\and
        Dirk Heumann\orcid{0000-0002-7671-0047}\inst{\ref{inst14}}\and
        Luis C. Ho (\cntext{何子山})\orcid{0000-0001-6947-5846}\inst{\ref{inst83},\ref{inst84}}\and
        Paul Ho\orcid{0000-0002-3412-4306}\inst{\ref{inst11},\ref{inst21},\ref{inst20}}\and
        Dan Hoak\inst{\ref{inst1}}\and
        Mareki Honma\orcid{0000-0003-4058-9000}\inst{\ref{inst79},\ref{inst85},\ref{inst86}}\and
        Chih-Wei L. Huang\orcid{0000-0001-5641-3953}\inst{\ref{inst11}}\and
        Lei Huang (\cntext{黄磊})\orcid{0000-0002-1923-227X}\inst{\ref{inst40},\ref{inst77}}\and
        David H. Hughes\inst{\ref{inst75}}\and
        Shiro Ikeda\orcid{0000-0002-2462-1448}\inst{\ref{inst2},\ref{inst87},\ref{inst88},\ref{inst89}}\and
        C. M. Violette Impellizzeri\orcid{0000-0002-3443-2472}\inst{\ref{inst90},\ref{inst62}}\and
        Makoto Inoue\orcid{0000-0001-5037-3989}\inst{\ref{inst11}}\and
        Sara Issaoun\orcid{0000-0002-5297-921X}\inst{\ref{inst10},\ref{inst50}}\and
        David J. James\orcid{0000-0001-5160-4486}\inst{\ref{inst91},\ref{inst92}}\and
        Buell T. Jannuzi\orcid{0000-0002-1578-6582}\inst{\ref{inst14}}\and
        Michael Janssen\orcid{0000-0001-8685-6544}\inst{\ref{inst29},\ref{inst6}}\and
        Britton Jeter\orcid{0000-0003-2847-1712}\inst{\ref{inst11}}\and
        Wu Jiang (\cntext{江悟})\orcid{0000-0001-7369-3539}\inst{\ref{inst40}}\and
        Alejandra Jiménez-Rosales\orcid{0000-0002-2662-3754}\inst{\ref{inst29}}\and
        Michael D. Johnson\orcid{0000-0002-4120-3029}\inst{\ref{inst10},\ref{inst3}}\and
        Svetlana Jorstad\orcid{0000-0001-6158-1708}\inst{\ref{inst93}}\and
        Adam C. Jones\inst{\ref{inst19}}\and
        Abhishek V. Joshi\orcid{0000-0002-2514-5965}\inst{\ref{inst17}}\and
        Taehyun Jung\orcid{0000-0001-7003-8643}\inst{\ref{inst42},\ref{inst94}}\and
        Ramesh Karuppusamy\orcid{0000-0002-5307-2919}\inst{\ref{inst6}}\and
        Tomohisa Kawashima\orcid{0000-0001-8527-0496}\inst{\ref{inst95}}\and
        Garrett K. Keating\orcid{0000-0002-3490-146X}\inst{\ref{inst10}}\and
        Mark Kettenis\orcid{0000-0002-6156-5617}\inst{\ref{inst96}}\and
        Dong-Jin Kim\orcid{0000-0002-7038-2118}\inst{\ref{inst97}}\and
        Jae-Young Kim\orcid{0000-0001-8229-7183}\inst{\ref{inst98}}\and
        Jongsoo Kim\orcid{0000-0002-1229-0426}\inst{\ref{inst42}}\and
        Junhan Kim\orcid{0000-0002-4274-9373}\inst{\ref{inst99}}\and
        Motoki Kino\orcid{0000-0002-2709-7338}\inst{\ref{inst2},\ref{inst100}}\and
        Jun Yi Koay\orcid{0000-0002-7029-6658}\inst{\ref{inst11}}\and
        Prashant Kocherlakota\orcid{0000-0001-7386-7439}\inst{\ref{inst3},\ref{inst10}}\and
        Yutaro Kofuji\inst{\ref{inst79},\ref{inst86}}\and
        Patrick M. Koch\orcid{0000-0003-2777-5861}\inst{\ref{inst11}}\and
        Shoko Koyama\orcid{0000-0002-3723-3372}\inst{\ref{inst101},\ref{inst11}}\and
        Carsten Kramer\orcid{0000-0002-4908-4925}\inst{\ref{inst25}}\and
        Joana A. Kramer\orcid{0009-0003-3011-0454}\inst{\ref{inst6}}\and
        Michael Kramer\orcid{0000-0002-4175-2271}\inst{\ref{inst6}}\and
        Thomas P. Krichbaum\orcid{0000-0002-4892-9586}\inst{\ref{inst6}}\and
        Cheng-Yu Kuo\orcid{0000-0001-6211-5581}\inst{\ref{inst102},\ref{inst11}}\and
        Noemi La Bella\orcid{0000-0002-8116-9427}\inst{\ref{inst29}}\and
        Deokhyeong Lee\orcid{0009-0003-2122-9437}\inst{\ref{inst103}}\and
        Sang-Sung Lee\orcid{0000-0002-6269-594X}\inst{\ref{inst42}}\and
        Aviad Levis\orcid{0000-0001-7307-632X}\inst{\ref{inst22}}\and
        Zhiyuan Li (\cntext{李志远})\orcid{0000-0003-0355-6437}\inst{\ref{inst104},\ref{inst105}}\and
        Rocco Lico\orcid{0000-0001-7361-2460}\inst{\ref{inst106},\ref{inst5}}\and
        Greg Lindahl\orcid{0000-0002-6100-4772}\inst{\ref{inst107}}\and
        Michael Lindqvist\orcid{0000-0002-3669-0715}\inst{\ref{inst13}}\and
        Mikhail Lisakov\orcid{0000-0001-6088-3819}\inst{\ref{inst108}}\and
        Jun Liu (\cntext{刘俊})\orcid{0000-0002-7615-7499}\inst{\ref{inst6}}\and
        Kuo Liu\orcid{0000-0002-2953-7376}\inst{\ref{inst40},\ref{inst41}}\and
        Elisabetta Liuzzo\orcid{0000-0003-0995-5201}\inst{\ref{inst109}}\and
        Wen-Ping Lo\orcid{0000-0003-1869-2503}\inst{\ref{inst11},\ref{inst110}}\and
        Andrei P. Lobanov\orcid{0000-0003-1622-1484}\inst{\ref{inst6}}\and
        Laurent Loinard\orcid{0000-0002-5635-3345}\inst{\ref{inst111},\ref{inst3},\ref{inst112}}\and
        Colin J. Lonsdale\orcid{0000-0003-4062-4654}\inst{\ref{inst1}}\and
        Amy E. Lowitz\orcid{0000-0002-4747-4276}\inst{\ref{inst14}}\and
        Ru-Sen Lu (\cntext{路如森})\orcid{0000-0002-7692-7967}\inst{\ref{inst40},\ref{inst41},\ref{inst6}}\and
        Nicholas R. MacDonald\orcid{0000-0002-6684-8691}\inst{\ref{inst6}}\and
        Jirong Mao (\cntext{毛基荣})\orcid{0000-0002-7077-7195}\inst{\ref{inst113},\ref{inst114},\ref{inst115}}\and
        Nicola Marchili\orcid{0000-0002-5523-7588}\inst{\ref{inst109},\ref{inst6}}\and
        Sera Markoff\orcid{0000-0001-9564-0876}\inst{\ref{inst116},\ref{inst117}}\and
        Daniel P. Marrone\orcid{0000-0002-2367-1080}\inst{\ref{inst14}}\and
        Alan P. Marscher\orcid{0000-0001-7396-3332}\inst{\ref{inst93}}\and
        Iván Martí-Vidal\orcid{0000-0003-3708-9611}\inst{\ref{inst4},\ref{inst12}}\and
        Satoki Matsushita\orcid{0000-0002-2127-7880}\inst{\ref{inst11}}\and
        Lynn D. Matthews\orcid{0000-0002-3728-8082}\inst{\ref{inst1}}\and
        Lia Medeiros\orcid{0000-0003-2342-6728}\inst{\ref{inst118}}\and
        Karl M. Menten\orcid{0000-0001-6459-0669}\inst{\ref{inst6},\ref{inst119}}\and
        Izumi Mizuno\orcid{0000-0002-7210-6264}\inst{\ref{inst20},\ref{inst21}}\and
        Yosuke Mizuno\orcid{0000-0002-8131-6730}\inst{\ref{inst58},\ref{inst120},\ref{inst47}}\and
        Joshua Montgomery\orcid{0000-0003-0345-8386}\inst{\ref{inst81},\ref{inst19}}\and
        James M. Moran\orcid{0000-0002-3882-4414}\inst{\ref{inst10},\ref{inst3}}\and
        Kotaro Moriyama\orcid{0000-0003-1364-3761}\inst{\ref{inst47},\ref{inst79}}\and
        Monika Moscibrodzka\orcid{0000-0002-4661-6332}\inst{\ref{inst29}}\and
        Wanga Mulaudzi\orcid{0000-0003-4514-625X}\inst{\ref{inst116}}\and
        Cornelia Müller\orcid{0000-0002-2739-2994}\inst{\ref{inst6},\ref{inst29}}\and
        Hendrik Müller\orcid{0000-0002-9250-0197}\inst{\ref{inst6}}\and
        Alejandro Mus\orcid{0000-0003-0329-6874}\inst{\ref{inst69},\ref{inst106},\ref{inst121},\ref{inst122}}\and
        Gibwa Musoke\orcid{0000-0003-1984-189X}\inst{\ref{inst116},\ref{inst29}}\and
        Ioannis Myserlis\orcid{0000-0003-3025-9497}\inst{\ref{inst123}}\and
        Hiroshi Nagai\orcid{0000-0003-0292-3645}\inst{\ref{inst2},\ref{inst85}}\and
        Neil M. Nagar\orcid{0000-0001-6920-662X}\inst{\ref{inst16}}\and
        Dhanya G. Nair\orcid{0000-0001-5357-7805}\inst{\ref{inst16},\ref{inst6}}\and
        Masanori Nakamura\orcid{0000-0001-6081-2420}\inst{\ref{inst124},\ref{inst11}}\and
        Gopal Narayanan\orcid{0000-0002-4723-6569}\inst{\ref{inst30}}\and
        Iniyan Natarajan\orcid{0000-0001-8242-4373}\inst{\ref{inst10},\ref{inst3}}\and
        Antonios Nathanail\orcid{0000-0002-1655-9912}\inst{\ref{inst125},\ref{inst47}}\and
        Santiago Navarro Fuentes\inst{\ref{inst123}}\and
        Joey Neilsen\orcid{0000-0002-8247-786X}\inst{\ref{inst126}}\and
        Chunchong Ni\orcid{0000-0003-1361-5699}\inst{\ref{inst27},\ref{inst28},\ref{inst26}}\and
        Michael A. Nowak\orcid{0000-0001-6923-1315}\inst{\ref{inst127}}\and
        Junghwan Oh\orcid{0000-0002-4991-9638}\inst{\ref{inst96}}\and
        Hiroki Okino\orcid{0000-0003-3779-2016}\inst{\ref{inst79},\ref{inst86}}\and
        Héctor Raúl Olivares Sánchez\orcid{0000-0001-6833-7580}\inst{\ref{inst128}}\and
        Tomoaki Oyama\orcid{0000-0003-4046-2923}\inst{\ref{inst79}}\and
        Feryal Özel\orcid{0000-0003-4413-1523}\inst{\ref{inst129}}\and
        Daniel C. M. Palumbo\orcid{0000-0002-7179-3816}\inst{\ref{inst3},\ref{inst10}}\and
        Georgios Filippos Paraschos\orcid{0000-0001-6757-3098}\inst{\ref{inst6}}\and
        Jongho Park\orcid{0000-0001-6558-9053}\inst{\ref{inst130},\ref{inst11}}\and
        Harriet Parsons\orcid{0000-0002-6327-3423}\inst{\ref{inst20},\ref{inst21}}\and
        Nimesh Patel\orcid{0000-0002-6021-9421}\inst{\ref{inst10}}\and
        Ue-Li Pen\orcid{0000-0003-2155-9578}\inst{\ref{inst11},\ref{inst26},\ref{inst131},\ref{inst132},\ref{inst133}}\and
        Dominic W. Pesce\orcid{0000-0002-5278-9221}\inst{\ref{inst10},\ref{inst3}}\and
        Vincent Piétu\inst{\ref{inst25}}\and
        Alexander Plavin\orcid{0000-0003-2914-8554}\inst{\ref{inst3},\ref{inst10},\ref{inst6}}\and
        Aleksandar PopStefanija\inst{\ref{inst30}}\and
        Oliver Porth\orcid{0000-0002-4584-2557}\inst{\ref{inst116},\ref{inst47}}\and
        Ben Prather\orcid{0000-0002-0393-7734}\inst{\ref{inst17}}\and
        Giacomo Principe\orcid{0000-0003-0406-7387}\inst{\ref{inst134},\ref{inst135},\ref{inst106}}\and
        Dimitrios Psaltis\orcid{0000-0003-1035-3240}\inst{\ref{inst129}}\and
        Hung-Yi Pu\orcid{0000-0001-9270-8812}\inst{\ref{inst136},\ref{inst137},\ref{inst11}}\and
        Alexandra Rahlin\orcid{0000-0003-3953-1776}\inst{\ref{inst19}}\and
        Venkatessh Ramakrishnan\orcid{0000-0002-9248-086X}\inst{\ref{inst16},\ref{inst138},\ref{inst139}}\and
        Ramprasad Rao\orcid{0000-0002-1407-7944}\inst{\ref{inst10}}\and
        Mark G. Rawlings\orcid{0000-0002-6529-202X}\inst{\ref{inst140},\ref{inst20},\ref{inst21}}\and
        Luciano Rezzolla\orcid{0000-0002-1330-7103}\inst{\ref{inst47},\ref{inst141},\ref{inst142}}\and
        Angelo Ricarte\orcid{0000-0001-5287-0452}\inst{\ref{inst3},\ref{inst10}}\and
        Luca Ricci\orcid{0000-0002-4175-3194}\inst{\ref{inst143}}\and
        Bart Ripperda\orcid{0000-0002-7301-3908}\inst{\ref{inst131},\ref{inst144},\ref{inst132},\ref{inst26}}\and
        Jan Röder\orcid{0000-0002-2426-927X}\inst{\ref{inst5}}\and
        Freek Roelofs\orcid{0000-0001-5461-3687}\inst{\ref{inst29}}\and
        Cristina Romero-Cañizales\orcid{0000-0001-6301-9073}\inst{\ref{inst11}}\and
        Eduardo Ros\orcid{0000-0001-9503-4892}\inst{\ref{inst6}}\and
        Arash Roshanineshat\orcid{0000-0002-8280-9238}\inst{\ref{inst14}}\and
        Helge Rottmann\inst{\ref{inst6}}\and
        Alan L. Roy\orcid{0000-0002-1931-0135}\inst{\ref{inst6}}\and
        Ignacio Ruiz\orcid{0000-0002-0965-5463}\inst{\ref{inst123}}\and
        Chet Ruszczyk\orcid{0000-0001-7278-9707}\inst{\ref{inst1}}\and
        Kazi L. J. Rygl\orcid{0000-0003-4146-9043}\inst{\ref{inst109}}\and
        León D. S. Salas\orcid{0000-0003-1979-6363}\inst{\ref{inst116}}\and
        Salvador Sánchez\orcid{0000-0002-8042-5951}\inst{\ref{inst123}}\and
        David Sánchez-Argüelles\orcid{0000-0002-7344-9920}\inst{\ref{inst75},\ref{inst76}}\and
        Miguel Sánchez-Portal\orcid{0000-0003-0981-9664}\inst{\ref{inst123}}\and
        Mahito Sasada\orcid{0000-0001-5946-9960}\inst{\ref{inst145},\ref{inst79},\ref{inst146}}\and
        Kaushik Satapathy\orcid{0000-0003-0433-3585}\inst{\ref{inst14}}\and
        Saurabh\orcid{0000-0001-7156-4848}\inst{\ref{inst6}}\and
        Tuomas Savolainen\orcid{0000-0001-6214-1085}\inst{\ref{inst147},\ref{inst139},\ref{inst6}}\and
        F. Peter Schloerb\inst{\ref{inst30}}\and
        Jonathan Schonfeld\orcid{0000-0002-8909-2401}\inst{\ref{inst10}}\and
        Karl-Friedrich Schuster\orcid{0000-0003-2890-9454}\inst{\ref{inst25}}\and
        Lijing Shao\orcid{0000-0002-1334-8853}\inst{\ref{inst84},\ref{inst6}}\and
        Zhiqiang Shen (\cntext{沈志强})\orcid{0000-0003-3540-8746}\inst{\ref{inst40},\ref{inst41}}\and
        Sasikumar Silpa\orcid{0000-0003-0667-7074}\inst{\ref{inst16}}\and
        Des Small\orcid{0000-0003-3723-5404}\inst{\ref{inst96}}\and
        Randall Smith\orcid{0000-0003-4284-4167}\inst{\ref{inst10}}\and
        Bong Won Sohn\orcid{0000-0002-4148-8378}\inst{\ref{inst42},\ref{inst94},\ref{inst43}}\and
        Jason SooHoo\orcid{0000-0003-1938-0720}\inst{\ref{inst1}}\and
        Kamal Souccar\orcid{0000-0001-7915-5272}\inst{\ref{inst30}}\and
        Joshua S. Stanway\orcid{0009-0003-7659-4642}\inst{\ref{inst148}}\and
        He Sun (\cntext{孙赫})\orcid{0000-0003-1526-6787}\inst{\ref{inst149},\ref{inst150}}\and
        Fumie Tazaki\orcid{0000-0003-0236-0600}\inst{\ref{inst151}}\and
        Alexandra J. Tetarenko\orcid{0000-0003-3906-4354}\inst{\ref{inst152}}\and
        Paul Tiede\orcid{0000-0003-3826-5648}\inst{\ref{inst10},\ref{inst3}}\and
        Remo P. J. Tilanus\orcid{0000-0002-6514-553X}\inst{\ref{inst14},\ref{inst29},\ref{inst90},\ref{inst153}}\and
        Michael Titus\orcid{0000-0001-9001-3275}\inst{\ref{inst1}}\and
        Kenji Toma\orcid{0000-0002-7114-6010}\inst{\ref{inst154},\ref{inst155}}\and
        Pablo Torne\orcid{0000-0001-8700-6058}\inst{\ref{inst123},\ref{inst6}}\and
        Teresa Toscano\orcid{0000-0003-3658-7862}\inst{\ref{inst5}}\and
        Efthalia Traianou\orcid{0000-0002-1209-6500}\inst{\ref{inst5},\ref{inst6}}\and
        Tyler Trent\inst{\ref{inst14}}\and
        Sascha Trippe\orcid{0000-0003-0465-1559}\inst{\ref{inst156},\ref{inst157}}\and
        Matthew Turk\orcid{0000-0002-5294-0198}\inst{\ref{inst45}}\and
        Ilse van Bemmel\orcid{0000-0001-5473-2950}\inst{\ref{inst158}}\and
        Huib Jan van Langevelde\orcid{0000-0002-0230-5946}\inst{\ref{inst96},\ref{inst90},\ref{inst159}}\and
        Daniel R. van Rossum\orcid{0000-0001-7772-6131}\inst{\ref{inst29}}\and
        Sebastiano D. von Fellenberg\orcid{0000-0002-9156-2249}\inst{\ref{inst131},\ref{inst6}}\and
        Jesse Vos\orcid{0000-0003-3349-7394}\inst{\ref{inst160}}\and
        Jan Wagner\orcid{0000-0003-1105-6109}\inst{\ref{inst6}}\and
        Derek Ward-Thompson\orcid{0000-0003-1140-2761}\inst{\ref{inst148}}\and
        John Wardle\orcid{0000-0002-8960-2942}\inst{\ref{inst161}}\and
        Jasmin E. Washington\orcid{0000-0002-7046-0470}\inst{\ref{inst14}}\and
        Jonathan Weintroub\orcid{0000-0002-4603-5204}\inst{\ref{inst10},\ref{inst3}}\and
        Andrew T. West\orcid{0000-0002-5471-4709}\inst{\ref{inst14}}\and
        Robert Wharton\orcid{0000-0002-7416-5209}\inst{\ref{inst6}}\and
        Maciek Wielgus\orcid{0000-0002-8635-4242}\inst{\ref{inst5}}\and
        Kaj Wiik\orcid{0000-0002-0862-3398}\inst{\ref{inst162},\ref{inst138},\ref{inst139}}\and
        Gunther Witzel\orcid{0000-0003-2618-797X}\inst{\ref{inst6}}\and
        Michael F. Wondrak\orcid{0000-0002-6894-1072}\inst{\ref{inst29},\ref{inst163}}\and
        George N. Wong\orcid{0000-0001-6952-2147}\inst{\ref{inst164},\ref{inst35}}\and
        Jompoj Wongphexhauxsorn\orcid{0000-0002-7730-4956}\inst{\ref{inst143},\ref{inst6}}\and
        Qingwen Wu (\cntext{吴庆文})\orcid{0000-0003-4773-4987}\inst{\ref{inst165}}\and
        Nitika Yadlapalli\orcid{0000-0003-3255-4617}\inst{\ref{inst22}}\and
        Paul Yamaguchi\orcid{0000-0002-6017-8199}\inst{\ref{inst10}}\and
        Aristomenis Yfantis\orcid{0000-0002-3244-7072}\inst{\ref{inst29}}\and
        Doosoo Yoon\orcid{0000-0001-8694-8166}\inst{\ref{inst116}}\and
        André Young\orcid{0000-0003-0000-2682}\inst{\ref{inst29}}\and
        Ziri Younsi\orcid{0000-0001-9283-1191}\inst{\ref{inst166},\ref{inst47}}\and
        Wei Yu (\cntext{于威})\orcid{0000-0002-5168-6052}\inst{\ref{inst10}}\and
        Feng Yuan (\cntext{袁峰})\orcid{0000-0003-3564-6437}\inst{\ref{inst167}}\and
        Ye-Fei Yuan (\cntext{袁业飞})\orcid{0000-0002-7330-4756}\inst{\ref{inst168}}\and
        Ai-Ling Zeng (\cntext{曾艾玲})\orcid{0009-0000-9427-4608}\inst{\ref{inst5}}\and
        J. Anton Zensus\orcid{0000-0001-7470-3321}\inst{\ref{inst6}}\and
        Shuo Zhang\orcid{0000-0002-2967-790X}\inst{\ref{inst169}}\and
        Guang-Yao Zhao\orcid{0000-0002-4417-1659}\inst{\ref{inst6},\ref{inst5}}\and
        Shan-Shan Zhao (\cntext{赵杉杉})\orcid{0000-0002-9774-3606}\inst{\ref{inst40}}
        Ryan Berthold\inst{\ref{inst20}, \ref{inst21}}\and
        Shu-Hao Chang\inst{\ref{inst11}}\and
        Ryan Chilson\inst{\ref{inst23}}\and
        Chih-Chiang Han\inst{\ref{inst11}}\and
        David M. Gale\inst{\ref{inst75}}\and
        Gertie Geertsema\orcid{0000-0003-3933-0069}\inst{\ref{inst170}}\and
        José Luis Hernández-Rebollar\inst{\ref{inst75}}\and
        Yau-De Huang\orcid{0000-0001-8783-6211}\inst{\ref{inst11}}\and
        Ryan P. Keenan\orcid{0000-0003-1859-9640}\inst{\ref{inst14}}\and
        Derek Kubo\orcid{0000-0002-9793-2933}\inst{\ref{inst23}}\and
        Kuan-Yu Liu\orcid{0000-0002-9699-9685}\inst{\ref{inst20}, \ref{inst21}}\and
        Pierre L. Martin-Cocher\orcid{0000-0002-2702-9802}\inst{\ref{inst11}}\and
        Daniel Michalik\orcid{0000-0002-7618-6556}\inst{\ref{inst19}}\and
        Alfredo Montaña\inst{\ref{inst75}}\and
        Andrew Nadolski\orcid{0000-0001-9479-9957}\inst{\ref{inst19}}\and
        Peter Oshiro\inst{\ref{inst23}}\and
        Philippe A. Raffin\orcid{0000-0001-8111-9051}\inst{\ref{inst23}}\and
        Iván Rodríguez-Montoya\inst{\ref{inst76}, \ref{inst75}}\and
        Ta-Shun Wei\inst{\ref{inst11}}\and
        Matthew R. Young\orcid{0000-0003-1090-1673}\inst{\ref{inst18},\ref{inst32}}
}
\institute{
        Massachusetts Institute of Technology Haystack Observatory, 99 Millstone Road, Westford, MA 01886, USA\label{inst1}\and
        National Astronomical Observatory of Japan, 2-21-1 Osawa, Mitaka, Tokyo 181-8588, Japan\label{inst2}\and
        Black Hole Initiative at Harvard University, 20 Garden Street, Cambridge, MA 02138, USA\label{inst3}\and
        Departament d'Astronomia i Astrofísica, Universitat de València, C. Dr. Moliner 50, E-46100 Burjassot, València, Spain\label{inst4}\and
        Instituto de Astrofísica de Andalucía-CSIC, Glorieta de la Astronomía s/n, E-18008 Granada, Spain\label{inst5}\and
        Max-Planck-Institut für Radioastronomie, Auf dem Hügel 69, D-53121 Bonn, Germany\label{inst6}\and
        Department of Physics, Faculty of Science, Universiti Malaya, 50603 Kuala Lumpur, Malaysia\label{inst7}\and
        Department of Physics \& Astronomy, The University of Texas at San Antonio, One UTSA Circle, San Antonio, TX 78249, USA\label{inst8}\and
        Physics \& Astronomy Department, Rice University, Houston, TX 77005-1827, USA\label{inst9}\and
        Center for Astrophysics $|$ Harvard \& Smithsonian, 60 Garden Street, Cambridge, MA 02138, USA\label{inst10}\and
        Institute of Astronomy and Astrophysics, Academia Sinica, 11F of Astronomy-Mathematics Building, AS/NTU No. 1, Sec. 4, Roosevelt Rd., Taipei 106216, Taiwan, R.O.C.\label{inst11}\and
        Observatori Astronòmic, Universitat de València, C. Catedrático José Beltrán 2, E-46980 Paterna, València, Spain\label{inst12}\and
        Department of Space, Earth and Environment, Chalmers University of Technology, Onsala Space Observatory, SE-43992 Onsala, Sweden\label{inst13}\and
        Steward Observatory and Department of Astronomy, University of Arizona, 933 N. Cherry Ave., Tucson, AZ 85721, USA\label{inst14}\and
        Yale Center for Astronomy \& Astrophysics, Yale University, 52 Hillhouse Avenue, New Haven, CT 06511, USA\label{inst15}\and
        Astronomy Department, Universidad de Concepción, Casilla 160-C, Concepción, Chile\label{inst16}\and
        Department of Physics, University of Illinois, 1110 West Green Street, Urbana, IL 61801, USA\label{inst17}\and
        Fermi National Accelerator Laboratory, MS209, P.O. Box 500, Batavia, IL 60510, USA\label{inst18}\and
        Department of Astronomy and Astrophysics, University of Chicago, 5640 South Ellis Avenue, Chicago, IL 60637, USA\label{inst19}\and
        East Asian Observatory, 660 N. A'ohoku Place, Hilo, HI 96720, USA\label{inst20}\and
        \textit{James Clerk Maxwell} Telescope (JCMT), 660 N. A'ohoku Place, Hilo, HI 96720, USA\label{inst21}\and
        California Institute of Technology, 1200 East California Boulevard, Pasadena, CA 91125, USA\label{inst22}\and
        Institute of Astronomy and Astrophysics, Academia Sinica, 645 N. A'ohoku Place, Hilo, HI 96720, USA\label{inst23}\and
        Department of Physics and Astronomy, University of Hawaii at Manoa, 2505 Correa Road, Honolulu, HI 96822, USA\label{inst24}\and
        Institut de Radioastronomie Millimétrique (IRAM), 300 rue de la Piscine, F-38406 Saint Martin d'Hères, France\label{inst25}\and
        Perimeter Institute for Theoretical Physics, 31 Caroline Street North, Waterloo, ON N2L 2Y5, Canada\label{inst26}\and
        Department of Physics and Astronomy, University of Waterloo, 200 University Avenue West, Waterloo, ON N2L 3G1, Canada\label{inst27}\and
        Waterloo Centre for Astrophysics, University of Waterloo, Waterloo, ON N2L 3G1, Canada\label{inst28}\and
        Department of Astrophysics, Institute for Mathematics, Astrophysics and Particle Physics (IMAPP), Radboud University, P.O. Box 9010, 6500 GL Nijmegen, The Netherlands\label{inst29}\and
        Department of Astronomy, University of Massachusetts, Amherst, MA 01003, USA\label{inst30}\and
        Instituto de Astronomia, Geofísica e Ciências Atmosféricas, Universidade de São Paulo, R. do Matão, 1226, São Paulo, SP 05508-090, Brazil\label{inst31}\and
        Kavli Institute for Cosmological Physics, University of Chicago, 5640 South Ellis Avenue, Chicago, IL 60637, USA\label{inst32}\and
        Department of Physics, University of Chicago, 5720 South Ellis Avenue, Chicago, IL 60637, USA\label{inst33}\and
        Enrico Fermi Institute, University of Chicago, 5640 South Ellis Avenue, Chicago, IL 60637, USA\label{inst34}\and
        Princeton Gravity Initiative, Jadwin Hall, Princeton University, Princeton, NJ 08544, USA\label{inst35}\and
        Data Science Institute, University of Arizona, 1230 N. Cherry Ave., Tucson, AZ 85721, USA\label{inst36}\and
        Program in Applied Mathematics, University of Arizona, 617 N. Santa Rita, Tucson, AZ 85721, USA\label{inst37}\and
        Department of Physics, University of Maryland, 7901 Regents Drive, College Park, MD 20742, USA\label{inst38}\and
        Cornell Center for Astrophysics and Planetary Science, Cornell University, Ithaca, NY 14853, USA\label{inst39}\and
        Shanghai Astronomical Observatory, Chinese Academy of Sciences, 80 Nandan Road, Shanghai 200030, People's Republic of China\label{inst40}\and
        Key Laboratory of Radio Astronomy and Technology, Chinese Academy of Sciences, A20 Datun Road, Chaoyang District, Beijing, 100101, People’s Republic of China\label{inst41}\and
        Korea Astronomy and Space Science Institute, Daedeok-daero 776, Yuseong-gu, Daejeon 34055, Republic of Korea\label{inst42}\and
        Department of Astronomy, Yonsei University, Yonsei-ro 50, Seodaemun-gu, 03722 Seoul, Republic of Korea\label{inst43}\and
        WattTime, 490 43rd Street, Unit 221, Oakland, CA 94609, USA\label{inst44}\and
        Department of Astronomy, University of Illinois at Urbana-Champaign, 1002 West Green Street, Urbana, IL 61801, USA\label{inst45}\and
        Instituto de Astronomía, Universidad Nacional Autónoma de México (UNAM), Apdo Postal 70-264, Ciudad de México, México\label{inst46}\and
        Institut für Theoretische Physik, Goethe-Universität Frankfurt, Max-von-Laue-Straße 1, D-60438 Frankfurt am Main, Germany\label{inst47}\and
        Institute of Astrophysics, Central China Normal University, Wuhan 430079, People's Republic of China\label{inst48}\and
        Department of Astrophysical Sciences, Peyton Hall, Princeton University, Princeton, NJ 08544, USA\label{inst49}\and
        NASA Hubble Fellowship Program, Einstein Fellow\label{inst50}\and
        Dipartimento di Fisica ``E. Pancini'', Università di Napoli ``Federico II'', Compl. Univ. di Monte S. Angelo, Edificio G, Via Cinthia, I-80126, Napoli, Italy\label{inst51}\and
        INFN Sez. di Napoli, Compl. Univ. di Monte S. Angelo, Edificio G, Via Cinthia, I-80126, Napoli, Italy\label{inst52}\and
        Wits Centre for Astrophysics, University of the Witwatersrand, 1 Jan Smuts Avenue, Braamfontein, Johannesburg 2050, South Africa\label{inst53}\and
        Department of Physics, University of Pretoria, Hatfield, Pretoria 0028, South Africa\label{inst54}\and
        Centre for Radio Astronomy Techniques and Technologies, Department of Physics and Electronics, Rhodes University, Makhanda 6140, South Africa\label{inst55}\and
        LESIA, Observatoire de Paris, Université PSL, CNRS, Sorbonne Université, Université de Paris, 5 place Jules Janssen, F-92195 Meudon, France\label{inst56}\and
        JILA and Department of Astrophysical and Planetary Sciences, University of Colorado, Boulder, CO 80309, USA\label{inst57}\and
        Tsung-Dao Lee Institute, Shanghai Jiao Tong University, Shengrong Road 520, Shanghai, 201210, People’s Republic of China\label{inst58}\and
        National Astronomical Observatories, Chinese Academy of Sciences, 20A Datun Road, Chaoyang District, Beijing 100101, PR China\label{inst59}\and
        Las Cumbres Observatory, 6740 Cortona Drive, Suite 102, Goleta, CA 93117-5575, USA\label{inst60}\and
        Department of Physics, University of California, Santa Barbara, CA 93106-9530, USA\label{inst61}\and
        National Radio Astronomy Observatory, 520 Edgemont Road, Charlottesville, VA 22903, USA\label{inst62}\and
        Department of Electrical Engineering and Computer Science, Massachusetts Institute of Technology, 32-D476, 77 Massachusetts Ave., Cambridge, MA 02142, USA\label{inst63}\and
        Google Research, 355 Main St., Cambridge, MA 02142, USA\label{inst64}\and
        Institut für Theoretische Physik und Astrophysik, Universität Würzburg, Emil-Fischer-Str. 31, D-97074 Würzburg, Germany\label{inst65}\and
        Department of History of Science, Harvard University, Cambridge, MA 02138, USA\label{inst66}\and
        Department of Physics, Harvard University, Cambridge, MA 02138, USA\label{inst67}\and
        NCSA, University of Illinois, 1205 W. Clark St., Urbana, IL 61801, USA\label{inst68}\and
        Dipartimento di Fisica, Università degli Studi di Cagliari, SP Monserrato-Sestu km 0.7, I-09042 Monserrato (CA), Italy\label{inst69}\and
        INAF - Osservatorio Astronomico di Cagliari, via della Scienza 5, I-09047 Selargius (CA), Italy\label{inst70}\and
        INFN, sezione di Cagliari, I-09042 Monserrato (CA), Italy\label{inst71}\and
        Institute for Mathematics and Interdisciplinary Center for Scientific Computing, Heidelberg University, Im Neuenheimer Feld 205, Heidelberg 69120, Germany\label{inst72}\and
        Institut f\"ur Theoretische Physik, Universit\"at Heidelberg, Philosophenweg 16, 69120 Heidelberg, Germany\label{inst73}\and
        CP3-Origins, University of Southern Denmark, Campusvej 55, DK-5230 Odense, Denmark\label{inst74}\and
        Instituto Nacional de Astrofísica, Óptica y Electrónica. Apartado Postal 51 y 216, 72000. Puebla Pue., México\label{inst75}\and
        Consejo Nacional de Humanidades, Ciencia y Tecnología, Av. Insurgentes Sur 1582, 03940, Ciudad de México, México\label{inst76}\and
        Key Laboratory for Research in Galaxies and Cosmology, Chinese Academy of Sciences, Shanghai 200030, People's Republic of China\label{inst77}\and
        Graduate School of Science, Nagoya City University, Yamanohata 1, Mizuho-cho, Mizuho-ku, Nagoya, 467-8501, Aichi, Japan\label{inst78}\and
        Mizusawa VLBI Observatory, National Astronomical Observatory of Japan, 2-12 Hoshigaoka, Mizusawa, Oshu, Iwate 023-0861, Japan\label{inst79}\and
        Department of Physics, McGill University, 3600 rue University, Montréal, QC H3A 2T8, Canada\label{inst80}\and
        Trottier Space Institute at McGill, 3550 rue University, Montréal,  QC H3A 2A7, Canada\label{inst81}\and
        NOVA Sub-mm Instrumentation Group, Kapteyn Astronomical Institute, University of Groningen, Landleven 12, 9747 AD Groningen, The Netherlands\label{inst82}\and
        Department of Astronomy, School of Physics, Peking University, Beijing 100871, People's Republic of China\label{inst83}\and
        Kavli Institute for Astronomy and Astrophysics, Peking University, Beijing 100871, People's Republic of China\label{inst84}\and
        Department of Astronomical Science, The Graduate University for Advanced Studies (SOKENDAI), 2-21-1 Osawa, Mitaka, Tokyo 181-8588, Japan\label{inst85}\and
        Department of Astronomy, Graduate School of Science, The University of Tokyo, 7-3-1 Hongo, Bunkyo-ku, Tokyo 113-0033, Japan\label{inst86}\and
        The Institute of Statistical Mathematics, 10-3 Midori-cho, Tachikawa, Tokyo, 190-8562, Japan\label{inst87}\and
        Department of Statistical Science, The Graduate University for Advanced Studies (SOKENDAI), 10-3 Midori-cho, Tachikawa, Tokyo 190-8562, Japan\label{inst88}\and
        Kavli Institute for the Physics and Mathematics of the Universe, The University of Tokyo, 5-1-5 Kashiwanoha, Kashiwa, 277-8583, Japan\label{inst89}\and
        Leiden Observatory, Leiden University, Postbus 2300, 9513 RA Leiden, The Netherlands\label{inst90}\and
        ASTRAVEO LLC, PO Box 1668, Gloucester, MA 01931, USA\label{inst91}\and
        Applied Materials Inc., 35 Dory Road, Gloucester, MA 01930, USA\label{inst92}\and
        Institute for Astrophysical Research, Boston University, 725 Commonwealth Ave., Boston, MA 02215, USA\label{inst93}\and
        University of Science and Technology, Gajeong-ro 217, Yuseong-gu, Daejeon 34113, Republic of Korea\label{inst94}\and
        National Institute of Technology, Ichinoseki College, Takanashi, Hagisho, Ichinoseki, Iwate, 021-8511, Japan\label{inst95}\and
        Joint Institute for VLBI ERIC (JIVE), Oude Hoogeveensedijk 4, 7991 PD Dwingeloo, The Netherlands\label{inst96}\and
        CSIRO, Space and Astronomy, PO Box 76, Epping, NSW 1710, Australia\label{inst97}\and
        Department of Physics, Ulsan National Institute of Science and Technology (UNIST), Ulsan 44919, Republic of Korea\label{inst98}\and
        Department of Physics, Korea Advanced Institute of Science and Technology (KAIST), 291 Daehak-ro, Yuseong-gu, Daejeon 34141, Republic of Korea\label{inst99}\and
        Kogakuin University of Technology \& Engineering, Academic Support Center, 2665-1 Nakano, Hachioji, Tokyo 192-0015, Japan\label{inst100}\and
        Graduate School of Science and Technology, Niigata University, 8050 Ikarashi 2-no-cho, Nishi-ku, Niigata 950-2181, Japan\label{inst101}\and
        Physics Department, National Sun Yat-Sen University, No. 70, Lien-Hai Road, Kaosiung City 80424, Taiwan, R.O.C.\label{inst102}\and
        Department of Astronomy, Kyungpook National University, 80 Daehak-ro, Buk-gu, Daegu 41566, Republic of Korea\label{inst103}\and
        School of Astronomy and Space Science, Nanjing University, Nanjing 210023, People's Republic of China\label{inst104}\and
        Key Laboratory of Modern Astronomy and Astrophysics, Nanjing University, Nanjing 210023, People's Republic of China\label{inst105}\and
        INAF-Istituto di Radioastronomia, Via P. Gobetti 101, I-40129 Bologna, Italy\label{inst106}\and
        Common Crawl Foundation, 9663 Santa Monica Blvd. 425, Beverly Hills, CA 90210 USA\label{inst107}\and
        Instituto de Física, Pontificia Universidad Católica de Valparaíso, Casilla 4059, Valparaíso, Chile\label{inst108}\and
        INAF-Istituto di Radioastronomia \& Italian ALMA Regional Centre, Via P. Gobetti 101, I-40129 Bologna, Italy\label{inst109}\and
        Department of Physics, National Taiwan University, No. 1, Sec. 4, Roosevelt Rd., Taipei 106216, Taiwan, R.O.C\label{inst110}\and
        Instituto de Radioastronomía y Astrofísica, Universidad Nacional Autónoma de México, Morelia 58089, México\label{inst111}\and
        David Rockefeller Center for Latin American Studies, Harvard University, 1730 Cambridge Street, Cambridge, MA 02138, USA\label{inst112}\and
        Yunnan Observatories, Chinese Academy of Sciences, 650011 Kunming, Yunnan Province, People's Republic of China\label{inst113}\and
        Center for Astronomical Mega-Science, Chinese Academy of Sciences, 20A Datun Road, Chaoyang District, Beijing, 100012, People's Republic of China\label{inst114}\and
        Key Laboratory for the Structure and Evolution of Celestial Objects, Chinese Academy of Sciences, 650011 Kunming, People's Republic of China\label{inst115}\and
        Anton Pannekoek Institute for Astronomy, University of Amsterdam, Science Park 904, 1098 XH, Amsterdam, The Netherlands\label{inst116}\and
        Gravitation and Astroparticle Physics Amsterdam (GRAPPA) Institute, University of Amsterdam, Science Park 904, 1098 XH Amsterdam, The Netherlands\label{inst117}\and
        Center for Gravitation, Cosmology and Astrophysics, Department of Physics, University of Wisconsin–Milwaukee, P.O. Box 413, Milwaukee, WI 53201, USA\label{inst118}\and
        Deceased\label{inst119}\and
        School of Physics and Astronomy, Shanghai Jiao Tong University, 800 Dongchuan Road, Shanghai, 200240, People’s Republic of China\label{inst120}\and
        SCOPIA Research Group, University of the Balearic Islands, Dept. of Mathematics and Computer Science, Ctra. Valldemossa, Km 7.5, Palma 07122, Spain\label{inst121}\and
        Artificial Intelligence Research Institute of the Balearic Islands (IAIB), Palma 07122, Spain\label{inst122}\and
        Institut de Radioastronomie Millimétrique (IRAM), Avenida Divina Pastora 7, Local 20, E-18012, Granada, Spain\label{inst123}\and
        National Institute of Technology, Hachinohe College, 16-1 Uwanotai, Tamonoki, Hachinohe City, Aomori 039-1192, Japan\label{inst124}\and
        Research Center for Astronomy, Academy of Athens, Soranou Efessiou 4, 115 27 Athens, Greece\label{inst125}\and
        Department of Physics, Villanova University, 800 Lancaster Avenue, Villanova, PA 19085, USA\label{inst126}\and
        Physics Department, Washington University, CB 1105, St. Louis, MO 63130, USA\label{inst127}\and
        Departamento de Matemática da Universidade de Aveiro and Centre for Research and Development in Mathematics and Applications (CIDMA), Campus de Santiago, 3810-193 Aveiro, Portugal\label{inst128}\and
        School of Physics, Georgia Institute of Technology, 837 State St NW, Atlanta, GA 30332, USA\label{inst129}\and
        School of Space Research, Kyung Hee University, 1732, Deogyeong-daero, Giheung-gu, Yongin-si, Gyeonggi-do 17104, Republic of Korea\label{inst130}\and
        Canadian Institute for Theoretical Astrophysics, University of Toronto, 60 St. George Street, Toronto, ON M5S 3H8, Canada\label{inst131}\and
        Dunlap Institute for Astronomy and Astrophysics, University of Toronto, 50 St. George Street, Toronto, ON M5S 3H4, Canada\label{inst132}\and
        Canadian Institute for Advanced Research, 180 Dundas St West, Toronto, ON M5G 1Z8, Canada\label{inst133}\and
        Dipartimento di Fisica, Università di Trieste, I-34127 Trieste, Italy\label{inst134}\and
        INFN Sez. di Trieste, I-34127 Trieste, Italy\label{inst135}\and
        Department of Physics, National Taiwan Normal University, No. 88, Sec. 4, Tingzhou Rd., Taipei 116, Taiwan, R.O.C.\label{inst136}\and
        Center of Astronomy and Gravitation, National Taiwan Normal University, No. 88, Sec. 4, Tingzhou Road, Taipei 116, Taiwan, R.O.C.\label{inst137}\and
        Finnish Centre for Astronomy with ESO, University of Turku, FI-20014 Turun Yliopisto, Finland\label{inst138}\and
        Aalto University Metsähovi Radio Observatory, Metsähovintie 114, FI-02540 Kylmälä, Finland\label{inst139}\and
        Gemini Observatory/NSF NOIRLab, 670 N. A’ohōkū Place, Hilo, HI 96720, USA\label{inst140}\and
        Frankfurt Institute for Advanced Studies, Ruth-Moufang-Strasse 1, D-60438 Frankfurt, Germany\label{inst141}\and
        School of Mathematics, Trinity College, Dublin 2, Ireland\label{inst142}\and
        Julius-Maximilians-Universität Würzburg, Fakultät für Physik und Astronomie, Institut für Theoretische Physik und Astrophysik, Lehrstuhl für Astronomie, Emil-Fischer-Str. 31, D-97074 Würzburg, Germany\label{inst143}\and
        Department of Physics, University of Toronto, 60 St. George Street, Toronto, ON M5S 1A7, Canada\label{inst144}\and
        Department of Physics, Tokyo Institute of Technology, 2-12-1 Ookayama, Meguro-ku, Tokyo 152-8551, Japan\label{inst145}\and
        Hiroshima Astrophysical Science Center, Hiroshima University, 1-3-1 Kagamiyama, Higashi-Hiroshima, Hiroshima 739-8526, Japan\label{inst146}\and
        Aalto University Department of Electronics and Nanoengineering, PL 15500, FI-00076 Aalto, Finland\label{inst147}\and
        Jeremiah Horrocks Institute, University of Central Lancashire, Preston PR1 2HE, UK\label{inst148}\and
        National Biomedical Imaging Center, Peking University, Beijing 100871, People’s Republic of China\label{inst149}\and
        College of Future Technology, Peking University, Beijing 100871, People’s Republic of China\label{inst150}\and
        Tokyo Electron Technology Solutions Limited, 52 Matsunagane, Iwayado, Esashi, Oshu, Iwate 023-1101, Japan\label{inst151}\and
        Department of Physics and Astronomy, University of Lethbridge, Lethbridge, Alberta T1K 3M4, Canada\label{inst152}\and
        Netherlands Organisation for Scientific Research (NWO), Postbus 93138, 2509 AC Den Haag, The Netherlands\label{inst153}\and
        Frontier Research Institute for Interdisciplinary Sciences, Tohoku University, Sendai 980-8578, Japan\label{inst154}\and
        Astronomical Institute, Tohoku University, Sendai 980-8578, Japan\label{inst155}\and
        Department of Physics and Astronomy, Seoul National University, Gwanak-gu, Seoul 08826, Republic of Korea\label{inst156}\and
        SNU Astronomy Research Center, Seoul National University, Gwanak-gu, Seoul 08826, Republic of Korea\label{inst157}\and
        ASTRON, Oude Hoogeveensedijk 4, 7991 PD Dwingeloo, The Netherlands\label{inst158}\and
        University of New Mexico, Department of Physics and Astronomy, Albuquerque, NM 87131, USA\label{inst159}\and
        Centre for Mathematical Plasma Astrophysics, Department of Mathematics, KU Leuven, Celestijnenlaan 200B, B-3001 Leuven, Belgium\label{inst160}\and
        Physics Department, Brandeis University, 415 South Street, Waltham, MA 02453, USA\label{inst161}\and
        Tuorla Observatory, Department of Physics and Astronomy, University of Turku, FI-20014 Turun Yliopisto, Finland\label{inst162}\and
        Radboud Excellence Fellow of Radboud University, Nijmegen, The Netherlands\label{inst163}\and
        School of Natural Sciences, Institute for Advanced Study, 1 Einstein Drive, Princeton, NJ 08540, USA\label{inst164}\and
        School of Physics, Huazhong University of Science and Technology, Wuhan, Hubei, 430074, People's Republic of China\label{inst165}\and
        Mullard Space Science Laboratory, University College London, Holmbury St. Mary, Dorking, Surrey, RH5 6NT, UK\label{inst166}\and
        Center for Astronomy and Astrophysics and Department of Physics, Fudan University, Shanghai 200438, People's Republic of China\label{inst167}\and
        Astronomy Department, University of Science and Technology of China, Hefei 230026, People's Republic of China\label{inst168}\and
        Department of Physics and Astronomy, Michigan State University, 567 Wilson Rd, East Lansing, MI 48824, USA\label{inst169}\and
        Royal Netherlands Meteorological Institute, Utrechtseweg 297, 3731 GA, De Bilt, The Netherlands\label{inst170}
}